\pdfoutput=1
\documentclass[]{jkpaper}
\usepackage{tikz}
\usetikzlibrary{decorations.markings}
\usetikzlibrary{decorations.pathreplacing}
\usetikzlibrary{arrows.meta}
\usetikzlibrary{patterns}
\usetikzlibrary{decorations.pathmorphing}

\usepackage{aas_macros}
\usepackage{soul}
\usepackage{accents}
\usepackage{simplewick}
\usepackage{cancel}
\usepackage{cellspace}
\usepackage{placeins}
\usepackage{tikz-cd}

\newtheoremstyle{spaced}
{.8em}   
{.2em}   
{\itshape} 
{}        
{\bfseries} 
{.}       
{.5em}    
{}        

\theoremstyle{spaced}

\newcommand{\be}{
\begin{equation}}
  \newcommand{\ee}{
\end{equation}}
\newcommand{\baa}{
\begin{align}}
  \newcommand{\eaa}{
\end{align}}
\newcommand{\bea}{
\begin{eqnarray}}
  \newcommand{\eea}{
\end{eqnarray}}
\newcommand{\pa}{\partial}

\newcommand{\rd}{\mathrm{d}}

\newcommand{\GN}{\ensuremath{G_{\text{N}}}}

\newcommand{\Diff}{\ensuremath{\operatorname{Diff}}}

\newcommand{\SLTR}{\ensuremath{\operatorname{SL}(2,\RR)}}
\newcommand{\Greens}[2]{\mathcal{G}(#1,#2)}

\makeatletter
\newcommand{\normord@delim}{%
  \mspace{1mu}%
  \vphantom{|}
  \smash{%
    \vcenter{%
      \offinterlineskip
      \hbox{\scalebox{0.55}{$\scriptscriptstyle\bullet$}}%
      \vskip .7ex
      \hbox{\scalebox{0.55}{$\scriptscriptstyle\bullet$}}%
    }%
  }%
  \mspace{1mu}%
}

\newcommand{\normord}[1]{%
  \vphantom{#1}\mathord{%
    \vphantom{|}
    \smash{%
      \mathinner{%
        \mathopen{\normord@delim}%
        \mkern1mu#1\mkern1mu%
        \mathclose{\normord@delim}%
      }%
    }%
  }%
}
\newcommand{\revnormord@delim}{%
  \mspace{1mu}%
  \vphantom{|}%
  \smash{%
    \vcenter{%
      \offinterlineskip
      \hbox{\scalebox{0.8}{$\scriptscriptstyle\boldsymbol{\times}$}}%
      \vskip .2ex
      \hbox{\scalebox{0.8}{$\scriptscriptstyle\boldsymbol{\times}$}}%
    }%
  }%
  \mspace{1mu}%
}

\newcommand{\revnormord}[1]{%
  \vphantom{#1}\mathord{%
    \vphantom{|}%
    \smash{%
      \mathinner{%
        \mathopen{\revnormord@delim}%
        \mkern0.6mu#1\mkern0.6mu%
        \mathclose{\revnormord@delim}%
      }%
    }%
  }%
}

\newcommand{\partcovnormord@delim}{%
  \mspace{1.5mu}%
  \vphantom{|}%
  \smash{%
    \vcenter{%
      \offinterlineskip
      \hbox{\scalebox{1.1}{$\scriptscriptstyle\star$}}%
      \vskip .35ex
      \hbox{\hspace*{0.21ex}\scalebox{0.55}{$\scriptscriptstyle\bullet$}}%
      \vskip .05ex
    }%
  }%
  \mspace{1.5mu}%
}

\newcommand{\partcovnormord}[1]{%
  \vphantom{#1}\mathord{%
    \vphantom{|}%
    \smash{%
      \mathinner{%
        \mathopen{\partcovnormord@delim}%
        #1%
        \mathclose{\partcovnormord@delim}%
      }%
    }%
  }%
}

\newcommand{\covnormord@delim}{%
  \mspace{1.5mu}%
  \vphantom{|}%
  \smash{%
    \vcenter{%
      \offinterlineskip
      \hbox{\scalebox{1.1}{$\scriptscriptstyle\star$}}%
      \vskip .35ex
      \hbox{\scalebox{1.1}{$\scriptscriptstyle\star$}}%
      \vskip .05ex
    }%
  }%
  \mspace{1.5mu}%
}

\newcommand{\covnormord}[1]{%
  \vphantom{#1}\mathord{%
    \vphantom{|}%
    \smash{%
      \mathinner{%
        \mathopen{\covnormord@delim}%
        #1%
        \mathclose{\covnormord@delim}%
      }%
    }%
  }%
}
\makeatother

\newcommand{\leftdelta}{\ensuremath{\accentset{\leftarrow}{\delta}}}
\newcommand{\rightdelta}{\ensuremath{\accentset{\rightarrow}{\delta}}}

\newcommand{\db}[2]{\pb{#1}{#2}_{\mathrm{D}}}

\usepackage{expl3,xparse}

\ExplSyntaxOn
\NewDocumentCommand{\Operator}{m}
{
  \tl_if_single:nTF { #1 }
  { \hat{#1} }
  { \widehat{#1} }
}

\NewDocumentCommand{\OperatorGaugeFixed}{m}
{
  \tl_if_single:nTF { #1 }
  { \hat{\bm{#1}} }
  { \widehat{\bm{#1}} }
}
\ExplSyntaxOff

\newcommand{\mv}{\mathrm{v}}

\makeatletter
\newcommand\footnoteref[1]{\protected@xdef\@thefnmark{\ref{#1}}\@footnotemark}
\makeatother

\titleformat{\paragraph}
{\normalfont\normalsize\bfseries}{\theparagraph}{1em}{}
\titlespacing*{\paragraph}
{0pt}{3.25ex plus 1ex minus .2ex}{1.5ex plus .2ex}

\newcommand{\Schwarzian}[2]{\left\{#1,#2\right\}}

\newcommand{\Segment}{\mathcal{I}}

\newcommand{\Vv}{\bar{V}}
\newcommand{\Dstar}{\mathbin{\star_{\mkern1mu\Dmap}}}
\newcommand{\hatstar}{\mathbin{\hat\star}}
\newcommand{\Cov}[1]{#1^{\star}}
\newcommand{\Covrad}[1]{#1^{\text{rad}\star}}

\newcommand{\rad}{\text{rad}}

\newcommand{\gf}[1]{#1_{\text{gf}}}

\newcommand{\Dmap}{\mathrm{D}}
\newcommand{\GFmap}{\mathrm{GF}}

\newcommand{\rec}[1]{\check{#1}}

\newauthornote{\jjvk}{JJVK}{red!80}

\title{Gravitational null rays: Covariant\texorpdfstring{\\}{ }Quantization and the Dressing Time }
\author{Laurent Freidel\texorpdfstring{\textsuperscript{a}}{} and Josh Kirklin\texorpdfstring{\textsuperscript{b}}{}}
\institution{Perimeter Institute for Theoretical Physics,\texorpdfstring{\\}{ }31 Caroline Street North, Waterloo, ON, N2L 2Y5, Canada}

\email{
  \textsuperscript{a}\emaillink{lfreidel@pitp.ca}\\
  \textsuperscript{b}\emaillink{jkirklin@pitp.ca}
}

\abstr{
  We quantize the degrees of freedom on a gravitational null ray segment in a fully gauge-invariant manner by using the dressing time as a quantum reference frame (QRF). Our work goes beyond previous models in that the QRF we employ is made out of the gravitational field itself, and accounts for the full group of diffeomorphisms along the ray, not just a locally compact subgroup. The key tool we introduce is covariant normal ordering, a QRF-dependent but background-independent renormalization prescription that restores diffeomorphism covariance at the quantum level. This enables the definition of a quantum dressing map whose image is the algebra of gauge-invariant observables. We find that this algebra carries the structure of a Virasoro crossed product, and that the dressing map induces a deformed product on gauge-fixed operators which can be understood as a quantization of the Dirac bracket, with consequences for the fluctuations of observables. We explain how to cancel anomalies in the physical Hilbert space representation of the gauge-invariant algebra by including a deformation at the classical level, thereby eliminating all spurious degrees of freedom at the quantum level. The physical Hilbert space admits a Page--Wootters reduction map to the perspective of the dressing time, and we show that the dressing time is non-ideal in the sense that its distinct coherent states have non-vanishing overlaps governed by the Teo-Takhtajan energy, i.e.\ the K\"ahler potential for Virasoro coadjoint orbits.
}

\begin{document}
\maketitleandtoc

\section{Introduction}

Quantum reference frames (QRFs)~\cite{Aharonov1984,Rovelli1991a,Bartlett2007,Girelli_2008,Palmer_2014,Hoehn2019,Hoehn2020,Hoehn2021,Hoehn2021a,Hoehn2022,Hoehn2023,Angelo2011,Loveridge2019,Hamette2020,Hamette2021} have rapidly become established as a powerful conceptual basis for understanding locality, subsystems, and information in quantum gravity. A QRF is a collection of degrees of freedom relative to which one measures other degrees of freedom, and an appropriate choice of QRF allows for a diffeomorphism-invariant definition of local observables~\cite{Isham1985,Giddings_2006,Marolf_2015,Donnelly:2016rvo,Donnelly:2016auv,Donnelly2016,brunetti2015quantumgravitypointview,Gambini2001,Husain_2012,Gary:2006mw,Baldazzi2022,Dittrich_2017,Donnelly2017,Giddings_2019,CastroRuiz2020,Giddings:2025xym,Kuchar2011,Thiemann2006,isham1995structuralissuesquantumgravity,Giesel_2010,Giesel:2020raf,Giesel:2024xtb}. This enables a consistent understanding of gravitational operators and generalized entropies in terms of crossed product algebras~\cite{CLPW,Jensen2023,Fewster2025,DeVuyst:2024pop,DeVuyst:2024uvd,kirklin2024generalisedsecondlawsemiclassical,KudlerFlam2025,Klinger2024,Klinger2024,Klinger:2026tws,KudlerFlam2025a,Faulkner2024,Witten2022,AliAhmad2024a,AliAhmad2024}. Taking the quantum properties of the QRF (such as its fluctuations and the way in which it is entangled) into account allows one to explore gravitational physics beyond the semiclassical regime~\cite{kirklin2024generalisedsecondlawsemiclassical}.

Up to now, progress has mostly been made with simplified models where the group of gauge symmetries is semisimple or locally compact, and the QRFs are simple entities such as clocks added externally to the system of study and modelled with quantum mechanical Hilbert spaces (e.g.~\cite{CLPW,DeVuyst:2024pop,kirklin2024generalisedsecondlawsemiclassical}). Such QRFs only account for tiny subgroups of the diffeomorphism gauge group of gravity; for instance, a clock accounts for a 1-dimensional time evolution subgroup. In this paper, we go beyond such toy models by using components of the dynamical fields themselves as a QRF, and thereby account for the full diffeomorphism group of the subsystem under consideration.

There are numerous technical challenges associated with this. The fields typically interact in complicated ways, while QRFs are best understood in the non-interacting setting. Quantum mechanical QRFs are typically Schr\"odinger-quantized (meaning they are quantized in terms of square-integrable configuration space wavefunctions). But in order to obtain a separable Hilbert space with finite energy states, fields must be K\"ahler-quantized (meaning they are quantized in terms of holomorphic phase space wavefunctions)~\cite{Woodhouse:1980pa,KahlerQRF}. The main example of this is Fock quantization, with a split into positive and negative frequency modes~\cite{Birrell:1982ix}. Because there are infinitely many modes, well-defined composite field operators must be renormalized~\cite{Polchinski,Hollands:2023txn}, for example with normal ordering.
However, ordinary normal ordering is not preserved by diffeomorphisms -- the positive/negative frequency modes of the fields are mixed by Bogoliubov transformations. Moreover, the choice of vacuum state (annihilated by positive frequency modes) explicitly breaks the diffeomorphism symmetry. This typically leads to anomalies in the diffeomorphism gauge constraints \cite{Polchinski, Bowick:1986rc, Bowick:1987pw, Klinger:2025hjp}. A related point is that the diffeomorphism gauge group is highly non-locally compact, which makes constructing an appropriate invariant measure very subtle \cite{Klinger:2025tvg}. Such invariant measures have played an important role in approaches to QRFs transforming under locally compact gauge groups, for example in the G-twirl used to construct relational observables, and in the group averaging procedure used to define physical Hilbert spaces~\cite{Giulini:1998kf,Kaplan2025,CLPW}.

The subsystem we focus on in this paper corresponds to a segment of a null ray, for which we will show how to solve all the challenges listed above.
In~\cite{Ciambelli:2023mir,Ciambelli_2024} we constructed the phase space of null surfaces and set up the framework of their quantization using the important property of ultralocalization inherent to Hadamard quantization~\cite{kay1991theorems, Schroer:2010bjq, Wall:2011hj, Jensen2023}. These works built on the construction of constraints~\cite{Sachs:1962zzb, Torre:1985rw, Gourgoulhon_2006,Reisenberger:2007pq,Reisenberger:2012zq,Reisenberger:2018xkn,Mars:2022gsa} and canonical structure on horizons~\cite{Parattu:2015gga, Lehner:2016vdi, Hopfmuller:2016scf, DePaoli:2017sar, Wieland:2017zkf, Hopfmuller:2018fni, Chandrasekaran:2018aop, Donnay:2019jiz, Chandrasekaran2021, Adami:2021nnf, Sheikh-Jabbari:2022mqi, Chandrasekaran:2021hxc, Adami:2021nnf, Sheikh-Jabbari:2022mqi, Ciambelli:2023mir,Chandrasekaran:2023vzb, ,Odak:2023pga, Rignon-Bret:2024wlu, Wieland:2024dop} and the recent understanding of their Carrollian geometric properties~\cite{Freidel:2022vjq,Freidel:2024emv,ciambelli2025foundationscarrolliangeometry}.
In~\cite{LocalizationAnomalous}
we described in detail the classical structure of the null ray subsystem, its localized gauge invariant observables and the classical anomaly functional. Here we quantize that structure.
In many ways, the system we study here is the \emph{simplest} possible non-toy gravitational subsystem one could consider, i.e.\ the `harmonic oscillator' of quantum gravity. The methods we establish provide a foundation for a broader understanding of field-based QRFs in more complicated subsystems. See also~\cite{Hoehn2024,fewster2020quantumfieldslocalmeasurements, Gary:2006mw,kaplan2025sitterquantumgravityemergence,Kolchmeyer:2024fly,Kabel2023,Fewster:2025ijg} for other approaches to using quantum fields as QRFs, and~\cite{Wieland:2024dop,Wieland:2025qgx} for a different approach to quantizing null gravitational surfaces.

The gauge group on the null ray is the set of its orientation-preserving diffeomorphisms $\Diff^+(\RR)$. The classical version of a quantum reference frame for such a group is a \emph{dressing field}~\cite{Donnelly:2016auv, Freidel:2020ayo,Freidel:2020svx, Freidel:2021dxw, Ciambelli:2021nmv,Carrozza2022,Carrozza2024,Goeller:2022rsx,Freidel:2020xyx,LocalizationAnomalous,Francois:2024rdm}, which is a field $V$ transforming like $V\to V\circ F$ under a diffeomorphism $F$. There is a very convenient and physically well-motivated choice of dressing field on the null ray, called the `dressing time'~\cite{Ciambelli:2023mir,Ciambelli_2024,LocalizationAnomalous}; this is the one we use here. One of the main reasons it is convenient is that in a linearized perturbative regime its interactions with the rest of the fields can be neglected (in future work, we would like to understand how to reintroduce interactions and go beyond this regime). In the classical theory, equipped with $V$ and given another field $\phi$, we can construct the \emph{dressed field} $\tilde\phi:= V_*\phi$, i.e.\ the pushforward of $\phi$ through $V$. The dressed field is gauge-invariant, and it is localized with respect to the dressing time reference frame.

To quantize this description one must quantize functionals of the dressed fields $\tilde\phi$. Such functionals are composite operators generally, and so must be renormalized. Doing this improperly can lead to anomalies. For example, the simplest kind of renormalization is normal ordering. But the normal ordered dressed field $\normord{\tilde\phi}$ typically \emph{breaks} gauge-invariance. The reason for this is that normal ordering is defined by splitting the fields into their positive and negative frequency components with respect to a background time, or equivalently with respect to a choice of vacuum state. The introduction of this background structure is what undermines gauge-invariance.

To counter this, we replace this background structure dependent renormalization with a renormalization defined relative to a quantum reference frame.
In particular, we introduce the main technical tool of the paper: \emph{covariant normal ordering}, which differs from ordinary normal ordering in that it does not have any dependence on a background time, and instead can be thought of as normal ordering with respect to the dressing time $V$. The covariantly normal ordered dressed field, denoted $\covnormord{\tilde\phi}$, is gauge-invariant in the quantum theory.

The kinematical fields on the null ray decompose into two independent sectors: the so-called `spin 0' sector, consisting of the dressing time $V$ itself and the area element $\Omega$ of the null surface, and the `radiative' sector, consisting of matter and spin 2 gravitational radiation through the null surface. In the classical theory, we showed in~\cite{LocalizationAnomalous} that the dressed area $\tilde\Omega = V_*\Omega$ (or more precisely its second derivative, denoted $\tilde\tau$), generates an action of the diffeomorphism group $\Diff^+(\RR)$. This action corresponds to `reorientations' of the dressing time reference frame \cite{Donnelly:2016auv}; it is distinct from and commutes with the gauge transformation action of $\Diff^+(\RR)$. Indeed, reorientations act on the left of the dressing time as $V\to F^{-1}\circ V$, whereas gauge transformations act through the right action $V\to V\circ F$.

It is relatively simple to covariantly normal order the dressed radiative fields, because the radiative fields commute with the dressing time. But covariant normal ordering of dressed spin 0 fields such as the dressed area $\tilde\Omega = V_*\Omega$ is more complicated, because the area $\Omega$ does not commute with $V$. We provide in this paper a definition of covariant normal ordering for arbitrary observables depending on both the spin 0 and radiative fields.

We find that the covariantly normal ordered dressed spin 0 fields generate a representation of the Virasoro group, i.e.\ the unique central extension of $\Diff^+(\RR)$. Thus, the reorientation group is promoted from $\Diff^+(\RR)$ to $\operatorname{Virasoro}$ in the quantum theory. Furthermore, the algebra of gauge-invariant operators on the null ray is the crossed product~\cite{Connes1994,Takesaki2003II,Takesaki2003III} of radiative operators by the reorientations:
\begin{equation}
  \mathcal{A}_{\text{rad}}\rtimes \operatorname{Vir}.
\end{equation}
This matches with the general structure for the algebra of operators of a system relative to a QRF found in many other contexts~\cite{CLPW,Jensen2023,Fewster2025,DeVuyst:2024pop,DeVuyst:2024uvd,kirklin2024generalisedsecondlawsemiclassical,KudlerFlam2025,Klinger2024,Klinger2024,Klinger:2026tws,KudlerFlam2025a,Faulkner2024,Witten2022}, except that the classical symmetry algebra possesses an anomaly at the quantum level. Here, the `system' is the radiative sector, and the $\mathcal{A}_{\text{rad}}$ factor above accounts for dressed radiative operators.

Classically, the construction of dressed observables can be formalized in terms of a dressing map $\Dmap_{\text{cl}}$, which maps a gauge-fixed classical observable $O$ to its gauge-invariant dressed counterpart $\tilde O=\Dmap_{\text{cl}}(O)$~\cite{LocalizationAnomalous} (for example $\Dmap_\text{cl}(\phi)=\tilde\phi$). Our results allow us to directly import this procedure into the quantum theory by defining the quantum dressing map $\Dmap$ as the combination of (1) a change from ordinary normal ordering to covariant normal ordering, and (2) the classical dressing map:
\begin{equation}
  \Dmap(\normord{O}) = \covnormord{\Dmap_{\text{cl}}(O)}.
\end{equation}
The covariance of covariant normal ordering, and the classical gauge-invariance of the classical dressed observable, implies the quantum gauge-invariance of the quantum dressed observable $\Dmap(\normord{O})$. In this way, our formalism gives a precise quantization of classical dressing in field theory, consistent with renormalization. Note that $\Dmap(\normord{O})$ is defined without ever needing to perform any (possibly quite subtle) integrals over the non-locally compact group $\Diff^+(\RR)$, unlike in some other approaches (e.g.~\cite{Klinger:2026tws}). All gauge-invariant classical observables can be obtained by classical dressing, and similarly all gauge-invariant quantum operators can be obtained by our quantum dressing procedure.

The dressing map can be viewed as defining a deformation of the ordinary product of gauge-fixed operators in the following sense. The composition of two dressed operators may always be written as the dressing of a third operator, which defines a star product $\Dstar$ via
\begin{equation}
  \Dmap(\normord{O_1})\Dmap(\normord{O_2})=\Dmap(\normord{O_1}\Dstar\normord{O_2}).
  \label{Equation: intro Dstar}
\end{equation}
As we explain in the paper, this deformation is the quantum analogue of the classical replacement of the Poisson bracket by the Dirac bracket, and fundamentally modifies the fluctuations of quantum operators, with potential phenomenological consequences.

To complete the quantization of the physical degrees of freedom, it remains to represent the gauge-invariant dressed operators on a physical Hilbert space $\mathcal{H}_{\text{phys}}$ satisfying the gauge constraint, which in this case is the Raychaudhuri equation. Here, we construct $\mathcal{H}_{\text{phys}}$ as the vacuum GNS representation~\cite{Haag:1996hvx} of the gauge-invariant operators. In order to remove spurious degrees of freedom, one must cancel the anomaly that appears in the Raychaudhuri constraint; here we do this by introducing a classical counterterm of the kind described in~\cite{LocalizationAnomalous}. This forces the central charge of the reorientation Virasoro representation to be equal to $c=M$, where $M$ is the number of radiative fields. We view our approach to imposing gauge-invariance on the quantum theory (in terms of dressed degrees of freedom) as a more physically intuitive alternative to other methods (such as the BRST formalism), and one that emphasizes the role played by the reference frame.

We provide some explanation of our notation in Subsection~\ref{Subsection: notation}, and a summary of our technical results in Subsection~\ref{Section: technical summary}. The rest of the paper is then structured as follows.

In Section~\ref{Section: kinematical quantization}, we recall the structure of the kinematical classical degrees of freedom on a gravitational null ray segment, and carry out their canonical quantization. We also describe the unitary operators implementing a projective representation of gauge transformations on the resulting kinematical space of states. Then, we describe in Section~\ref{Section: crossed product} the anomalous properties of operators in the quantum theory, explain how one must cancel these anomalies with an appropriate quantization prescription, and give more details on the Virasoro crossed product structure of the algebra of operators obtained in this way. We call the quantization prescription `covariant normal ordering', and describe it in Section~\ref{Section: covariant normal ordering}. In particular, we explain how ordinary normal ordering relies on a choice of background time (which is the origin of the anomalous behavior), and that our prescription gets rid of this background structure.
Combining classical dressing with covariant normal ordering allows us to define a non-anomalous quantum dressing map in Section~\ref{Section: dressed operators}, where we also give important examples of the resulting quantum dressed operators, e.g.\ the dressed area.
In Section~\ref{Section: three quantum diffeomorphisms}, we study the properties of the three important diffeomorphism actions identified in~\cite{LocalizationAnomalous} in the quantum theory: reparametrizations (gauge transformations), reorientations, and dressed reparametrizations.
The physical quantization is completed in Section~\ref{Section: physical representation}, where we construct the vacuum GNS representation of the dressed operators, and discuss how our setup connects to the perspective-neutral picture of QRFs.
We conclude in Section~\ref{Section: conclusion} with some discussion of open directions.

We also include some appendices. Appendix~\ref{Subsection: reparametrization algebra} provides a standard computation of the stress tensor bracket and central charge for reparametrizations. Appendix~\ref{Appendix: cocycle} provides an efficient derivation of the Virasoro cocycle arising from the exponentiation of the quantum stress tensors that appear in this paper, and of the K\"ahler potential that corresponds to the overlap of coherent states for the dressing time QRF. Appendix~\ref{Appendix: more on covariant normal ordering} gives a complete definition of covariant normal ordering (we relegate this to an appendix due to its technical nature; in the main text of the paper we only need the details given in Section~\ref{Section: covariant normal ordering}).

\subsection{Notation}
\label{Subsection: notation}
\begin{samepage}
  Before proceeding, let us provide a reference for our notation and conventions.
  Some important general rules are as follows:
  \begin{itemize}
    \item Unless otherwise stated, classical observables are written explicitly as functionals depending on the fields, e.g.\ $O[\Phi]$.
    \item This is to distinguish them from the corresponding quantum operators $O$, which we write without this dependence. In particular, $O=\normord{O[\Phi]}$ denotes the normal ordered quantization of $O[\Phi]$.
    \item A tilde $\sim$ denotes a classically gauge-invariant object. For example, $\tilde O[\Phi]$ is a gauge-invariant classical observable. Note that anomalies lead to violations of gauge-invariance in the normal-ordered quantum operator $\tilde O$ (see Section~\ref{Section: crossed product}).
    \item A superscript star $\Cov{}$ on a quantum operator denotes that it is obtained by \emph{covariantly} normal ordering the classical observable (see Section~\ref{Section: covariant normal ordering}). Note that $\Cov{\tilde O}$ is quantum gauge-invariant.
    \item We often use $v$ to denote a background time coordinate, and $\mv$ to denote a gauge-fixed time coordinate (e.g.\ in the target space of the dressing time $V$).
  \end{itemize}
  Other key details may be found in Table~\ref{Table: notation}.
\end{samepage}


\begin{table}[ht]
  \centering
  \setlength\cellspacetoplimit{0.75em}
  \setlength\cellspacebottomlimit{0.75em}

  \begin{tabular}{S{c}m{0.85\linewidth}}
    \toprule\addlinespace\addlinespace
    $\Phi$ & Shorthand for the collection of fields on the null ray: the spin 0 fields $\tau,V$, and the radiative fields $\varphi_i$.\\
    $\tilde{\Phi}$ & Shorthand for $X\triangleright{\Phi}$ the dressed fields, where $X=V^{-1}$ and $\triangleright$ denotes the diffeomorphism action, for instance.\\
    $U[F]$ & The unitary operator implementing a reparametrization $F\in\operatorname{Diff}^+(\RR)$. Similarly, $U^0[F],U^\text{rad}[F]$ are reparametrization operators restricted to the spin 0 and radiative fields.\\
    $O[\Phi]$ & A classical observable, i.e.\ a functional of the fields $\Phi$. \\
    $\normord{O[\Phi]}$ & The quantum operator obtained by ordinary normal ordering of the classical observable $O[\Phi]$, i.e.\ with positive frequency parts of the fields on the right, and negative frequency parts of the fields on the left. \\
    $O$ & Shorthand for $\normord{O[\Phi]}$. \\
    $\covnormord{O[\Phi]}$ & The quantum operator obtained by covariant normal ordering of the classical observable $O[\Phi]$. Covariant normal ordering is defined in Section~\ref{Section: covariant normal ordering} and Appendix~\ref{Appendix: more on covariant normal ordering}. \\
    $\Cov{O}$ & Shorthand for $\covnormord{O[\Phi]}$.  \\
    $\tilde O[\Phi]$ & A gauge-invariant classical observable. \\
    $\Cov{\tilde O}$ & Shorthand for $\covnormord{\tilde O[\Phi]}$. This is the most general quantum gauge-invariant observable, $U[F]\Cov{\tilde O}U[F^{-1}]=\Cov{\tilde O}$. \\
    $*$ & The product associated with ordinary normal ordering, i.e.\ $\normord{A*B[\Phi]} = AB$. \\
    $\star$ & The product associated with covariant normal ordering, i.e.\ $\covnormord{A\star B[\Phi]} = \Cov{A}\Cov{B}$. \\
    $\hatstar$ & The product relating ordinary and covariant normal ordering, i.e.\ $\Cov{A}\Cov{B}=\Cov{(A\hatstar B)}$.\\
    $\Dstar$ & The product associated with dressed operators,\ i.e.\ $\Dmap(A)\Dmap(B) = \Dmap(A\Dstar B)$. \\
    \bottomrule
  \end{tabular}
  \caption{A summary of the notation used throughout the paper.}
  \label{Table: notation}
\end{table}

\subsection{Synopsis of technical results}
\label{Section: technical summary}

We end the introduction with a whirlwind summary of our technical results.
\begin{itemize}
  \item \textbf{Quantization of kinematical null ray segment.} We perform a canonical quantization of the kinematical structure obtained in~\cite{LocalizationAnomalous}, obtaining a space of kinematical states
    \begin{equation}
      \mathcal{K}_{\text{kin}} = \mathcal{K}^0\otimes\mathcal{H}^\text{rad}\otimes\mathcal{H}^\text{edge}.
    \end{equation}
    This space is equipped with an indefinite Hermitian inner product. The three tensor factors correspond to spin 0, edge mode, and radiative degrees of freedom respectively.
  \item \textbf{Quantum reparametrizations and their cocycle.} We show that the normal ordered Raychaudhuri stress tensor $T=\normord{T}$ generates a projective (pseudo-)unitary representation of the gauge group $\Diff^+(\RR)$ on $\mathcal{K}_{\text{kin}}$. In other words it gives a Virasoro representation at a certain central charge $c_T=2+M$, where $M$ is the number of radiative fields. We exponentiate this representation, finding (in Appendix~\ref{Appendix: cocycle}) the explicit group cocycle that appears (which we write in terms of the usual representative of the Bott-Thurston cocycle~\cite{bott1977characteristic, oblak2017berry,Alekseev:2022efp} and a coboundary related to integrals of the Kirillov-Kostant-Souriau potential~\cite{Guillemin:1990ew,Woodhouse:1980pa}).
  \item \textbf{Background dependence in normal ordering.} Normal ordered operators $\normord{O[\Phi]}$ are subject to diffeomorphism anomalies, and we find a precise general formula for the anomalous transformation law. This takes the form
    \begin{equation}
      U[F]\normord{O[\Phi]}U[F]^\dagger = \normord{e^{\hbar \mathscr{A}[F]}(F\triangleright O[\Phi])}
      \label{Equation: technical summary A}
    \end{equation}
    where $U[F]$ is the operator implementing the quantum reparametrization $F\in\Diff^+(\RR)$, and $F\,\triangleright$ is the classical action of $F$. Crucially, $\mathscr{A}[F]$ is a second order differential operator in the fields which produces the anomalous behavior in the quantum theory (and which we give explicitly in~\eqref{Equation: normal ordering anomaly}). It arises from the dependence of normal ordering on background structures.
  \item \textbf{Covariant normal ordering.} We define a new notion of normal ordering which we call \emph{covariant normal ordering} $\covnormord{O[\Phi]}$, which we show satisfies background-independent covariance:
    \begin{equation}
      U[F]\covnormord{O[\Phi]}U[F]^\dagger = \covnormord{F\triangleright O[\Phi]}.
    \end{equation}
    We find formulae relating covariant normal ordering to ordinary normal ordering, of the form
    \begin{equation}
      \covnormord{O[\Phi]} = \normord{\mathfrak{N}(O[\Phi])},
    \end{equation}
    where $\mathfrak{N}$ is an operator acting on the classical fields. This operator may be decomposed as $\mathfrak{N}=\mathfrak{N}^0\circ \mathfrak{N}^\text{rad}$, where $\mathfrak{N}^\text{rad}$, $\mathfrak{N}^0$ should be understood as carrying out covariant normal ordering on the radiative and spin 0 fields respectively. The former $\mathfrak{N}^\text{rad}$ may be written as an exponential of a second order differential operator~\eqref{Equation: covariant normal order varphi}, and essentially normal orders the radiative fields with respect to the dressing time. The latter $\mathfrak{N}^0$ is somewhat more complicated but is given in Appendix~\ref{Appendix: more on covariant normal ordering}; in particular it is determined by the map $\mathcal{D}$ given in~\eqref{Equation: spin 0 covnormord operator}, or in the presence of a classical central charge (see below and Subsection~\ref{Section: anomaly shift}) the map $\mathcal{D}_{c_{\text{cl}}}$ given in~\eqref{gentau}. For most of the paper, we only need to know how to covariantly normal order operators to $\order{\hbar}$, for which a formula is given in~\eqref{Equation: covariant normal ordering linear}.
  \item \textbf{Dressing map for quantum observables.}
    Equipped with covariant normal ordering, we define a dressing map $\Dmap$ on quantum operators which produces gauge-invariant quantum operators. The dressing procedure induces a deformed product $\Dstar$ at the gauge-fixed level~\eqref{Equation: intro Dstar}. We show that the classical limit of the $\Dstar$-commutator reproduces the Dirac bracket found in~\cite{LocalizationAnomalous}:
    \begin{equation}
      \lim_{\hbar\to 0}\frac1{i\hbar}\comm{\mathcal{O}_1}{\mathcal{O}_2}_{\Dstar} = \db{\mathcal{O}_1}{\mathcal{O}_2},
    \end{equation}
    where
    \begin{equation}
      \comm{\mathcal{O}_1}{\mathcal{O}_2}_{\Dstar}:= \mathcal{O}_1\Dstar\mathcal{O}_2 - \mathcal{O}_2\Dstar\mathcal{O}_1.
    \end{equation}
    Hence, quantum dressed operators provide a quantization of the classically reduced phase space.

    Additionally, as explained above, the dressed/gauge-invariant operators form a crossed product $\mathcal{A}_{\text{rad}}\rtimes \operatorname{Vir}$ of the radiative degrees of freedom by the Virasoro group. There is a simple formula~\eqref{Equation: dressed not Pi dependent} for the dressing map applied to radiative operators which shows that it is an isomorphism restricted to them, but the dressing of spin 0 operators is more complicated.
  \item \textbf{Quantum reorientations and reparametrizations.}
    We show that the quantum dressed stress tensor of the spin 0 fields
    \begin{equation}
      \Cov{\tilde\tau} = \covnormord{\tilde\tau} = \Dmap(\tau)
    \end{equation}
    generates a Virasoro representation at central charge $c_{\tilde\tau}=24$, which we refer to as `reorientations'. This is the Virasoro action which contributes to the crossed product structure $\mathcal{A}_{\text{rad}}\rtimes \operatorname{Vir}$. Also, we show that the quantum dressed stress tensor of \emph{all} the fields, i.e.\ the dressed Raychaudhuri constraint
    \begin{equation}
      \Cov{\tilde T} = \covnormord{\tilde T} = \Dmap(T),
    \end{equation}
    generates a Virasoro representation at central charge $c_{\tilde T}=24-M$, which we refer to as `dressed reparametrizations'. Reorientations and reparametrizations have the same cocycle structure described in Appendix~\ref{Appendix: cocycle}.
  \item \textbf{Shifting the central charges.} By introducing a deformation of the theory parametrized by a classical central charge $c_{\text{cl}}$ (of the kind previously described in~\cite{LocalizationAnomalous}), we show that one can shift the central charges of the three important Virasoro representations as
    \begin{equation}
      c_T \to 2 + M + c_{\text{cl}}, \qquad c_{\tilde \tau} \to 24-c_{\text{cl}}, \qquad c_{\tilde T} \to 24-M - c_{\text{cl}}.
    \end{equation}
    The structure of covariant normal ordering and quantum dressing extends to this deformed setup.
  \item \textbf{Canceling the anomaly in the physical Hilbert space.} We formulate a physical Hilbert space $\mathcal{H}_{\text{phys}}$ for the quantum null ray by constructing the GNS representation of the gauge-invariant / dressed operators in the kinematical vacuum state, showing that it is unitarily equivalent to the tensor product of $\mathcal{H}^{\text{rad}}\otimes \mathcal{H}^{\text{edge}}$ and a Virasoro Verma module at central charge $c_{\tilde{T}}$. The Verma module contains a spurious set of anomalous degrees of freedom, but by setting $c_{\text{cl}}=24-M$, we cancel the dressed reparametrization central charge: $c_{\tilde T}=0$. This removes all the spurious degrees of freedom, thus giving a consistent quantization of the gauge-invariant null ray, and fixes the central charge of the reorientations (and thus the Virasoro crossed product) to $c_{\tilde \tau}=M$.
  \item \textbf{Properties of dressing time as a quantum reference frame.} Finally, we study further the way in which the dressing time may be understood as a QRF. We demonstrate that the physical Hilbert space $\mathcal{H}_{\text{phys}}$ is isomorphic to a `reduced Hilbert space' $\mathcal{H}_{\text{red}}$, with the two being related by a Page--Wootters reduction map $\mathcal{R}:\mathcal{H}_{\text{phys}}\to\mathcal{H}_{\text{red}}$ given in~\eqref{Equation: reduction map}, thus connecting with the perspective-neutral formalism. The reduction map $\mathcal{R}$ conditions on the dressing time agreeing with the background time, $V(v)=v$. We also describe how the dressing time QRF may be thought of as `Heisenberg ideal' in the sense that the dressing map $\Dmap$ is an isomorphism when restricted to radiative operators, but `Schr\"odinger non-ideal' in the sense that its coherent states have non-vanishing overlap (in the future we would like to understand the operational interpretation of this). This overlap is characterized by the K\"ahler potential of Virasoro coadjoint orbits, also known as the Teo-Takhtajan energy~\cite{Takhtajan:2003hm, Alekseev:2022efp} (of which we provide a derivation in Appendix~\ref{Appendix: Kahler}).
\end{itemize}

\section{Kinematical degrees of freedom on the null ray}
\label{Section: kinematical quantization}

Our starting point is the classical symplectic potential for a gravitational null ray segment $\Segment$ in the presence of linearized radiation. An appropriate choice of variables puts this in the remarkably simple form~\cite{LocalizationAnomalous}
\begin{equation}
  \bm\Theta_{\Segment} = \int_\Segment \dd{v}\qty[\tau\frac{\delta V}{\partial_v V} +  \sum_i\partial_v\varphi_i\delta\varphi_i] + \sum_a\epsilon_a \omega_a\delta q_a.
  \label{Equation: classical symplectic potential}
\end{equation}
We use $v$ as a coordinate on $\Segment=[v_0,v_1]$. Let us explain the various terms appearing here.
\begin{itemize}
  \item $\{\varphi_i\}$ is a collection of scalar fields labelled by $i$. This collection includes the two polarizations of perturbative gravitational radiation. They are `half-densitized fields', defined as $\varphi_i=\sqrt{\Omega}\phi_i$, where $\phi_i$ are the original fields, and $\Omega$ is the area element on a null surface containing $\Segment$.
  \item $V:\Segment \to I$, where $I=[0,1]$, is the \emph{dressing time}. It plays the role of a reference frame with which we dress other quantities. It is defined to obey the boundary conditions $V(v_a)=a$, and the second order equation
    \begin{equation}
      \frac{\partial_v^2 V}{\partial_v V} = \beta:= \kappa - \frac12\theta + 4\pi \GN\frac{1}{\sqrt{\Omega}}\partial_v\qty\Big(\sqrt{\Omega}\sum_i\phi_i^2),
    \end{equation}
    where $\kappa$ and $\theta$ are the inaffinity and expansion of the null ray associated with the background time coordinate $v$.
  \item $\tau$ is the `spin $0$ stress tensor', defined by
    \begin{equation}
      \tau = \frac1{8\pi \GN} \partial_v V\partial_v\qty(\frac{\partial_v\Omega}{\partial_vV}) = \frac1{8\pi \GN}\left( \partial_v^2\Omega-\beta\partial_v\Omega \right).
    \end{equation}
  \item $q_a$, $a=0,1$ is a gauge-invariant edge mode measuring the relative boost between the dressing time and an external embedding field used to localize the segment $\Segment$. Its conjugate momentum is the area element evaluated at the corresponding endpoint
    \begin{equation}
      \omega_a = \frac1{8\pi\GN}\Omega(v_a).
    \end{equation}
    The factor $\epsilon_a$, with $\epsilon_0=-1$ and $\epsilon_1=1$, accounts for the orientations of the endpoints.
\end{itemize}
More details may be found in~\cite{LocalizationAnomalous}.

In Section~\ref{Subsection: canonical quantization}, we carry out the quantization of these degrees of freedom. The result is a space of kinematical states of the form
\begin{equation}
  \mathcal{K}_{\text{kin}} = \mathcal{K}^0\otimes\mathcal{H}^{\text{rad}}\otimes\mathcal{H}^{\text{edge}}.
\end{equation}
Here $\mathcal{K}^0$, $\mathcal{H}^{\text{rad}}$, $\mathcal{H}^{\text{edge}}$ are the spaces of states for the spin 0 fields $\tau,V$, radiative fields $\varphi_i$ and edge modes $\omega_a,q_a$ respectively.
We use the notation $\mathcal{K}$ to indicate a space which is \emph{not} a Hilbert space because its inner product is not positive-definite (note that there is no issue with kinematical states having an indefinite inner product, so long as \emph{physical} states have a positive-definite inner product, c.f.~Gupta-Bleuler quantization).

The space $\mathcal{K}_{\text{kin}}$ defines the quantum null ray kinematically. To achieve the physical quantization of the null ray, the remaining steps are to unitarily\footnote{Actually, because of the indefinite inner product, it is more correct to say that we have a \emph{pseudo}-unitary representation of the gauge group. But to avoid being too verbose we here mostly just use the word `unitary' to refer to pseudo-unitary things as well.} represent the gauge group, and impose invariance under the gauge transformations. We do the former in Section~\ref{Section: quantum diffeos}; the latter is the subject of the rest of the paper.

\subsection{Canonical quantization}
\label{Subsection: canonical quantization}

The space of states $\mathcal{K}^{0}$ for the spin 0 degrees of freedom $\tau,V$ has an indefinite inner product. We formulate this space by canonically quantizing the classical bracket
\begin{equation}
  \pb{\Pi(v)}{\Vv(v')} = \delta(v-v'), \qq{where} \Pi := \frac{\tau}{\partial_v V}, \quad \Vv := V-v. \label{Pitau}
\end{equation}
The brackets of $\Pi$ and $\Vv$ with themselves vanish.

Although the fields $\tau,V$ are originally defined on the segment $\Segment$, it is convenient to extend $\Pi,\Vv$ to $v\in\RR$, with falloffs $\Pi,\Vv\to 0$ (so $\tau\to 0$, $V\to v$) as $\abs{v}\to\infty$. Eventually we can restrict to field operators that only have support on $\Segment$.

The canonical quantization proceeds by formulating the corresponding quantum commutators
\begin{equation}
  [\Pi(v),\Vv(v')] = -i\hbar\delta(v-v'), \qquad [\Pi(v),\Pi(v')]=[\Vv(v),\Vv(v')]=0,
\end{equation}
and defining a vacuum state $\ket{0}\in\mathcal{K}^{0}$ satisfying
\begin{equation}
  P_+\Pi(v)\ket{0} = P_+\Vv(v)\ket{0} = 0,
  \label{Equation: standard vacuum}
\end{equation}
where $P_\pm$ project onto positive/negative frequencies respectively (as usual the limit $\epsilon\to0^+$ is taken at the end):
\begin{equation}
  P_+\phi(v) = \int_\RR\frac{\dd{u}}{2\pi i} \frac{\phi(u)}{v-u-i\epsilon},\qquad
  P_-\phi(v) = \int_\RR\frac{\dd{u}}{2\pi i} \frac{\phi(u)}{u-v-i\epsilon}.
\end{equation}
The image of $P_-$ is holomorphic in the upper-half plane (UHP) while that of
$P_+$ is holomorphic in the lower-half plane (LHP). These projectors satisfy the following properties:
\begin{equation}
  P_++P_-= \mathrm{Id}, \qquad \int_\RR \dd{v} f P_+(g)= \int_\RR \dd{v} P_-(f) g, \qquad
  [P_+(f)]^*= P_-(f^*),
\end{equation}
and
\begin{equation}
  P_\pm(e^{-ikv})=
  \begin{cases}
    e^{-ikv} & \text{if }\pm k > 0,\\
    0 & \text{if }\pm k<0.
  \end{cases}
\end{equation}
The rest of the state space $\mathcal{K}^{0}$ is generated by the action of the fields on the vacuum, and we define the Hermitian inner product on $\mathcal{K}^{0}$ by demanding that $\Pi,\Vv$ be self-adjoint:
\begin{equation}
  \bra{0}=(\ket{0})^\dagger,\qquad \braket{0}{0} = 1, \qquad \Pi(v)^\dagger = \Pi(v), \qquad \Vv(v)^\dagger = \Vv(v).
\end{equation}
Using $\expval{\cdot}$ to denote vacuum expectation values, one finds the two point functions
\begin{align}\label{vacexp0}
  \expval{\Pi(v)\Vv(v')} &= \expval{\comm{P_+\Pi(v)}{\Vv(v')}} = -\frac\hbar{2\pi}\frac1{v-v'-i\epsilon}, \\
  \expval{\Vv(v)\Pi(v')} &= \expval{\comm{P_+\Vv(v)}{\Pi(v')}} = \frac\hbar{2\pi}\frac1{v-v'-i\epsilon},
\end{align}
with the two point functions of $\Pi,\Vv$ with themselves vanishing. Higher order correlators may be computed with Wick's theorem.

The result of this procedure is a Fock space whose creation/annihilation operators are the negative/positive frequencies respectively of $\Pi,\Vv$. To see that the inner product is indefinite, one can consider the state
\begin{equation}
  \ket{f,g} = \qty(\Pi(f)+\Vv(g))\ket{0}, \qq{where} \Pi(f) = \int_\RR\dd{v}f \Pi, \quad \Vv(g) =\int_\RR\dd{v}g \Vv,
\end{equation}
with $f,g$ some functions. One then has
\begin{align}
  \norm{\ket{f,g}}^2 &= \int_{\RR^2}\dd{v}\dd{v'}\qty(f^*(v)g(v')\expval{\Pi(v)\Vv(v')} + g^*(v)f(v')\expval{\Vv(v)\Pi(v')})\\
  &= -i\hbar\int_\RR\dd{v}\qty(f^* P_+ g - g^* P_+ f)\\
  &= \frac\hbar4\int_\RR\dd{v}\qty((P_+(f-ig))^* P_+ (f-ig) - (P_+(f+ig))^*P_+(f+ig)) \\
  &= \frac\hbar4\norm{P_+(f-ig)}^2 -\frac\hbar4\norm{P_+(f+ig)}^2,
\end{align}
where the norms on the last line are taken with respect to the $L^2$ norm for functions on $\RR$. This is clearly of indefinite sign; indeed it is positive/negative for $f=\mp ig$ respectively.

The other two factors $\mathcal{H}^{\text{edge}}$, $\mathcal{H}^{\text{rad}}$ in $\mathcal{K}_{\text{kin}}$ are genuine Hilbert spaces. The first is the Hilbert space for the edge modes $\omega_a,q_a$. We can simply formulate this as
\begin{equation}
  \mathcal{H}^{\text{edge}} = L^2(\RR^2),
\end{equation}
and promote $q_a$ to position operators on this space, with $\omega_a$ their conjugate momenta:
\begin{equation}
  [q_a,\omega_b] = i\hbar\epsilon_a\delta_{ab}.
\end{equation}
The remaining Hilbert space $\mathcal{H}^{\text{rad}}$ is the Fock space for the radiative fields $\varphi_i$, defined similarly to spin 0, i.e.\ we formulate the canonical commutator
\begin{equation}
  [\partial_v\varphi_i(v),\partial_v\varphi_i(v')] = \frac{i\hbar}{2}\partial_v\delta(v-v')
\end{equation}
and generate $\mathcal{H}^{\text{rad}}$ by the action of these operators on a vacuum state $\ket{0}_{\text{rad}}$ satisfying
\begin{equation}
  P_+\partial_v\varphi_i(v)\ket{0}_{\text{rad}}=0.
\end{equation}
The two point function is then given by
\begin{equation}
  \expval{\partial_v\varphi_i(v)\partial_v\varphi_i(v')} = \expval{\comm{P_+\partial_v\varphi_i(v)}{\partial_v\varphi_i(v')}} = -\frac{\hbar}{4\pi}\frac1{(v-v'-i\epsilon)^2}.\label{vacexp1}
\end{equation}
The inner product on $\mathcal{H}^{\text{rad}}$ is fixed by requiring $\partial_v\varphi_i$ to be self-adjoint. As expected for states of scalar fields, the inner product is positive-definite; for example, defining the state
\begin{equation}
  \ket{f} = \int_\RR\dd{v} f \partial_v\varphi\ket{0}_{\text{rad}},
\end{equation}
one has
\begin{align}
  \norm{\ket{f}}^2 &= -\frac{\hbar}{4\pi}\int_{\RR^2}\dd{v}\dd{v'}f^*(v) f(v')\partial_{v'}\frac{1}{v-v'-i\epsilon} \\
  &= \frac{\hbar}{4\pi}\int_{\RR^2}\dd{v}\dd{v'}f^*(v) \partial_{v'}f(v')\frac{1}{v-v'-i\epsilon}\\
  &= \frac{\hbar}{4\pi}\int_\RR\dd{v}f^* 2\pi iP_+(\partial_{v}f)\\
  &= \frac\hbar2 \int_\RR\dd{v}(P_+f)^* \qty(i\partial_v)P_+f.
\end{align}
In the space of positive frequency functions $e^{-ikv}$, with $k>0$, $i\partial_v$ is a positive operator, so this is positive as required.

\subsection{Quantum reparametrizations}
\label{Section: quantum diffeos}

The gauge group of the degrees of freedom on a null ray $\RR$ is the group $\Diff^+(\RR)$ of orientation-preserving \emph{reparametrizations} of the fields with respect to the background time $v$. The classical generator of reparametrizations is the Raychaudhuri stress tensor
\begin{equation}
  T = \tau + \sum_i(\partial_v\varphi_i)^2.
\end{equation}
We quantize this stress tensor using normal ordering: $T = \tau + T^{\text{rad}}$, where
\begin{equation}
  \tau=\normord{\Pi \partial_v V}, \quad T^\text{rad} = \sum_i\normord{(\partial_v\varphi)^2}.
\end{equation}
Here $\tau$ and $T^{\text{rad}}$ are the quantum stress tensors for the spin 0 and radiative degrees of freedom respectively.
Normal ordering means putting all vacuum-annihilating modes $(\Pi_+, V_+, \varphi_{i+})$\footnote{We denote $\Phi_\pm := P_{\pm}(\Phi)$ for any field $\Phi=(\Pi,V,\varphi_i)$. When we write $V_\pm = P_\pm V$ we really mean $V_\pm = v + P_\pm \Vv$.} to the right. Given a function $f(v)$ which decays as $v\to \infty$, we define
\begin{equation}
  T_f = \tau_f + T^{\text{rad}}_f, \qquad \tau_f = \int_\RR\dd{v}f\tau, \qquad T^{\text{rad}}_f = \int_\RR\dd{v}fT^{\text{rad}},
\end{equation}
one then has
\begin{equation}
  \frac1{i\hbar}[\varphi_i,T_f]=f\partial_v \varphi_i, \qquad
  \frac1{i\hbar}[V,T_f]=f\partial_v V, \qquad
  \frac1{i\hbar}[\Pi,T_f]=\partial_v(f\Pi),
\end{equation}
which are the appropriate actions of reparametrization on the scalars $V,\varphi_i$ and 1-form $\Pi$.

As discussed in~\cite{Ciambelli_2024}, the normal ordering in the stress tensor leads to anomalies in the stress tensor commutators.
By the reader's favorite method\footnote{Our favorite way, detailed in Appendix \ref{Subsection: reparametrization algebra}, is to use that the quantum anomalies come from double Wick contractions between the quadratic terms in the stress tensors. }
they will find that
\begin{align}
  \frac1{i\hbar}[\tau_f,\tau_g] &= \tau_{[f,g]} + \frac{c_{0}\hbar}{48\pi}\int_\RR\dd{v}\qty(\pa_vf\partial_v^2 g-\pa_vg\partial_v^2 f)\label{Equation: T 0 algebra}\\
  \frac1{i\hbar}[T^{\text{rad}}_f,T^{\text{rad}}_g] &= T^{\text{rad}}_{[f,g]} + \frac{c_{\text{rad}}\hbar}{48\pi}\int_\RR\dd{v}\qty(\pa_vf\partial_v^2g- \pa_vg\partial_v^2 f),
\end{align}
where $[f,g]= f\pa_vg - g \pa_vf$ is the Witt product, and the central charges are $c_{0}=2$ and $c_{\text{rad}}=M$, where $M$ is the number of radiative fields (so the index $i$ runs from $1$ to $M$). Thus, at the quantum level, the group of reparametrizations is deformed from $\Diff^+(\RR)$ to Virasoro.

These stress tensors exponentiate to projective (pseudo-)unitary representations:
\begin{align}
  U_{0}[F] = \exp(i\tau_f/\hbar), \qquad
  U_{\text{rad}}[F] = \exp(iT_f^{\text{rad}}/\hbar),
\end{align}
where $F=\exp(f\partial_v)$.
The overall representation of Virasoro is given by the tensor product
\begin{equation}
  U = U_{0}\otimes U_{\text{rad}}\otimes\mathds{1}_{\text{edge}},
\end{equation}
where $\mathds{1}_{\text{edge}}$ is the identity on $\mathcal{H}^{\text{edge}}$.
This representation is projective because of additional phases appearing in the composition:
\begin{equation}
  U[F]U[G] = U[F\circ G] \exp(-\frac{ic_T}{24}C(F,G))
\end{equation}
The total central charge is $c_T=c_0+c_{\text{rad}}=2+M$, and $C$ is a 2-cocycle. This means it
satisfies the 2-cocycle identity
\begin{equation}\label{eq:cocycle}
  C(F,G)+C(F\circ G,H)=C(F,G\circ H)+C(G,H);
\end{equation}
the property $U[F]^{-1}=U[F^{-1}]$
also implies that $C$ satisfies
the normalization conditions
\begin{equation}\label{sym}
  C(F,F^{-1})=C(F,1)= C(1,F)=0,
\end{equation}
the skew-symmetry property
\begin{equation}\label{skew}
  C(F,G^{-1})=-C(G,F^{-1}),
\end{equation}
and the cyclicity property: given $F,G,H \in \Diff^+(\mathbb{R})$ which are such that $FGH=1$ we have
\begin{equation}\label{cycl}
  C(F,G)=C(G,H)=C(H,F).
\end{equation}
Finally, $C$ satisfies the limiting behavior
\begin{equation}\label{limit}
  \left.\pa_s\pa_tC(F_s,G_t)\right|_{s=t=0} =\frac1{2\pi}\int\pa_vf\pa_v^2g \dd{v},
\end{equation}
where $F_s =\exp(sf\pa_v)$ and $G_t=\exp(tg\pa_v)$.
In Appendix \ref{Appendix: cocycle}, we evaluate explicitly this cocycle, and find that it is given by
\begin{equation}
  C(F,G) = \mathrm{B}(F,G) - b(F\circ G) + b(F) + b(G),
  \label{Equation:cocycle0}
\end{equation}
where
\begin{equation}
  \mathrm{B}(F,G) := \frac1{2\pi} \int_\RR\ln \qty(\partial_vF\circ G) \dd{\ln\partial_v G}
\end{equation}
is the Bott-Thurston cocycle \cite{bott1977characteristic, oblak2017berry,Alekseev:2022efp}, which also satisfies all the properties (\ref{eq:cocycle}-\ref{limit}).
The 1-cocycle $b(F)$ is
obtained by integrating  the Kirillov-Kostant-Souriau symplectic potential~\cite{Guillemin:1990ew,Woodhouse:1980pa} along the path from the identity diffeomorphism to $F_s := \exp(sf\partial_v)$  for $s\in[0,1]$
\begin{equation}
  b(F) := \int_0^1 \Theta_s(f) \rd s, \qquad
  \Theta_s(f) :=\frac1{2\pi}\int_{\mathbb{R}}\dd{v}\pa_s(\ln\partial_vF_s) \partial_v\ln\partial_v F_s \label{KKS}.
\end{equation}
This cocycle satisfies $b(F^{-1})=-b(F)$.

The adjoint action of $U[F]$ on the fields generates the diffeomorphism transformation.
One has
\begin{equation}
  U[F] \varphi_i U[F]^\dagger = \varphi_i\circ F, \qquad
  U[F] V U[F]^\dagger = V\circ F, \qquad U[F] \Pi U[F]^\dagger = \partial_v F \Pi\circ F
\end{equation}
so the basic fields $\varphi_i$, $V$, $\Pi$ transform as expected under a finite reparametrization -- but due to normal ordering this is not true for all operators. For example, the stress tensors transform anomalously:
\begin{align}
  U[F]\tau U[F]^\dagger &= (\partial_vF)^2 \tau\circ F - \frac{c^0\hbar}{24\pi}\Schwarzian{F}{v},\\
  U[F]T^{\text{rad}} U[F]^\dagger &= (\partial_vF)^2 T^{\text{rad}}\circ F - \frac{c_{\text{rad}}\hbar}{24\pi}\Schwarzian{F}{v}, \label{Ttransform}
\end{align}
where\footnote{We use the shorthand $F'=\pa_vF$.}
\begin{equation} \Schwarzian{F}{v}=\partial_v\qty(\frac{\partial_v^2F}{\partial_vF})-\frac12\qty(\frac{\partial_v^2F}{\partial_vF})^2= \left(\frac{F'''}{F'}\right)- \frac32 \left(\frac{F''}{F'}\right)^2,\label{Schwarz}
\end{equation} is the Schwarzian derivative of $F$, which satisfies the composition law
\begin{equation}
  \Schwarzian{G\circ F}{v}= (\pa_vF)^2\Schwarzian{G}{F(v)}+ \Schwarzian{F}{v}.
\end{equation}

\section{Gauge invariance, normal ordering and the crossed product}
\label{Section: crossed product}

Section~\ref{Section: kinematical quantization} describes the \emph{kinematical} quantization of the degrees of freedom on $\Segment$, and the reparametrization action on them. To obtain the \emph{physical} (i.e.\ gauge-invariant) quantization, we can write down the algebra of all gauge-invariant operators acting on the kinematical degrees of freedom, and then find an appropriate Hilbert space representation of these operators. The first challenge is therefore to construct explicitly the gauge-invariant operators.

To construct gauge-invariant observables in the classical theory one can use `dressing', which means employing some fields as dynamical coordinates relative to which one measures other fields. The resulting observables are known as dressed observables. In~\cite{LocalizationAnomalous}, we described how the full set of gauge-invariant classical observables of the null ray segment $\Segment$ can be obtained as observables dressed to the dressing time $V$. These include the dressed radiative fields $\tilde\varphi_i$, defined as the pushforward of $\varphi_i$ through $V$:
\begin{equation}
  \tilde\varphi_i = \varphi_i\circ X,
\end{equation}
and the dressed area element $\tilde\Omega$, defined similarly:
\begin{equation}
  \tilde\Omega = \Omega\circ X.
\end{equation}
Here and in the rest of the paper $X=V^{-1}$. Classically, these observables are gauge-invariant due to the dressing, because under a reparametrization $\varphi_i\to\varphi_i\circ F$, $\Omega\to \Omega\circ F$ is compensated for by $X\to F^{-1}\circ X$.

With this in mind, an appropriate strategy for the physical quantization of the theory is to quantize the classical dressed observables. This is the strategy we employ in the present paper.

For most of the rest of the paper we leave the edge modes $\omega_a,q_a$ out of the discussion, because they don't have the same subtleties (described below) as the fields. Indeed, they are simple quantum mechanical degrees of freedom, having a Hilbert space $L^2(\RR^2)$.\footnote{This is the case because we have restricted to a single null ray. In higher than 2 spacetime dimensions, accounting for the full null surface would require treating the edge modes as codimension 2 fields.} Moreover, they are already gauge-invariant, so their properties do not change upon imposing gauge-invariance. So from now on, unless otherwise stated, we only discuss the field degrees of freedom $\Phi=(\Pi,V,\varphi_i)$.

\subsection{Dealing with anomalies in dressed observables}

Let's first consider observables depending only on $\tilde\varphi_i$. The quantization of such observables is reasonably straightforward, because, unlike $\Omega$, the radiative fields $\varphi_i$ commute with the reference frame field $V$.

The simplest example is $\tilde\varphi_i$ itself.
Since there are no ordering ambiguities or divergences it is simple to see that one can directly promote $\tilde\varphi_i$ to a gauge-invariant quantum operator
\begin{equation}
  U(F) \tilde{\varphi}_i U(F)^\dagger = \tilde{\varphi}_i.
\end{equation}
The situation is more nuanced when considering composite observables such as $(\tilde\varphi_i)^2$, since one needs to use a renormalization procedure to get a well-defined quantum operator. Usually this renormalization is done with  normal ordering, or through the OPE expansion. This means e.g.\ $\normord{\varphi_i^2(x)}=\lim_{x\to y} \left[\varphi_i(x)\varphi_i(y)- G(x,y) \right] $ where $G(x,y)$ is taken to be the two point function of the field in a given state (the vacuum for normal ordering)~\cite{Polchinski} or is given by the Hadamard parametrix (for OPE regularization)~\cite{Hollands:2023txn}. Unfortunately, the ordinary normal ordering
\begin{align}
  \normord{(\tilde\varphi_i)^2} &= \qty((P_+\varphi_i)\circ X)^2 + \qty((P_-\varphi_i)\circ X)^2 + 2\qty((P_-\varphi_i)\circ X)\qty((P_+\varphi_i)\circ X)
\end{align}
is not suitable, because\footnote{One efficient way to show this is to consider
  \begin{equation}
    \normord{\tilde\varphi_i(u)\tilde\varphi_i(v)}=\tilde\varphi_i(u)\tilde\varphi_i(v) - [(P_+\varphi_i)(X(u)),\varphi_i(X(v))]=\frac\hbar{4\pi}\ln(X(u)-X(v)-i\epsilon),
  \end{equation}
  which satisfies
  \begin{align}
    \MoveEqLeft U[F]\normord{\tilde\varphi_i(u)\tilde\varphi_i(v)}U[F]^\dagger - \normord{\tilde\varphi_i(u)\tilde\varphi_i(v)}
    = \frac\hbar{4\pi}\ln(\frac{F^{-1}(X(u))-F^{-1}(X(v))}{X(u)-X(v)}).
  \end{align}
One recovers~\eqref{Equation: tilde varphi squared anomaly} in the limit $u\to v$.}
\begin{equation}
  U[F]\normord{(\tilde\varphi_i)^2}U[F]^\dagger
  = \normord{(\tilde\varphi_i)^2} +\frac\hbar{4\pi}\ln(\partial_v F^{-1})\circ X,
  \label{Equation: tilde varphi squared anomaly}
\end{equation}
so gauge-invariance is not maintained. This is an \emph{anomaly} in the quantization of the dressed operator.

To account for this, one can introduce an additional quantum correction / improvement term that depends on the dressing time as follows:
\begin{equation}
  \covnormord{(\tilde\varphi_i)^2} := \normord{(\tilde\varphi_i)^2} - \frac\hbar{4\pi}\ln\partial_\mv X.
\end{equation}
One may check that
\begin{equation}
  U[F]\covnormord{(\tilde\varphi_i)^2} U[F]^\dagger = \covnormord{(\tilde\varphi_i)^2},
\end{equation}
so $\covnormord{(\tilde\varphi_i)^2}$ is a quantization of $(\tilde\varphi_i)^2$ that preserves gauge-invariance.

We refer to $\covnormord{(\tilde\varphi_i)^2}$ as the `covariant normal ordering' of $(\tilde\varphi_i)^2$, a concept we describe in more detail in Section~\ref{Section: covariant normal ordering} and Appendix~\ref{Appendix: more on covariant normal ordering}. Its nature is made more obvious by writing it in the form
\begin{align}
  \covnormord{(\tilde\varphi_i)^2} &= \qty(P_+\tilde\varphi_i)^2 + \qty(P_-\tilde \varphi_i)^2 + 2\qty(P_-\tilde\varphi_i)\qty(P_+\tilde\varphi_i).
\end{align}
In other words, covariant normal ordering is just normal ordering applied to the dressed field $\tilde\varphi_i$.

We can extend this to any functional $\mathcal{F}[\tilde\varphi]$ of the dressed radiative fields, by defining $\covnormord{\mathcal{F}[\tilde\varphi]}$ as the operator obtained by ordering $P_+\tilde\varphi_i$ to the right of $P_-\tilde\varphi_i$. This always gives a gauge-invariant operator.
Since $\tilde\varphi_i = U_{\rad}[V]^\dagger\varphi_iU_{\rad}[V]$, one has the convenient formula
\begin{equation}
  \covnormord{\mathcal{F}[\tilde\varphi]} = U_{\rad}[V]^\dagger \normord{\mathcal{F}[\varphi]} U_{\rad}[V].
  \label{Equation: dressed field operator}
\end{equation}
The operator $U_{\rad}[V]$ appearing here is well-defined because $V$ commutes with itself and the radiative fields, so we can just treat it as a commuting variable when we substitute $F\to V$ in $U_{\rad}[F]$. It is useful to define a \emph{dressing map} $\Dmap$ with
\begin{equation}
  \Dmap(a) = U_{\rad}[V]^\dagger a U_{\rad}[V] \qq{where} a\in\mathcal{B}(\mathcal{H}^{\text{rad}}),
  \label{Equation: D a}
\end{equation}
so that $\Dmap(\normord{\mathcal{F}[\varphi]}) = \covnormord{\mathcal{F}[\tilde\varphi]}$.

For the spin zero field, the situation is more involved. Suppose we consider
the simplest dressed classical observable involving the spin 0 fields which is the dressed spin 0 stress tensor
\begin{equation}
  \tilde\tau = (\partial_\mv X)^2\tau\circ X = \partial_\mv^2\tilde\Omega.
\end{equation}
Like $(\tilde\varphi)^2$, this is a composite observable, and for reasons explained in Section~\ref{Section: covariant normal ordering}, the normal ordered version of this operator is anomalous, meaning it breaks gauge-invariance:
\begin{equation}
  U[F]\normord{\tilde\tau}U[F]^\dagger \ne \normord{\tilde\tau}.
\end{equation}
However, just as for the dressed radiative fields $\tilde\varphi_i$, there is a notion of covariant normal ordering which removes this anomaly and restores gauge-invariance:
\begin{equation}
  U[F]\covnormord{\tilde\tau}U[F]^\dagger = \covnormord{\tilde\tau}.
\end{equation}
It is useful to extend the definition of the dressing map such that $\Dmap(\tau):=\covnormord{\tilde\tau}$. More details are given in Section~\ref{Section: dressed operators}.

Spin 0 covariant normal ordering is not quite as easily defined as for the radiative fields, because the dressing time $V$ itself can no longer be treated as a commutative variable, since, for example, it has non-trivial brackets with $\tau$. We give more details on how spin 0 covariant normal ordering is defined in Section~\ref{Section: covariant normal ordering} and Appendix~\ref{Appendix: more on covariant normal ordering}.

\subsection{The gauge-invariant algebra is a Virasoro crossed product}

As explained in Section~\ref{Section: dressed operators}, the full algebra of gauge-invariant operators along the null ray (excluding the edge modes) is generated by $\Dmap(\tau)$ and dressed radiative operators $\Dmap(a)$.
The dressing map $\Dmap$ is an algebra morphism when restricted to the radiative fields: given $a, b \in \mathcal{B}(\mathcal{H}^{\text{rad}})$ we have
\begin{equation}
  \Dmap(a) \Dmap(b)= \Dmap(ab).
\end{equation}
Moreover, $\Dmap(\tau)$ itself generates a Virasoro algebra at a certain central charge $c_{\tilde\tau}$, and its bracket with dressed radiative operators $\Dmap(a)$ is equivalent to a bracket with $-T^\text{rad}$ on the bare operator $a$:
\begin{equation}
  [\Dmap(\tau),\Dmap(a)] = -\Dmap([T^\text{rad},a]).
  \label{Equation: reorientation inverse reparametrization}
\end{equation}
In Sections~\ref{Section: anomaly shift} and~\ref{Section: physical representation} we explain how a proper cancellation of gauge anomalies forces $c_{\tilde\tau}=M$, where $M$ is the number of radiative fields.

These properties imply that the algebra of gauge-invariant operators is the \emph{crossed product}~\cite{Connes1994,Takesaki2003II,Takesaki2003III} of the algebra of radiative operators $\mathcal{B}(\mathcal{H}^\text{rad})$ by the Virasoro group at central charge $c_{\tilde\tau}=M$:\footnote{One can also understand this as a `\emph{twisted} crossed product' of $\mathcal{B}(\mathcal{H}^\text{rad})$ by $\Diff^+(\RR)$. A twisted crossed product by a group $G$ differs from an ordinary crossed product by explicitly including a cocycle in the representation of $G$ (see~\cite{Sutherland1980}). We thank Marc Klinger and Shadi Ali Ahmad for pointing this out to us.}
\begin{equation}
  \mathcal{B}(\mathcal{H}^{\text{rad}})\rtimes \operatorname{Vir}_M = \{\Dmap(\tau),\,\Dmap(a)\,\mid\,a\in\mathcal{B}(\mathcal{H}^{\text{rad}})\}''.
\end{equation}
In Section~\ref{Section: physical representation} we explicitly construct the Hilbert space $\mathcal{H}_{\text{phys}}$ on which this algebra is represented and on which the identity $\Dmap(\tau)=-\Dmap(T^{\text{rad}})$ is implemented.

Given an algebra of `system' operators $\mathcal{A}$ and a group $G$ acting by automorphisms on $\mathcal{A}$, the crossed product $\mathcal{A}\rtimes G$ is the appropriate algebra for describing the system in the presence of a quantum reference frame with gauge group $G$. The crossed product is generated by two sets of operators: dressed system operators (here the `system' is the radiative sector $\mathcal{H}^\text{rad}$, so the dressed system operators are $\Dmap(a)$), and `reorientations'. In the present case, $\Dmap(\tau)$ is the generator of reorientations, which are transformations which affect the reference frame degrees of freedom (here the dressing time $V$) without acting on the system.

Usually a semisimple or locally compact group is used when constructing crossed products. Here we are leaving that setting by generalizing to the Virasoro group, which is not locally compact.

\subsubsection{Crossed products for Lie groups}
\label{Subsubsection: Lie group crossed product}
In order to contextualize our results, let us give some more details on the crossed product construction in the case of a locally compact Lie group $G$.

The canonical example of a crossed product is $G=\RR$ corresponding to time evolution (the case where this is modular time evolution for a system corresponding to a QFT subregion has led to much recent progress in understanding gravitational entropies~\cite{CLPW,Jensen2023,Fewster2025,DeVuyst:2024pop,DeVuyst:2024uvd,kirklin2024generalisedsecondlawsemiclassical,KudlerFlam2025,Klinger2024,Klinger2024,Klinger:2026tws,KudlerFlam2025a,Faulkner2024}). Then the dressed system operators take the form
\begin{equation}
  \Dmap(a):= e^{i\hat{T}H_S} a e^{-i\hat{T}H_S}, \qquad a \in \mathcal{A},
\end{equation}
for some system algebra $\mathcal{A}$. Here, $\hat{T}$ is an operator for the time given by the QRF (which in this case is simply a clock), while $H_S$ is the Hamiltonian of the system. Reorientations are generated by the Hamiltonian $H_C$ of the clock.

For a more general $G$,  one has a unitary action $U_S[g]:\mathcal{H}_S\to\mathcal{H}_S$ of $G$ on the system Hilbert space $\mathcal{H}_S$. Given a basis $\sigma_i$ for the Lie algebra, with structure constants $F_{ij}{}^k$ defined by $[\sigma_i,\sigma_j]=F_{ij}{}^k\sigma_k$, we can write this in the form (for group elements connected to the identity)
\begin{equation}
  U_S\qty[\exp (\alpha^i\sigma_i)] = \exp (\alpha^iP_i),
\end{equation}
where $P_i:=\tr(P \sigma_i)$, for some Lie algebra valued operator $P$, are the corresponding Noether charges (which we take here to be anti-Hermitian). The Lie algebra then acts on system observables via the adjoint action $\mathrm{ad}P_i$, which satisfies
\begin{equation}
  [\mathrm{ad}P_i,\mathrm{ad}P_j]= F_{ij}{}^k \mathrm{ad}P_k.
\end{equation}
The QRF has as Hilbert space $\mathcal{H}_R=L^2(G)$, and carries operators $(\pi, \hat{g})$, where
$\pi$ is Lie algebra valued (and is taken to be anti-Hermitian), and $\hat g$ is group valued. On $\mathcal{H}_R$, $\hat{g}$ acts by multiplication, while $\pi$ acts by left invariant derivation.
If we denote $\pi_i :=\tr(\pi \sigma_i)$, then
\begin{equation}
  [\pi_i, \hat{g} ]=  \hat{g} \sigma_i, \qquad
  [\pi_i, \pi_j ]= f_{ij}{}^k \pi_k.
\end{equation}
The gauge constraint is the generator of the joint action of the group on the system and QRF, and is given by
\begin{equation}\label{constraint}
  C_i = \pi_i +P_i.
\end{equation}

The dressed system operators are defined by
\begin{equation}
  \Dmap(a)= U_S[\hat{g}]^\dagger a  U_S[\hat{g}].
  \label{Equation: dressed system operator}
\end{equation}
They obey
\begin{equation}
  \Dmap(a)\Dmap(b) = \Dmap(ab);
  \label{Equation: G cross 1}
\end{equation}
in other words, dressing preserves the algebraic relations of the system operators.
To see that dressed system operators are gauge-invariant we use that $\qty[\pi_i,U_S[\hat g]] = U_S[\hat g]P_i$,\footnote{This follows from
  \begin{equation}
    \exp(\alpha^i\pi_i)U_S[\hat g]\exp(-\alpha^i\pi_i) = U_S\qty[\hat g \exp(\alpha^i\sigma_i)] = U_S[\hat g] U_S\qty[\exp(\alpha^i\sigma_i)] = U_S[\hat g] \exp(\alpha^iP_i).
\end{equation}} so
\begin{align}
  [\pi_i, \Dmap(a)]
  = -\left[ P_i , \Dmap(a) \right],
\end{align}
which implies $[C_i, \Dmap(a)]=0$.

We can also dress the QRF momentum operator $\pi$ to obtain
\begin{equation}
  \tilde{\pi}_i := \tr(  \hat{g} \pi \hat{g}^{-1} \sigma_i).
  \label{Equation: dressed momentum operator}
\end{equation}
This operator is such that
\begin{gather}
  [\pi_i,\tilde{\pi}_j]=0, \quad
  [\tilde{\pi}_i, \hat{g}]= \sigma_i g,\\
  [\tilde{\pi}_i, \tilde{\pi}_j ]= -F_{ij}{}^k \tilde{\pi}_k
  \label{Equation: G cross 2}
\end{gather}
The first equality implies that $[C_i,\tilde{\pi}_j]=0$, so that $\tilde{\pi}_i$ is gauge-invariant. The second equality, and the minus sign in the last equality, imply that $\tilde\pi$ acts by right invariant derivation on the QRF state; such transformations are the \emph{reorientations} of the $L^2(G)$ QRF. Using $[\tilde\pi_i,U_S[\hat g]] = P_i U_S[\hat g]$,\footnote{This follows from
  \begin{equation}
    \exp(\alpha^i\tilde\pi_i)U_S[\hat g]\exp(-\alpha^i\tilde\pi_i) = U_S\qty[ \exp(\alpha^i\sigma_i)\hat g] = U_S\qty[\exp(\alpha^i\sigma_i)]U_S[\hat g]  = \exp(\alpha^iP_i)U_S[\hat g] .
\end{equation}} the action of the reorientation generator $\tilde{\pi}_i$ on a dressed system operator is (mirroring~\eqref{Equation: reorientation inverse reparametrization})
\begin{equation}
  [\tilde{\pi}_i, \Dmap(a)]= - \Dmap([P_i, a]);
  \label{Equation: G cross 3}
\end{equation}
in other words, a reorientation acts as an inverse gauge transformation on the `bare' operator $a$.

Together, (\ref{Equation: G cross 1}, \ref{Equation: G cross 2}, \ref{Equation: G cross 3}) define the algebra generated by $\tilde\pi_i, \Dmap(a)$ for $a\in\mathcal{A}$, where $\mathcal{A}\subseteq\mathcal{B}(\mathcal{H}_S)$, as the crossed product algebra
\begin{equation}
  \mathcal{A}\rtimes G := \{\tilde\pi_i,\,\Dmap(a)\,\mid\, a\in\mathcal{A}\}''.
\end{equation}
One may show that this algebra accounts for all gauge-invariant operators in $\mathcal{A}\otimes\mathcal{B}(\mathcal{H}_R)$:
\begin{equation}
  \mathcal{A}\rtimes G = \qty(\mathcal{A}\otimes\mathcal{B}(\mathcal{H}_R))^G.
\end{equation}
Here the superscript ${}^G$ notation indicates a restriction to operators invariant under the gauge group $G$.

Note that there is another common isomorphic presentation of the crossed product~\cite{Takesaki2003II,Takesaki2003III,Connes1994,CLPW,Jensen2023}:
\begin{equation}
  \mathcal{A}\rtimes G \simeq \{\tilde\pi_i-P_i,\,a\,\mid\, a\in\mathcal{A}\}''.
  \label{Equation: undressed crossed product G}
\end{equation}
This is not manifestly gauge-invariant, but simpler to employ in some contexts due to the lack of explicit dressing map $\Dmap$ acting on system operators. In the perspective-neutral formalism it corresponds to the crossed product `in the perspective of' the QRF~\cite{delaHamette:2021oex,Hoehn2023,Hoehn2021a,Hoehn2021,AliAhmad2022,CastroRuiz2020,Vanrietvelde2020,Hoehn2022,Suleymanov:2023wio,DeVuyst:2024pop,DeVuyst:2024uvd,kirklin2024generalisedsecondlawsemiclassical}. The isomorphism is
\begin{equation}
  \tilde\pi_i \leftrightarrow \tilde\pi_i - P_i,
  \qquad
  \Dmap(a) \leftrightarrow a, \quad a\in\mathcal{A},
\end{equation}
and is explicitly implemented by conjugation by $U_S[\hat g]$ (for $\Dmap(a)$ this follows from~\eqref{Equation: dressed system operator}, for $\tilde\pi_i$ it follows from $[\tilde\pi_i,U_S[\hat g]] = P_i U_S[\hat g]$).

\subsubsection{Dressing as an algebraic deformation}

The dressed system operators~\eqref{Equation: dressed system operator} and dressed QRF momentum operators~\eqref{Equation: dressed momentum operator} were defined in slightly different ways above, but they are physically the same kind of object: gauge-invariant operators obtained by conjugating kinematical operators with the orientation $\hat{g}$ of the QRF. With this in mind, it is useful to extend the definition of the dressing map $\Dmap$ so that it accounts for both kinds of operators, i.e.\ define $\Dmap(A)$ where $A$ is any arbitrary combination of system operators and QRF momenta $\pi_i$. The dressing map must then be understood as a \emph{deformation} of the algebraic relations of the operators it acts on, meaning there is some non-trivial deformed product $\Dstar$ such that
\begin{equation}
  \Dmap(A)\Dmap(B) = \Dmap(A\Dstar B).
\end{equation}
We illustrate this deformation below for the $L^2(G)$ frame, but in Section~\ref{Section: dressed operators}, we explain how it works for the gravitational null ray, and show that it may be understood as the quantum analogue of the deformation from the Poisson bracket to the Dirac bracket in the classical theory.

Let's first define the action of the dressing map $\Dmap$ on the QRF momentum:
\begin{equation}
  \Dmap(\pi_i) := \tilde\pi_i.
\end{equation}
Similarly, we can define the action of $\Dmap$ on arbitrary functions of $\pi$ by setting
\begin{equation}
  \Dmap\qty(\exp(\alpha^i\pi_i)) = \exp(\alpha^i\tilde\pi_i)
\end{equation}
and requiring $\Dmap$ to be linear. Note that this means $\Dmap$ is not an algebraic isomorphism when acting on functions of $\pi_i$. Rather it is an anti-morphism, meaning the order of operators is reversed in the following way:
\begin{equation}
  \Dmap\qty(\exp(\alpha^i\pi_i)\exp(\beta^i\pi_i)) =
  \Dmap\qty(\exp(\beta^i\pi_i))\Dmap\qty(\exp(\alpha^i\pi_i)),
\end{equation}
or for a more simple example
\begin{equation}
  \Dmap(\pi_i\pi_j) = \Dmap(\pi_j)\Dmap(\pi_i).
\end{equation}
This follows from the minus sign in~\eqref{Equation: G cross 2}.

To obtain a unified picture, we need to define the action of $\Dmap$ on a more general operator made up of the QRF momentum $\pi_i$ and arbitrary system operators $a\in\mathcal{A}$, but there is not generally a unique way to do this. One needs an ordering prescription to account for the fact that $[\pi_i,a]=0$ but $[\tilde\pi_i,\Dmap(a)]\ne 0$.

For illustrative purposes, we can choose the following ordering prescription:
\begin{equation}
  \Dmap\qty(\exp(\alpha^i\pi_i)a) := \Dmap\qty(\exp(\alpha^i\pi_i))\Dmap(a) = \exp(\alpha^i\tilde\pi_i) U_S[\hat g]^\dagger a U_S[\hat g].
  \label{Equation: total dressing map}
\end{equation}
Then for $a,a'\in\mathcal{A}$ one has
\begin{align}
  \Dmap(\pi_i a)\Dmap (\pi_j a') &= \tilde\pi_i \Dmap(a)\tilde\pi_j \Dmap(a') \\
  &= \tilde\pi_i\tilde\pi_j\Dmap (a)\Dmap(a') + \tilde\pi_i\Dmap([P_j,a])\Dmap(a') \\
  &= \Dmap(\pi_j\pi_i a a' + \pi_i [P_j,a] a'),
\end{align}
so
\begin{equation}
  (\pi_i a)\Dstar (\pi_j a') = \pi_j\pi_i a a' + \pi_i [P_j,a] a'.
\end{equation}
The reversal of indices on the momenta accounts for the fact that the dressing map is an anti-morphism on $\pi$, while the final term accounts for the fact that the dressing map induces non-trivial brackets between the momenta and the system operators, thereby representing the crossed product.

The construction above assumed a locally compact Lie group.
There are various complications associated with extending it to infinite-dimensional gauge groups acting on quantum fields, which we now describe. The anomalies in normal ordering described earlier are one such challenge. Related is that the QFT vacuum explicitly breaks diffeomorphism covariance. Also, we have described in this section \emph{ideal} reference frames, with the QRF transforming in the regular representation. But there is no canonical regular representation for non-locally compact groups like $\Diff^+(\RR)$, which means QRFs for such a group must be fundamentally non-ideal (in a sense we describe in Section~\ref{Subsection: nonideal}).

In Section~\ref{Section: dressed operators}, we define the dressing map for the quantum fields on the null ray, where the ordering prescription for $\Dmap$ is based on covariant normal ordering. The anomalous behavior of the quantum fields leads to important differences with the case of the finite dimensional Lie group discussed above. For example, the dressing map restricted to the spin $0$ field is no longer an antimorphism, but rather a `projective antimorphism', due to the different central charges of the spin 0 stress tensor and radiative stress tensors:
\begin{equation}
  [\Dmap(\tau_f),\Dmap(\tau_g)] = \Dmap([T_f^\text{rad},T_g^\text{rad}]) = -\Dmap([\tau_f,\tau_g]) + i\frac{(c_{\text{rad}}-c_{0})\hbar^2}{48\pi}\int\dd{v}\qty(\partial_vf\partial_v^2g-\partial_vg\partial_v^2f).
  \label{eq:tautau}
\end{equation}
The difference comes from the central charges, which are $c_{\text{rad}}=M$ for $T^{\text{rad}}$ and $\Dmap(\tau)$,\footnote{After cancelling the anomaly in physical states, as described in Sections~\ref{Section: three quantum diffeomorphisms} and~\ref{Section: physical representation}.} but $c_0=2$ for $\tau$.

In this paper, we are treating the null ray essentially as a two-dimensional system, which leads to anomalies characterized by finite central charges. Accounting for a complete null surface in higher than two spacetime dimensions leads to divergent central charges due to the infinite number of null rays~\cite{Wall:2011hj}. So an analogous dressing map for the full null surface would only be well-defined if we have some prescription for regularizing the central charges as described in~\cite{Ciambelli_2024}. We leave an exploration of this to future work.

\section{A covariant prescription for quantizing observables}
\label{Section: covariant normal ordering}

In the previous section, we showed that ordinary normal ordering fails to respect diffeomorphism gauge symmetry. Here, we give more details on the origin of this failure, and define the covariant normal ordering prescription that we can use instead.

Rather than immediately going to the gauge-invariant operators, let us first establish a more general result: the non-anomalous construction of gauge-\emph{covariant} operators, meaning the following. Suppose we have an arbitrary classical observable $O[\Phi]$ which changes to $F\triangleright O[\Phi] := O[F\triangleright\Phi]$ under a reparametrization $F\in\Diff^+(\RR)$. We could quantize $O[\Phi]$ with ordinary normal ordering to form the operator $O:=\normord{O[\Phi]}$. But this would break covariance, meaning that in general
\begin{equation}
  U[F]\normord{O[\Phi]}U[F]^\dagger \ne \normord{F\triangleright O[\Phi]}.
\end{equation}
As we describe in Section~\ref{Subsection: ordinary normord background-dependent}, this anomalous behavior stems from the dependence of ordinary normal ordering on the use of background time $v$. We then define \emph{covariant normal ordering} $\Cov{O}:=\covnormord{O[\Phi]}$ in Sections~\ref{Subsection: covnormord} and~\ref{Subsection: covnormord 0}, which removes this background-dependence and has the property
\begin{equation}
  U[F]\covnormord{O[\Phi]
  }U[F]^\dagger = \covnormord{F\triangleright O[\Phi]}.
  \label{Equation: covnormord covariance}
\end{equation}
Note that~\eqref{Equation: covnormord covariance} means that at this point we can simply write
\begin{equation}
  U[F]\Cov{O} U[F]^\dagger=  F\triangleright \Cov{O},
\end{equation}
with the understanding that $\Cov{O}$ transforms just like its classical counterpart (so for example if $O[\Phi]$ is a rank $n$ tensor, then so is $\Cov{O}$; this is not true of $O=\normord{O[\Phi]}$ which instead transforms anomalously).

To be clear, ordinary normal ordering \emph{can} be thought of as covariant, but only in a way that depends on the background time, as described in Section~\ref{Subsection: ordinary normord background-dependent}. On the other hand covariant normal ordering is \emph{background-independently} covariant. We avoid a fully descriptive name like `background-independently covariant normal ordering' since it would be too much of a mouthful.

We describe below how $\covnormord{O[\Phi]}$ is defined. For ease of reading we only give the most relevant details here, postponing a full explanation to Appendix~\ref{Appendix: more on covariant normal ordering}. For now, suffice it to say that the map between classical observables and quantum operators is 1-to-1, in the sense that every classical observable may be covariantly normal ordered to form a quantum operator, and every quantum operator may be written as a covariantly normal ordered classical observable.

In Section~\ref{Section: dressed operators}, we restrict to classical observables satisfying $O[\Phi]=F\triangleright O[\Phi]$ for all $F$; then $\Cov{O}$ is a gauge-invariant operator.

\subsection{Ordinary normal ordering is background-dependent}
\label{Subsection: ordinary normord background-dependent}

The usual quantization prescription for observables in QFT is normal ordering. However, normal ordering introduces additional background structure into the theory, which complicates the construction of reparametrization-invariant operators. In any case, theories of quantum gravity ought to be background-independent, so one should avoid introducing background structures if possible. At the very least, whatever background structures one does introduce should ultimately be discarded.

The extra background structure involved in normal ordering is the vacuum state. Recall that the vacuum $\ket{0}$ from the previous section is defined~\eqref{Equation: standard vacuum} such that it is annihilated by the positive frequency components of the fields $\Pi,\Vv$. Normal ordering is then defined such that positive frequency components of these fields always appear to the right of negative frequency components. But the notion of positive and negative frequency itself depends on a choice of \emph{time}; in $\ket{0}$ and this normal ordering, we use a \emph{background time} $v$.

We could equally well use a different background time coordinate $v'$ to define a different vacuum and normal ordering. Indeed, suppose $v'$ is related to $v$ by a diffeomorphism $F$:
\begin{equation}
  v' = F(v), \qquad F\in\Diff^+(\RR).
\end{equation}
Then we can define a new vacuum state $\ket{F}=U[F]^\dagger\ket{0}$, which one may straightforwardly confirm is annihilated by the positive frequency \emph{with respect to $v'$} components of the fields. Here positive/negative frequency with respect to $v'$ are defined by
\begin{equation}
  P^F_\pm\phi = F\triangleright P_\pm(F^{-1}\triangleright \phi).
\end{equation}
Explicitly, we have for example
\begin{equation}
  P^F_+\Pi(v) = \int_\RR\frac{\dd{u}}{2\pi i} \frac{\Pi(u)\partial_vF(v)}{F(v)-F(u)-i\epsilon}, \qquad
  P^F_+\Vv(v) = \int_\RR\frac{\dd{u}}{2\pi i} \frac{\Vv(u)\partial_uF(u)}{F(v)-F(u)-i\epsilon}.
\end{equation}
This similarly defines a different normal ordering in which $P_+^F\phi$ always appears to the right of $P_-^F\phi$. Let us denote the normal orderings defined with respect to the background time $v$ and the new time $v'=F(v)$ with $\normord{O[\Phi]}$ and $\normord{O[\Phi]}_{F}$ respectively.

We thus have inequivalent prescriptions for quantizing a given classical observable $O[\Phi]$, depending on the choice of vacuum / background time:
\begin{equation}
  \normord{O[\Phi]} \ne
  \normord{O[\Phi]}_{F}.
\end{equation}
The fact that these two operators are inequivalent can be made clear by writing $\normord{O[\Phi]}_{F}$ as an operator normal ordered with respect to $v$. It turns out that there is a concise general formula for this change of normal ordering:
\begin{equation}
  \normord{O[\Phi]}_F = \normord{\exp(-\frac\hbar{2\pi}\int_{\RR^2}\dd{u}\dd{v}G_{F}(u,v)\qty(\partial_u\fdv{\Pi(u)}\fdv{\Vv(v)}-\frac14\sum_i\fdv{\varphi_i(u)}\fdv{\varphi_i(v)}))O[\Phi]}.
  \label{Equation: change of time in normal ordering}
\end{equation}
The key ingredient in this formula is the propagator
\begin{equation}
  G_{F}(u,v) = \ln(\frac{F(u)-F(v)}{u-v}),
\end{equation}
which accounts for the difference between the two-point functions for the two times $v$ and $F(v)$, and appears in the above formula every time one reorders the various components of the fields past one another. This propagator enters the relationship between the expectation values of the same operator in different vacuums related to each other by a change of time. If we denote $|F\rangle= U(F)^\dagger|0\rangle$ we have
\begin{equation}
  \langle F|\varphi_i(u) \varphi_i(v)|F\rangle = \langle 0|\varphi_i(u) \varphi_i(v)|0\rangle - \frac{\hbar}{4\pi} G_F(v,u).
\end{equation}
The propagator $G_F(u,v)$ is regular in the limit $u\to v$. It is even analytical, and we have the expansion~\cite{Bianchi_2014,PhysRevD.13.2720,Birrell:1982ix}
\begin{equation}
  G_F(u,v) = \ln(F'(v)) + \frac12(u-v)\frac{F''(v)}{F'(v)} + \frac16(u-v)^2\qty(\frac{F'''(v)}{F'(v)}-\frac34\qty(\frac{F''(v)}{F'(v)})^2) + \order{(u-v)^3},
  \label{Equation: G expansion}
\end{equation}
To understand the general structure  of this expansion it is  convenient to consider the Hadamard propagator,~\cite{kay1991theorems, Janssen:2022vux} which is obtained through the derivative
\begin{equation}
  H_F(u,v): = \pa_u\pa_v G_F(u,v) =
  \frac{\pa_u F(u) \pa_vF(v)}{(F(u)-F(v))^2} -\frac{1}{(u-v)^2}.
  \label{Equation: H}
\end{equation}
This derivative can be shown to be invariant under M\"obius transformation $F \to \frac{aF+b}{cF+d}$ and therefore its expansion can be achieved entirely in terms of the Schwarzian derivative \eqref{Schwarz}~\cite{Bianchi_2014,PhysRevD.13.2720,Birrell:1982ix}:
\begin{align}
  H_F(u,v)
  &= \frac{1}{6}
  \{F,v\}+
  \frac{(u-v)}{12}  \pa_v \{F,v\} + \frac{(u-v)^2}{40} \left( \pa_v^2\{F,v\} +\frac{2}{3} \{F,v\}^2\right)
  +\order{(u-v)^3}.  \label{Equation: H expansion}
\end{align}

For example, consider the stress tensor $T= \Pi\partial_vV +\sum_i(\partial_v\varphi_i)^2$. Using this expansion
one finds
\begin{align}
  \normord{T(u)}_F &= \normord{T(u)} + \frac\hbar{4\pi}(2+M)\partial_u\partial_w G_F(u,w)|_{w=u} \\
  &= \normord{T(u)} + \frac{c\hbar}{24\pi}\Schwarzian{F}{u}.\label{Equation: stress tensor different normord}
\end{align}
where $c=c^0+c^\text{rad}=2+M$~\cite{Polchinski,DiFrancesco:1997nk}.
One explicitly sees that different choices of normal ordering correspond to different stress tensors (although note that the different stress tensors differ by constants, and so generate via commutators the same actions on other operators).

One may also confirm the following action of diffeomorphisms on normal ordered operators:
\begin{align}
  U[F]\normord{O[\Phi]}_{F\circ G}\, U[F]^\dagger
  & = \normord{F\triangleright O[\Phi]}_G.
  \label{Equation: ordinary normal ordering covariance}
\end{align}
This equation provides the covariant transformation law of normal-ordered observables, and makes the background-dependence of normal ordering particularly explicit. One sees that the action of the quantum reparametrization (i.e.\ conjugation by $U[F]$) only agrees with its action on the classical observable (i.e.\ within the normal ordering), if one \emph{also} acts on the background time  which defines normal ordering.
As a consistency check one may take the conjugation of~\eqref{Equation: stress tensor different normord} with~\eqref{Equation: ordinary normal ordering covariance} to  obtain the anomalous transformation law for the stress tensor \eqref{Ttransform}.

More generally, combining~\eqref{Equation: change of time in normal ordering} with~\eqref{Equation: ordinary normal ordering covariance}, one has
\begin{align}
  \MoveEqLeft U[F]\normord{O[\Phi]}U[F]^\dagger\\
  &= \normord{\exp(-\frac\hbar{2\pi}\int_{\RR^2}\dd{u}\dd{w}G_{F^{-1}}(u,w)\qty(\partial_u\fdv{\Pi(u)}\fdv{\Vv(w)}-\frac14\sum_i\fdv{\varphi_i(u)}\fdv{\varphi_i(w)}))F\triangleright O[\Phi]}\\
  &= \normord{F\triangleright\exp(\frac\hbar{2\pi}\int_{\RR^2}\dd{u}\dd{w}G_{F}(u,w)\qty(\partial_u\fdv{\Pi(u)}\fdv{\Vv(w)}-\frac14\sum_i\fdv{\varphi_i(u)}\fdv{\varphi_i(w)}))O[\Phi]}.
  \label{Equation: normal ordering anomaly}
\end{align}
The exponent here should be identified with $\hbar\mathscr{A}[F]$ appearing in~\eqref{Equation: technical summary A}. This reveals the precise general form of the reparametrization anomalies arising from a fixed normal ordering. For any observable that depends on both $\Pi$ and $V$, or quadratically on $\varphi_i$, the quantum action of a reparametrization on a normal ordered operator does not agree with the classical action of the same reparametrization on the corresponding classical observable -- there are quantum corrections contained in the exponential above.

This is a facet of the problem of time in quantum gravity that is uniquely field theoretical, and the foundational origin of diffeomorphism anomalies~\cite{Ciambelli_2024}. To quantize field operators, we bring in a notion of background time to define normal ordering -- but this goes against the desired background-independence of the theory.
Note that the \emph{violation of background independence} inherent to QFT is due to the fact that normal ordering is ultimately associated with a choice of vacuum, by definition one has $\langle 0|\normord{O[\Phi]}|0\rangle=O[0]$ and the $F$ normal ordering is associated with a different vacuum
\begin{equation}
  \langle F | \normord{O[\Pi,V,\varphi_i]}_F |F\rangle =O[0,F, 0] \qquad |F\rangle : = U[F]^\dagger|0\rangle.
\end{equation}
Note that the OPE perspective of AQFT \cite{Hollands:2023txn, Crawford:2021adf} attempts to address this issue by defining the renormalization, not in terms of a normal order, i.e a choice of vacuum  state, but in terms of the choice of a Hadamard parametrix. The parametrix depends however on a choice of metric representative and a choice of time.  It therefore suffers the same drawbacks as normal ordering: it provides a covariant prescription which is not background independent.

In what follows, we get rid of this background time by introducing a quantization prescription for observables that is both covariant and background-independent.

Before moving on, it is worth commenting that more na\"ive choices of operator ordering \emph{are} actually covariant in a background-independent way. For example, suppose we quantize an observable $O[\Pi,V]$ of $\Pi$ and $V$ such that $\Pi$ always appears to the left of $V$, e.g.\ for the observable $O[\Pi,V]=\Pi(v)V(v')$ the corresponding quantum operator would be $\hat O=\Pi(v)V(v')$. Such operators are of course background-independently covariant, since for example
\begin{equation}
  U[F]\hat OU[F]^\dagger = U[F]\Pi(v)U[F]^\dagger\,U[F]V(v')U[F]^\dagger = F\triangleright \Pi(v)F\triangleright V(v').
\end{equation}
The caveat is that this prescription only gives well-defined operators for the algebra generated by the smeared fields
\begin{equation}
  \int_\RR\dd{v}f(v)\Pi(v), \qquad \int_\RR\dd{v}g(v)V(v)
\end{equation}
for functions $f$ and $g$, which is somewhat limited. For example, $\hat O=\Pi(v)V(v')$ is not well-defined when $v=v'$ (whereas $\normord{\Pi(v)V(v)}$ is). More generally this quantization prescription is invalid for any composite observables of $\Pi$ and $V$. Since most diffeomorphism-invariant observables are composite, such a quantization prescription would clearly be unsuitable for our purposes.

\subsection{Radiative covariant normal ordering}
\label{Subsection: covnormord}

Let us now explain how to carry out covariant normal ordering, starting with observables depending only on the radiative fields $\varphi_i$. To do so, we split these fields into their positive and negative \emph{dressing time} frequency parts:
\begin{equation}
  \varphi_i = P_+^V\varphi_i+P_-^V\varphi_i,
\end{equation}
where
\begin{equation}
  P_+^V\varphi_i(v) = \int_\RR\frac{\dd{u}}{2\pi i}\frac{\varphi_i(u)\partial_u V(u)}{V(v)-V(u)-i\epsilon},
  \qquad
  P_-^V\varphi_i(v) = \int_\RR\frac{\dd{u}}{2\pi i}\frac{\varphi_i(u)\partial_u V(u)}{V(u)-V(v)-i\epsilon}.
\end{equation}
We can directly convert $P_\pm^V\varphi$ to quantum operators.
Because $V$ and $\varphi_i$ commute, and because $V$ commutes with itself at different times, there are no ordering ambiguities entering the definitions of these operators. Moreover, since $V$ and $\varphi_i$ transform in the same way under diffeomorphisms, $P_\pm^V\varphi_i$ are covariant operators:
\begin{equation}
  U[F](P_\pm^V\varphi_i) U[F]^\dagger = (P_\pm^V\varphi_i)\circ F.
\end{equation}
The covariant normal ordering of radiative observables $O[\varphi_i]$, denoted $\covnormord{O[\varphi_i]}$, is defined by having $P_-^V\varphi_i$ on the left, and $P_+^V\varphi_i$ on the right. This completely resolves any ordering ambiguities since $[P_\pm^V\varphi_i(u),P^V_\pm\varphi_i(v)]=0$. Moreover, it gives a well-defined quantum operator free of singularities, as can be most clearly seen by the simple formula
\begin{equation}
  \covnormord{O[\varphi_i]} = U_{\rad}[V]^\dagger\normord{O[\varphi_i\circ V]}U_{\rad}[V],
  \label{Equation: covariant normal order varphi}
\end{equation}
To establish this we first use that
$(P_\pm^V\varphi)\circ V^{-1} = U_{\rad}[V]^\dagger (P_\pm\varphi_i) U_{\rad}[V]$, which follows from
\begin{align}\nonumber
  (P_+^V\varphi_i)\circ V^{-1}(\mv) &= \int_\RR\frac{\dd{u}}{2\pi i}\frac{\partial_u \hat{V}(u) \varphi_i(u)}{\mv-V(u)-i\epsilon} = \int_\RR\frac{\dd{\mathrm{u}}}{2\pi i}\frac{ (\varphi_i\circ V^{-1})(\mathrm{u})}{\mv-\mathrm{u}-i\epsilon} = U_{\rad}[V]^\dagger\int_\RR\frac{\dd{\mathrm{u}}}{2\pi i}\frac{ \varphi_i(\mathrm{u})}{\mv-\mathrm{u}-i\epsilon} U_{\rad}[V].
\end{align}
Then we use that $V$ commutes with $U_{\rad}[V]$.

Provided ordinary normal ordering yields well-defined composite operators,~\eqref{Equation: covariant normal order varphi} guarantees the same for covariant normal ordering.
This can also be seen through the  concise general formula:
\begin{equation}
  \covnormord{O[\varphi_i]} = \normord{\exp(\frac\hbar{
  8\pi}\int_{\RR^2}\dd{u}\dd{v}G_{V}(u,v)\sum_i\fdv{\varphi_i(u)}\fdv{\varphi_i(v)})O[\varphi_i]},
  \label{covnormordrad}
\end{equation}
which follows from substituting $F\to V$ in~\eqref{Equation: change of time in normal ordering}.

For example, consider the covariantly normal ordered radiative stress tensor
$\Covrad{T}$.
Using~\eqref{Equation: covariant normal order varphi} and the anomalous transformation \eqref{Ttransform} law for the radiative stress tensor we have
\begin{align}
  \Covrad{T} = \sum_i\covnormord{(\partial_v\varphi_i)^2} &= U_{\rad}[V]^\dagger\sum_i\normord{(\partial_v\varphi_i)^2\circ V}U_{\rad}[V] \\
  &= U_{\rad}[V]^\dagger\qty((\partial_vV)^2 T^\text{rad}\circ V) U_{\rad}[V]\\
  &= (\partial_vV)^2 \qty(U_{\rad}[V]^\dagger T^\text{rad}U_{\rad}[V])\circ V\\
  &= T^\text{rad} + \frac{c^{\text{rad}}\hbar}{24\pi}\Schwarzian{V}{v}.
  \label{Equation: covradT}
\end{align}
This follows also from~\eqref{covnormordrad} using the expansion~\eqref{Equation: H expansion}.

\subsubsection{Covariant star product}

It is well-known that the composition of two (ordinarily) normal ordered operators may be written in normal ordered form:
\begin{equation}
  O_1 O_2= \normord{O_1[\Phi]}\normord{O_2[\Phi]} = \normord{O_{1}*O_{2}[\Phi]},
\end{equation}
using Wick's theorem. This defines the Wick-Moyal `star product' $*$. Similarly, the product of two covariantly ordered operators defines a `covariant star product' $\star$ via
\begin{equation}
  \Cov{O_1}\Cov{O_2}=
  \covnormord{O_1[\Phi]}\covnormord{O_2[\Phi]} = \covnormord{O_{1}\star O_2[\Phi]},
\end{equation}
for any $O_1,O_2$.
One can thus understand the requirement of covariance as deforming the usual star product to the covariant one. We can express this deformation purely at the operator level. Composing the covariant star product with the inverse of the regular star product we obtain that the operator product of covariant operators can be viewed as a deformation of the operator product of ordinarily normal ordered operators:
\begin{equation}
  \Cov{O_1}\Cov{O_2}=
  (O_1\hatstar O_2)^{\star},
  \label{Equation: hatstar defn}
\end{equation}
We see that while quantization can be understood as a deformation of a commutative classical algebra, covariant quantization can be understood as a further deformation of the non-commutative quantum algebra.

For purely radiative observables $O_1[\varphi]$, $O_2[\varphi]$, we have
\begin{align}
  O_1\hatstar O_2
  = m \left[ \exp(\frac1{\pi\hbar}
      \int_{\RR^2}\dd{u}\dd{v}H_{V}(u,v)
      \sum_i  \text{ad}\varphi_i(u)\otimes
      \text{ad}\varphi_i(v)
    )
  O_1\otimes O_2\right]
  \label{Equation: hatstar formula}
\end{align}
where $m(O_1\otimes O_2)=O_1O_2$ denotes the multiplication of operators and $\text{ad}\varphi_i(u) O= [\varphi_i(u),O]$ denotes the adjoint action. To derive this, one can first use
\begin{equation}
  \text{ad}\partial_u\varphi_i(u)\normord{O[\Phi]} = -\frac{i\hbar}2\normord{\fdv{\varphi_i(u)}O[\Phi]}
\end{equation}
to rewrite~\eqref{covnormordrad} as
\begin{equation}
  \Cov{O} = \exp(-\frac1{
    2\pi\hbar}\int_{\RR^2}\dd{u}\dd{v}H_{V}(u,v)\sum_i\text{ad}\varphi_i(u)
  \text{ad}\varphi_i(v))O.
\end{equation}
Applying this, and the inverse formula
\begin{equation}
  O = \exp(\frac1{
    2\pi\hbar}\int_{\RR^2}\dd{u}\dd{v}H_{V}(u,v)\sum_i\text{ad}\varphi_i(u)
  \text{ad}\varphi_i(v))\Cov{O},
\end{equation}
to both sides of~\eqref{Equation: hatstar defn} yields~\eqref{Equation: hatstar formula}.

The covariant star product $O_1\star O_2[\varphi]$ of $O_1[\varphi]$, $O_2[\varphi]$ may be computed with Wick's theorem by summing over contractions with a leg in each of $O_1[\varphi]$ and $O_2[\varphi]$, of the following form:
\begin{equation}
  \contraction{}{\varphi_i}{(u)}{\varphi_j}
  \varphi_i(u)\varphi_j(v)
  = -\frac\hbar{4\pi}\delta_{ij}\ln(V(u)-V(v)-i\epsilon).
\end{equation}
These are the same as the contractions that would appear in the ordinary Wick's theorem, but dressed by $V$.

\subsection{Covariant normal ordering including spin \texorpdfstring{$0$}{0}}
\label{Subsection: covnormord 0}
Now let us address covariantly normal ordered observables of the spin 0 fields. In order to avoid any ambiguity, we temporarily write in this subsection $\Pi,V$ for the classical fields, and $\hat\Pi,\hat V$ for the quantum fields.

First, consider an observable $O[\Pi,V]=\mathcal{F}[V]$ depending on $V$ alone. We set
\begin{equation}
  \Cov{O} = \covnormord{O[\Pi,V]}=\mathcal{F}[\hat V].
  \label{Equation: covariant normal ordered F[V]}
\end{equation}
Since $\hat V$ commutes with itself, this is covariant and well-defined. Next, suppose $O[\Pi,V]$ depends linearly on $\Pi$. Without loss of generality let us set $O[\Pi,V]=\Pi(v) \mathcal{F}[V]$ (the definition given below extends to more general such linear in $\Pi$ observables by requiring linearity of the covariant normal ordering prescription, i.e. $\covnormord{\alpha A[\Phi]+\beta B[\Phi]}=\alpha\covnormord{A[\Phi]}+\beta\covnormord{B[\Phi]}$ for $\alpha,\beta\in\CC$). How should we covariantly normal order this observable? Setting
\begin{equation}
  \covnormord{O[\Pi,V]}\stackrel{?}{=}\hat\Pi(v)\mathcal{F}[\hat V]
\end{equation}
would be covariant -- but, as described above, it is not always well-defined, generally giving a divergent operator. To get something well-defined, we could use ordinary normal ordering
\begin{equation}
  \covnormord{O[\Pi,V]}\stackrel{?}{=}\normord{\Pi(v)\mathcal{F}[V]} = \hat\Pi_-(v)\mathcal{F}[\hat V]+\mathcal{F}[\hat V]\hat\Pi_+(v),
\end{equation}
but this would not be covariant. To see how to proceed, note that the ordinary normal ordering can be written in the form
\begin{align}
  \normord{\Pi(v)\mathcal{F}[V]} &= \hat \Pi(v)\mathcal{F}[\hat V] - \comm{\hat \Pi_+(v)}{\mathcal{F}[\hat V]} \\
  &= \hat \Pi(v)\mathcal{F}[\hat V] - \int\frac{\dd{u}}{2\pi i} \frac1{v-u-i\epsilon} \comm{\hat \Pi(u)}{\mathcal{F}[\hat V]} \\
  &= \hat \Pi(v)\mathcal{F}[\hat V] + \frac{\hbar}{2\pi}\int\dd{u}\frac1{v-u-i\epsilon} \fdv{\mathcal{F}[\hat V]}{\hat V(u)}.
  \label{Equation: linear normal ordering}
\end{align}
The purpose of the latter term is to subtract off the divergent contributions from the na\"ive first term, but it involves a projection onto positive frequencies in the \emph{background time}, which is the source of non-covariance. Motivated by this, we can define the covariant normal ordering of $O[\Pi,V]$ by simply replacing this projection by one onto positive frequencies in the \emph{dressing time}:
\begin{equation}
  \covnormord{O[\Pi,V]}:=\hat{\Pi}(v)\mathcal{F}[\hat{V}] + \frac{\hbar}{2\pi}\int\dd{u}\frac{\partial_v\hat V(v)}{\hat V(v)-\hat V(u)-i\epsilon} \fdv{\mathcal{F}[\hat V]}{\hat V(u)}.
  \label{Equation: linear covariant normal ordering}
\end{equation}
We can verify covariance as follows:\footnote{Note that $U[F]\fdv{\mathcal{F}[\hat V]}{\hat V}U[F^{-1}] = \fdv{\mathcal{F}[\hat V\circ F]}{(\hat V\circ F)} = \partial_v F \fdv{\mathcal{F}[\hat V\circ F]}{\hat V}\circ F$.}
\begin{align}
  \MoveEqLeft U[F]\covnormord{O[\Pi,V]}U[F^{-1}]\\
  &=\partial_v F(v)\hat \Pi(F(v))\mathcal{F}[\hat V\circ F] + \frac{\hbar}{2\pi}\int\dd{u}\frac{\partial_v(\hat V\circ F)(v)}{(\hat V(F(v))-\hat V(F(u))-i\epsilon)} \fdv{\mathcal{F}[\hat V\circ F]}{(\hat V\circ F)(u)}\\
  &=\partial_v F(v)\hat\Pi(F(v))\mathcal{F}[\hat V\circ F] + \frac{\hbar}{2\pi}\partial_v F(v)\int\dd{u}\frac{\partial_v\hat V(F(v))}{(\hat V(F(v))-\hat V(u)-i\epsilon)} \fdv{\mathcal{F}[\hat V\circ F]}{\hat V(u)}\label{Equation: linear covariance change of variables}\\
  &= \covnormord{\partial_vF(v)\Pi(F(v))\mathcal{F}[V\circ F]}\label{Equation: linear covariance by definition}\\
  &=\covnormord{O[F\triangleright \Pi, F\triangleright V]},
\end{align}
where~\eqref{Equation: linear covariance change of variables} follows from a change of variables $F(u)\to u$, and~\eqref{Equation: linear covariance by definition} follows from the definition~\eqref{Equation: linear covariant normal ordering}. We can verify well-definedness by combining~\eqref{Equation: linear normal ordering} and~\eqref{Equation: linear covariant normal ordering} to write
\begin{equation}
  \covnormord{\Pi(v)\mathcal{F}[V]} = \normord{\Pi(v)\mathcal{F}[V]} + \frac\hbar{2\pi}\int\dd{u}\partial_vG_{\hat V}(v,u)\fdv{\mathcal{F}[\hat V]}{\hat V(u)}.
  \label{Equation: covariant normal ordering linear}
\end{equation}
The first term on the left-hand side is well-defined by the properties of ordinary normal ordering, while the second term is well-defined since $\partial_vG_{\hat V}(v,u)$ is a smooth kernel.

The formulas for covariant normal ordering given above suffice for most of the paper, but in Appendix~\ref{Appendix: more on covariant normal ordering}, we explain how to extend covariant normal ordering to observables $O$ that depend arbitrarily on $\Pi,V,\varphi$. To order $\order{\hbar}$, the general formula is
\begin{equation}
  \covnormord{O[\Phi]} = \normord{O[\Phi]} -\frac\hbar{2\pi}\int_{\RR^2}\dd{u}\dd{w}\normord{G_{V}(u,w)\qty(\partial_u\fdv{\Pi(u)}\fdv{V(w)}-\frac14\sum_i\fdv{\varphi_i(u)}\fdv{\varphi_i(w)})O[\Phi]} +\order{\hbar^2}.
  \label{Equation: covnormord to order hbar}
\end{equation}
Comparing with~\eqref{Equation: change of time in normal ordering}, one sees that covariant normal ordering can approximately be thought of as normal ordering with respect to dressing time, i.e.
\begin{equation}
  \covnormord{O[\Phi]}\approx \normord{O[\Phi]}_{\hat V} + \order{\hbar^2}.
  \label{Equation: covnormord approx normord V}
\end{equation}
In fact, for purely radiative observables the $\order{\hbar^2}$ term in~\eqref{Equation: covnormord approx normord V} vanishes, as explained in Subsection~\ref{Subsection: covnormord}.

But for more general observables, one should be careful with this, because $\hat V$ is an operator, not a fixed diffeomorphism, so for general observables $O[\Phi]$ the notation $\normord{O[\Phi]}_{\hat V}$ does not quite make complete sense within the framework described previously. Moreover, the quadratic and higher order in $\hbar$ terms in $\covnormord{O[\Phi]}$ do not match with a na\"ive substitution of $F=\hat V$ into~\eqref{Equation: change of time in normal ordering}, essentially due to the non-commutativity of $\hat \Pi, \hat V$.

From now on, we drop the notation $\hat\Pi,\hat V$, and just use $\Pi,V$ for both classical and quantum fields; it should always be clear from the context which is meant.

\subsubsection{Covariant star product}

Let's now explain how the covariant star product extends to observables involving the spin 0 fields.

Using Wick's theorem, the ordinary star product of two general classical observables $O_1[\Phi]$, $O_2[\Phi]$ may be written in the form
\begin{align}
  \MoveEqLeft O_{1}*O_{2}[\Phi]
  \label{Equation: Wick's theorem}\\
  &=O_1[\Phi]\exp(\textstyle\frac\hbar{2\pi}\int_{\RR^2}\dd{u}\dd{v}\ln(u-v-i\epsilon)\qty(\partial_u\frac{\leftdelta}{\delta \Pi(u)}\frac{\rightdelta}{\delta V(v)}+\frac{\leftdelta}{\delta V(u)}\partial_v\frac{\rightdelta}{\delta \Pi(v)}-\frac12\sum_i\frac{\leftdelta}{\varphi_i(u)}\frac{\rightdelta}{\varphi_i(v)}))O_2[\Phi].\nonumber
\end{align}
The ${}^\leftarrow,{}^\rightarrow$ notation here denotes functional derivatives acting to the left and right respectively.
To order $\order{\hbar}$, one may compute the covariant star product by combining~\eqref{Equation: Wick's theorem} with~\eqref{Equation: covnormord to order hbar}. One finds
\begin{multline}
  O_1 \star O_2[\Phi] = O_1[\Phi]O_2[\Phi]\\
  +O_1[\Phi]\qty(\textstyle\frac\hbar{2\pi}\int_{\RR^2}\dd{u}\dd{v}\ln(V(u)-V(v)-i\epsilon)\qty(\partial_u\frac{\leftdelta}{\delta \Pi(u)}\frac{\rightdelta}{\delta V(v)}+\frac{\leftdelta}{\delta V(u)}\partial_v\frac{\rightdelta}{\delta \Pi(v)}-\frac12\sum_i\frac{\leftdelta}{\varphi_i(u)}\frac{\rightdelta}{\varphi_i(v)}))O_2[\Phi]\\
  + \order{\hbar^2}.
  \label{Equation: covariant Wick's theorem to order hbar}
\end{multline}
This is similar to~\eqref{Equation: Wick's theorem} at $\order{\hbar}$, except with the positive/negative background time frequency kernel $\ln(u-v-i\epsilon)$ being replaced by one with respect to the dressing time $V$. One might guess that the higher order terms in $\hbar$ in the covariant star product can similarly be obtained by making this same substitution in~\eqref{Equation: Wick's theorem}, but one must be careful about how the various $V$ functional derivatives interact with the $V$-dependent integration kernels. In Appendix~\ref{Appendix: more on covariant normal ordering} we address the higher order terms.

It is illuminating to change variables from $\Pi,V,\varphi_i$ to $\tilde\Pi,X,\tilde\varphi_i$. Using $P_\pm\fdv{O}{\tilde\Pi} = \fdv{O}{P_\mp\tilde\Pi}$ and the chain rule for variational derivatives
\begin{gather}
  \fdv{O}{\Pi} \to \fdv{O}{\tilde\Pi} \circ V,
  \qquad
  \fdv{O}{\varphi_i} \to \partial_vV \fdv{O}{\tilde\varphi_i} \circ V,
  \\
  \fdv{O}{V} \to \partial_vV\qty(\tilde\Pi\partial_\mv\fdv{O}{\tilde\Pi}-\sum_i\partial_\mv\tilde\varphi_i\fdv{O}{\tilde\varphi_i}-\partial_\mv X\fdv{O}{X})\circ V,
\end{gather}
one finds the first order expansion of the covariant star product can be written as
\begin{multline}
  O_1\star O_2[\Phi]
  = O_1[\Phi]O_2[\Phi] \\
  + i\hbar\int\dd{\mv}{\textstyle\qty(\fdv{O_1[\Phi]}{P_+\tilde\Pi}\tilde{\mathcal{L}}\qty(O_2[\Phi]) -\tilde{\mathcal{L}}\qty(O_1[\Phi])\fdv{O_2[\Phi]}{P_-\tilde\Pi}-\frac12\int\dd{\rm u}\ln({\rm u}-\mv-i\epsilon)\sum_i\fdv{O_1[\Phi]}{\tilde\varphi_i({\rm u})}\fdv{O_2[\Phi]}{\tilde\varphi_i({\rm v})})}
  + \order{\hbar^2},
  \label{Equation: other covariant Wick's theorem to order hbar}
\end{multline}
where
\begin{equation}
  \tilde{\mathcal{L}} = \partial_\mv X \fdv{X} + \sum_i\partial_\mv\tilde\varphi_i\fdv{\tilde\varphi_i} - \tilde\Pi \partial_\mv\fdv{\tilde\Pi}
\end{equation}
acts as Lie derivatives on the fields $\tilde\Pi,X,\tilde\varphi_i$, with $\tilde\Pi$ treated as a 1-form, and $X,\tilde\varphi_i$ as scalars. Denoting $\tilde \lie_{\tilde{f}}:= \int \dd{\mv} \tilde f \tilde{\mathcal{L}}$ we find
\begin{align}
  \tilde \lie_{\tilde{f}} \tilde\Pi=\partial_\mv(\tilde{f}\tilde\Pi),\qquad
  \tilde \lie_{\tilde{f}} \tilde\varphi_i=
  \partial_v\tilde{f}\partial_\mv\tilde\varphi_i  ,\qquad
  \tilde \lie_{\tilde{f}} X=\tilde{f}\partial_\mv X.
\end{align}
We see that~\eqref{Equation: other covariant Wick's theorem to order hbar} accounts for contractions with $\tilde\Pi$ acting as a generator of these infinitesimal diffeomorphisms, which we refer to as `reorientations' in \cite{LocalizationAnomalous}. There are also ordinary Wick contractions between the dressed radiative fields $\tilde\varphi_i$.

\subsection{Example: covariantly normal ordered stress tensor}

A noteworthy example of covariant normal ordering is the covariantly normal ordered stress tensor
\begin{align}
  \Cov{T} &= \Cov{\tau}+\Covrad{T} \\
  &=  \covnormord{\Pi\partial_vV} + \sum_i\covnormord{(\partial_v\varphi_i)^2}
\end{align}
For the first term above we can invoke~\eqref{Equation: covariant normal ordering linear} with the expansion~\eqref{Equation: G expansion}, finding\footnote{More generally, Given a function $\mathcal{F}(V^{0},\cdots, V^{n})$ on $\mathbb{R}^{n+1}$, We can define a local field functional $\mathcal{F}[V](u) := \mathcal{F}(V(u),V'(u), \cdots)$. We can evaluate the covariant normal order product to be
  \begin{align}
    \covnormord{\Pi(v) \mathcal{F}[V](u)} &= \normord{\Pi(v) \mathcal{F}[V](u)} + \frac{\hbar}{2\pi}\left(
    \sum_{n\geq 1} \pa^n_u\pa_vG_V(v,u) \frac{\pa \mathcal{F}}{\pa V^{(n)}}[V](u) \right)
  \end{align} Taking the limit $u=v$ and using the expansions (\ref{Equation: G expansion}, \ref{Equation: H expansion}) gives
  \begin{align}
    \covnormord{\Pi(v) \mathcal{F}[V](v)}=
    \frac{\hbar}{4\pi} \left(\frac{V''}{V'}\right)
    \frac{\pa \mathcal{F}}{\pa V}[V](v)
    +\frac{\hbar}{12\pi}
    \{V,v\} \frac{\pa \mathcal{F}}{\pa V'}[V](v)+\frac{\hbar}{24\pi}\pa_v\{V,v\} \frac{\pa \mathcal{F}}{\pa V''}[V](v) +\cdots
\end{align} we use }
\begin{equation}
  \Cov{\tau} = \covnormord{\Pi\partial_vV} = \normord{\Pi\partial_v V} +   \frac{\hbar}{12\pi}\Schwarzian{V}{v} = \tau+\frac{c^0\hbar}{24\pi}\Schwarzian{V}{v}.
\end{equation}
For the second term, we can use~\eqref{Equation: covradT}.
Altogether, we have
\begin{equation}
  \Cov{T} = {T} + \frac{c\hbar}{24\pi}\Schwarzian{V}{v},
  \label{Equation: covnormord T}
\end{equation}
where $c=c^0+c^\text{rad}=2+M$.
The second term here is responsible for cancelling the usual anomaly in the stress tensor. Indeed, under a reparametrization one may directly compute
\begin{align}
  U[F] \Cov{T} U[F]^\dagger &= U[F]\qty(T + \frac{c\hbar}{24\pi}\Schwarzian{V}{v})U[F]^\dagger\\
  &= (\partial_vF)^2T\circ F - \frac{c\hbar}{24\pi}\Schwarzian{F}{v} + \frac{c\hbar}{24\pi}\Schwarzian{V\circ F}{v}\\
  &= (\partial_vF)^2\qty(T+\frac{c\hbar}{24\pi}\Schwarzian{V}{v})\circ F\\
  & = (\partial_v F)^2\Cov{T}\circ F,
\end{align}
as required.

\section{Quantum dressed observables}
\label{Section: dressed operators}

Suppose that $\tilde{O}[\Phi]$ is a classical gauge-invariant functional depending on the fields $\Phi$. It then follows straightforwardly from~\eqref{Equation: covnormord covariance} that
the covariantly normal ordered operator $\Cov{\tilde{O}} := \covnormord{\tilde{O}[\Phi]}$ is gauge-invariant at the quantum level:
\begin{equation}
  \Cov{\tilde{O}} = U[F]\Cov{\tilde{O}}U[F]^\dagger.
\end{equation}
Due to the 1-to-1 correspondence between operators and classical observables $O[\Pi,V,\varphi_i]$, the converse is also true: \emph{any gauge-invariant operator may be written as the covariant normal ordering of a gauge-invariant classical observable}.

As described in~\cite{LocalizationAnomalous}, the full set of gauge-invariant classical observables of the null ray segment $\Segment$ consists of observables dressed to the dressing time $V$. For example, the dressed area element is $\tilde\Omega=\Omega\circ X$, where $X=V^{-1}$. Therefore, by combining this classical dressing technique with the covariant normal ordering prescription described in the previous section, one may construct all gauge-invariant operators. The purpose of this section is to explain in detail how this works.

Let's first review the basic structures underlying dressing in the classical theory, so that we can compare what changes in the quantum theory. Given a general (i.e.\ not necessarily gauge-invariant) classical observable $O[\Pi,V,\varphi_i]$, we want to construct a gauge-invariant \emph{dressed} observable $\tilde O[\Pi,V,\varphi_i]$. This can be implemented in two steps: (i) gauge-fixing, and (ii) dressing. Gauge-fixing proceeds by imposing a gauge-fixing condition, which  should be reachable from any field configuration by acting with an appropriate gauge transformation, and whose aim is to eliminate any remaining gauge freedom. Here, we simply gauge fix the dressing time to be equal to the background time, i.e.\ $V(v)=v$, and define the gauge-fixed observable
\begin{equation}
  \gf{O}[\Pi,\varphi_i] = O[\Pi,V,\varphi_i]\big|_{V(v)=v}.
  \label{Equation: classical gauge fix}
\end{equation}
Next, given any functional (possibly arising from gauge-fixing) $\mathcal{O}[\Pi,\varphi_i]$ of $\Pi$ and $\varphi_i$, we define the dressed observable $\tilde{\mathcal{O}}[\Pi,V,\varphi_i]$ as the unique gauge-invariant observable which agrees with $\mathcal{O}[\Pi,\varphi_i]$ after gauge-fixing:
\begin{equation}
  F\triangleright \tilde{\mathcal{O}}[\Pi,V,\varphi_i] = \tilde{\mathcal{O}}[\Pi,V,\varphi_i], \qquad \gf{\tilde{\mathcal{O}}}[\Pi,\varphi_i] = \mathcal{O}[\Pi,\varphi_i].
\end{equation}
It is clear that one can obtain $\tilde{\mathcal{O}}[\Pi,V,\varphi_i]$ by substituting the bare fields $\varphi_i$ and $\Pi$ for the dressed fields $\tilde\varphi_i=\varphi_i\circ X$ and
\begin{equation}
  \tilde{\Pi}:= (\pa_{\mv} X) \Pi\circ X = \pa_v X \frac{\tau\circ X}{\pa_vV\circ X} = (\pa_{\mv} X)^2 \tau\circ X =\tilde{\tau}
\end{equation}
(where we use the connection between $\Pi$ and $\tau$ in \eqref{Pitau}) in $\mathcal{O}[\Pi,\varphi_i]$:
\begin{equation}
  \tilde{\mathcal{O}}[\Pi,V,\varphi_i] = \mathcal{O}[\tilde\tau,\tilde\varphi_i].
  \label{Equation: most general gauge-invariant observable}
\end{equation}
In this way, the most general gauge-invariant classical observable may be written as a function of the dressed fields alone.

In the quantum theory, we cannot directly `set $V(v)=v$' to gauge fix a quantum operator $O$ without specifying an ordering prescription, because $V$ is an operator that does not commute with  $\Pi$. Let us use ordinary normal ordering $O=\normord{O[\Pi,V,\varphi_i]}$, and define the corresponding gauge-fixed operator to be $\gf{O}=\normord{\gf{O}[\Pi,\varphi_i]}$, with $\gf{O}[\Pi,\varphi_i]$ determined by~\eqref{Equation: classical gauge fix}. In this way we get a `gauge-fixing map'
\begin{equation}
  \begin{array}{rrcl}
    \GFmap\,: & \mathcal{A}_{\text{kin}} &\to & \gf{\mathcal{A}}\\
    & O &\mapsto& \gf{O}
  \end{array}
\end{equation}
taking a general operator $O\in\mathcal{A}_{\text{kin}}$ to its gauge-fixed counterpart $\gf{O}\in\gf{\mathcal{A}}$, where
\begin{equation}
  \mathcal{A}_{\text{kin}}=\mathcal{B}(\mathcal{K}^0\otimes\mathcal{H}^\text{rad})
\end{equation}
is the algebra of operators acting on the spin 0 and radiative fields, whereas
\begin{equation}
  \gf{\mathcal{A}} = \mathcal{A}_\Pi \otimes \mathcal{B}(\mathcal{H}^\text{rad}), \qquad \mathcal{A}_\Pi = \{\Pi\}'' = \{\text{functions of $\Pi$}\}
\end{equation}
is the algebra of gauge-fixed operators.\footnote{Note $\gf{\mathcal{A}}=\{\Pi\}' = \mathcal{A}_\Pi'$.} The map $\GFmap$ was defined above using normal ordering
\begin{equation}
  \GFmap(\normord{O[\Pi,V,\varphi_i]}) = \normord{\gf{O}[\Pi,\varphi_i]},
\end{equation}
but it is worth noting that using covariant normal ordering gives an equivalent definition:
\begin{equation}
  \GFmap(\covnormord{O[\Pi,V,\varphi_i]}) = \covnormord{\gf{O}[\Pi,\varphi_i]}.
  \label{Equation: gf map}
\end{equation}
This is because covariant normal ordering and ordinary normal ordering agree when we set $V(v)=v$.\footnote{This is simplest to see with the technology of Appendix~\ref{Appendix: more on covariant normal ordering}. The covariant normal ordering on the spin 0 fields is implemented by~\eqref{covnormord0}, with the map $\mathcal{D}$ acting trivially at $V(v)=v$, while the covariant normal ordering on the radiative fields is ordinary normal ordering with respect to $V$, which agrees with ordinary background time normal ordering when $V(v)=v$.}

The dressing of a general gauge-fixed operator $\mathcal{O}\in\gf{\mathcal{A}}$ can then be defined in the same way as the classical case: it is the unique gauge-invariant operator $\Dmap(\mathcal{O})$ whose gauge-fixing agrees with $\mathcal{O}$, i.e.
\begin{equation}
  U[F]\Dmap(\mathcal{O})U[F]^\dagger = \Dmap(\mathcal{O}),\qquad \GFmap(\Dmap(\mathcal{O})) = \mathcal{O}.
\end{equation}
By the covariance of the 1-to-1 correspondence between covariantly normal ordered operators and classical observables, the dressed operator $\Dmap(\mathcal{O})$ is well-defined and given by
\begin{equation}
  \Dmap(\mathcal{O}) = \Cov{\tilde{\mathcal{O}}} = \covnormord{\tilde{\mathcal{O}}[\Pi,V,\varphi_i]} :=\covnormord{\mathcal{O}[\tilde\tau,\tilde\varphi_i]}.
\end{equation}
In other words, we just need to substitute the dressed fields $\Pi\to\tilde\tau$, $\varphi_i\to\tilde\varphi_i$ in the covariantly normal ordered expression $\mathcal{O}=\covnormord{\mathcal{O}[\Pi,\varphi_i]}$. This defines the `dressing map'
\begin{equation}
  \begin{array}{rrcl}
    \Dmap\,: & \gf{\mathcal{A}} &\to & \mathcal{A}_{\text{kin}}^{\Diff^+(\RR)}\\
    & \mathcal{O} &\mapsto& \Cov{\tilde{\mathcal{O}}},
  \end{array}
\end{equation}
where $\mathcal{A}_{\text{kin}}^{\Diff^+(\RR)}$ denotes the set of operators invariant under the reparametrization group $\Diff^+(\RR)$. Note that $\GFmap\circ \Dmap$ is by definition the identity acting on $\gf{\mathcal{A}}$, but $\Dmap\circ\GFmap$ is not the identity -- rather it is a projection onto gauge-invariant operators $\mathcal{A}_{\text{kin}}^{\Diff^+(\RR)}\subset\mathcal{A}_{\text{kin}}$.\footnote{Occasionally we abuse notation in this paper to just write $\Dmap$ for $\Dmap\circ\GFmap$. For example $\Dmap(\tau)=\Dmap(\GFmap(\tau))=\Dmap(\Pi)$.}

Therefore, the most general gauge-invariant operator is a covariantly normal ordered functional of the dressed radiative fields $\tilde\varphi_i$ and of the reorientation generator $\tilde\tau$, of the form $\covnormord{\mathcal{O}[\tilde\tau,\tilde\varphi_i]}$. Moreover, a general gauge-invariant operator can be written as a combination of $\Dmap(\Pi)=\Cov{\tilde\tau}$ and dressed radiative operators $\Dmap(a)$, $a\in\mathcal{B}(\mathcal{H}_S)$.\footnote{This is implied by the fact that, if two bare operators $\mathcal{O}_1,\mathcal{O}_2$ are degree $m,n$ in $\Pi$ respectively, then $\mathcal{O}_1\Dstar\mathcal{O}_2$ is degree $m+n$ in $\Pi$, and moreover the leading order term in $\mathcal{O}_1\Dstar\mathcal{O}_2$ is $\mathcal{O}_1\mathcal{O}_2$, where $\Dstar$ is defined in Subsection~\ref{Subsection: Dirac deformation}.} This is the main lesson of the crossed product construction~\cite{Connes1994,Takesaki2003II,Takesaki2003III} (see Section~\ref{Section: crossed product}). Indeed, the full algebra of gauge-invariant operators may be written in the form
\begin{equation}
  \mathcal{A}_{\text{kin}}^{\Diff^+(\RR)} = \{\Cov{\tilde\tau},\, \Dmap(a)\,\mid\,a\in\mathcal{B}(\mathcal{H}^\text{rad})\}''.
  \label{Equation: crossed product}
\end{equation}
In Section~\ref{Section: three quantum diffeomorphisms}, we show that $\Cov{\tilde\tau}$ generates a representation of the Virasoro group.
The algebra~\eqref{Equation: crossed product} is therefore precisely the form of a crossed product $\mathcal{B}(\mathcal{H}^\text{rad})\rtimes\operatorname{Vir}$, with $\Cov{\tilde\tau}$ playing the role of the reorientation generator (as explained in Section~\ref{Subsection: reorientation algebra}).

\subsection{Dressing is an endomorphism for radiative field operators}
\label{Subsection: radiative endomorphism}

The goal of this section is to note that dressing is an algebraic endomorphism when restricted to purely radiative operators $\mathcal{O}_1,\mathcal{O}_2\in \mathcal{B}(\mathcal{H}^{\text{rad}})$, meaning that
\begin{equation}
  \Dmap(\mathcal{O}_1)\Dmap(\mathcal{O}_2) = \Dmap(\mathcal{O}_1\mathcal{O}_2).
  \label{Equation: Dstar rad}
\end{equation}
This follows directly from the simple formula (already given in~\eqref{Equation: dressed field operator} and~\eqref{Equation: D a})
\begin{equation}
  \mathcal{O}\in\mathcal{B}(\mathcal{H}^\text{rad}) \implies \Dmap(\mathcal{O}) = U_\text{rad}[V]^\dagger\mathcal{O}U_\text{rad}[V].
  \label{Equation: dressed not Pi dependent}
\end{equation}
The endomorphism thus maps $\mathcal{B}(\mathcal{H}^\text{rad})$ onto
\begin{equation}
  \Dmap\qty(\mathcal{B}(\mathcal{H}^\text{rad})) = U_\text{rad}[V]^\dagger\mathcal{B}(\mathcal{H}^\text{rad})U_\text{rad}[V] \subset \mathcal{B}(\mathcal{K}_{\text{kin}})^{\operatorname{Diff}^+(\RR)}.
\end{equation}

With~\eqref{Equation: dressed not Pi dependent} we may rewrite~\eqref{Equation: crossed product} as
\begin{equation}
  \mathcal{A}_{\text{kin}}^{\Diff^+(\RR)} = \{\Cov{\tilde\tau},\, U_\text{rad}[V]^\dagger a U_\text{rad}[V]\,\mid\,a\in\mathcal{B}(\mathcal{H}^\text{rad})\}'',
  \label{Equation: crossed product 2}
\end{equation}
which makes the crossed product nature of this algebra even more apparent.

To demonstrate~\eqref{Equation: dressed not Pi dependent}, we write the operator in its normal ordered form
\begin{equation}
  \mathcal{O} = \normord{\mathcal{O}[\varphi_i]},
\end{equation}
so that we have by definition
\begin{equation}
  \Dmap(\mathcal{O}) = \covnormord{\mathcal{O}[\tilde\varphi_i]}.
\end{equation}
Now we may invoke~\eqref{Equation: covariant normal order varphi} (as well as the fact that no particular ordering needs to be applied to $V$, since these observables have no dependence on $\Pi$) to write
\begin{align}
  \Dmap(\mathcal{O}) = U_\text{rad}[V]^\dagger\normord{\mathcal{O}[\varphi_i]}U_\text{rad}[V],
\end{align}
which implies~\eqref{Equation: dressed not Pi dependent}.

\subsection{Dressing is a deformation when including spin \texorpdfstring{$0$}{0} fields}

On the other hand,~\eqref{Equation: Dstar rad} \emph{does not} hold for more general operators that include the spin $0$ fields. In general, the dressing map is not an isomorphism. Instead, it provides a deformation of the algebraic structure. Indeed, covariant normal ordering has a non-trivial effect whenever $\mathcal{O}$ depends on $\Pi$.

Since the dressing map $\Dmap$ is one-to-one, and gauge-invariance is preserved by operator composition, two bare operators $\mathcal{O}_1$, $\mathcal{O}_2$ uniquely determine a third $\mathcal{O}_1\Dstar\mathcal{O}_2$ defined by
\begin{equation}
  \Dmap(\mathcal{O}_1)\Dmap(\mathcal{O}_2)=\Dmap(\mathcal{O}_1\Dstar\mathcal{O}_2),
\end{equation}
or simply
\begin{equation}
  \mathcal{O}_1\Dstar\mathcal{O}_2 = \GFmap\qty\Big(\Dmap(\mathcal{O}_1)\Dmap(\mathcal{O}_2)).
\end{equation}
This is a \emph{deformation} of the ordinary operator composition $\mathcal{O}_1\mathcal{O}_2$. For radiative operators $\mathcal{O}_1,\mathcal{O}_2\in\mathcal{B}(\mathcal{H}^\text{rad})$ it reduces to $\mathcal{O}_1\Dstar\mathcal{O}_2 = \mathcal{O}_1\mathcal{O}_2$, but for more general operators this is not true.

\subsection{Examples of dressed spin \texorpdfstring{$0$}{0} operators}

We give several key examples of dressed operators involving the spin $0$ fields below.

\subsubsection{Dressed area element}

The simplest example is the dressed $\Pi$ operator
\begin{equation}
  \Dmap(\Pi)
  =
  \Cov{\tilde\tau} = \tilde\tau + \frac{\hbar}{4\pi}\partial_\mv\qty(\frac{\partial_\mv^2X}{\partial_\mv X}).
  \label{Equation: dressed Pi}
\end{equation}
Covariant normal ordering produces the above correction term in the dressed $\tau$, relative to the ordinary normal ordering. This correction term is of the same form as the one that appears in the construction of  longitudinal DDF operators in string theory (see
\cite{DelGiudice:1972, Brower:1972, Goddard:1972,Bianchi:2019, Gebert:1995, Schreiber_2004, schreiber2004ddf,Biswas:2025} for key references on DDF operators).
The second equality in~\eqref{Equation: dressed Pi} follows from integrating $\Cov{\tilde\tau}$ against any function $g$ and applying~\eqref{Equation: G expansion}. In particular, one finds\footnote{We use that
  \begin{align}
    -\frac{\pa_v^2 V}{\pa_v V}=   \pa_v V \pa_v \left(\frac{1}{\pa_v V}\right)
    = \pa_v V \pa_v ((\pa_{\mv}X)\circ V )= (\pa_v V)^2 \pa_{\mv}^2 X\circ V = (\pa_v V)
    \left(\frac{\pa_{\mv}^2 X}{\pa_\mv X}\right)\circ V.
\end{align}}
\begin{align}
  \int\dd{\mv}g\Cov{\tilde\tau} &= \int\dd{\mv} g\covnormord{\partial_\mv X\Pi\circ X} = \int\dd{v}\covnormord{\Pi g(V)} \\
  &= \int\dd{v} \qty(\normord{\Pi g(V)} + \frac{\hbar}{4\pi}\frac{\partial_v^2V}{\partial_vV}\partial_vg(V))\\
  &= \int \dd{\mv}g \qty(\tilde\tau + \frac{\hbar}{4\pi}\partial_\mv\qty(\frac{\partial_\mv^2X}{\partial_\mv X})),
\end{align}
from which~\eqref{Equation: dressed Pi} follows.

From this we can recover the gauge-invariant dressed area operator $\Cov{\tilde\Omega} = \covnormord{\tilde\Omega}$. Indeed, classically
\begin{equation}
  \tilde\Pi=\tilde\tau = \frac1{8\pi\GN}\partial_\mv^2\tilde\Omega,
\end{equation}
and boundary conditions for solving this equation are provided by the edge modes $\omega_a = \tilde\Omega(a)/8\pi\GN$. Explicitly, for $\mv\in I=[0,1]$, one can integrate this equation to find
\begin{align}\label{area}
  \Cov{\tilde\Omega}(\mv) = 8\pi\GN\qty(\mv \omega_1+(1-\mv)\omega_0  + \int_{0}^1 \dd{\mv'}\Greens{\mv}{\mv'}\Cov{\tilde\tau}(\mv')),
\end{align}
where the Green's function is
\begin{equation}
  \Greens{\mv}{\mv'} = \frac12\qty\Big(\abs{\mv-\mv'} +2 \mv\mv'-\mv-\mv').
  \label{Equation: Green's}
\end{equation}
This is the quantum version of a similar formula appearing in~\cite{LocalizationAnomalous}.

\subsubsection{Dressed Raychaudhuri constraint}

In the construction of the physical Hilbert space presented in Section~\ref{Section: physical representation}, a particularly important dressed operator is the dressed Raychaudhuri constraint. By linearity this may simply be written
\begin{align}
  \Dmap(T) &=  \Cov{\tilde T} = \Cov{\tilde{\tau}} +  \Covrad{\tilde T} \\
  &= {\tilde\tau} -
  \frac{\hbar}{4\pi}\partial_\mv\qty(\frac{\partial_\mv^2X}{\partial_\mv X})
  + U_{\rad}[V]^\dagger T^\rad U_{\rad}[V]  \\
  &= \tilde\tau -
  \frac{\hbar}{4\pi}\partial_\mv\qty(\frac{\partial_\mv^2X}{\partial_\mv X})
  + \partial_\mv^2X
  T^\rad\circ X  - \frac{c^{\text{rad}}\hbar}{24\pi}\Schwarzian{X}{\mv}\\
  &= \tilde T - \frac\hbar{4\pi}\qty[\partial_\mv\qty(\frac{\partial_\mv^2X}{\partial_\mv X}) + \frac{c^{\text{rad}}}{6}\Schwarzian{X}{\mv}].
\end{align}
Note the distinction between this $\Dmap(T)=\Cov{\tilde T}$, and the $\Cov{T}$ defined in~\eqref{Equation: covnormord T}. These stress tensors are both covariantly normal ordered, and so transform under reparametrizations like their classical counterparts. The difference is that $\Cov{\tilde T}$ is dressed, so it is gauge-invariant, while $\Cov{T}$ is not dressed, so it transforms like a rank 2 tensor.

\subsubsection{Dressed matter fields}
Recall from~\cite{LocalizationAnomalous} and Section~\ref{Section: kinematical quantization} that the radiative fields $\varphi_i$ have been \emph{half-densitized}; that is, they are related (classically) to the actual matter fields $\phi_i$ by
\begin{equation}
  \phi_i = \varphi_i/\sqrt{\Omega}.
\end{equation}
Since $\phi_i$ depends on $\Pi$ through $\Omega$, the dressed quantum version
\begin{equation}
  \Dmap(\phi_i) = \Cov{\tilde\phi_i} = \covnormord{\tilde\varphi_i/\sqrt{\tilde\Omega}}
\end{equation}
picks up some non-trivial corrections from covariant normal ordering. We do not evaluate these corrections explicitly here.

\subsection{Dressing as a quantization of the Dirac bracket}
\label{Subsection: Dirac deformation}

Let us recap some of the above. Given a gauge-invariant classical observable $\tilde{O}[\Phi]$, we can construct a gauge-fixed observable $\mathcal{O}[\Pi, \varphi_i]:= \tilde{O}[\Pi, V, \varphi_i]|_{V(v)=v}$. Gauge invariance implies the equivalence between the two classical functionals
\begin{equation}
  \tilde{O}[\Pi,V,\varphi_i] = \mathcal{O}[\tilde{\tau}, \tilde{\varphi}_i].
\end{equation}
From this we can construct two different quantum operators. The first, denoted  $\tilde{O}^*$, corresponds to the covariant quantization of the gauge-invariant operator, while the second, denoted $\mathcal{O}$,  corresponds to the kinematical quantization of the gauge fixed operator:
\begin{equation}
  \tilde{O}^*:= \covnormord{\tilde{O}[\Pi,V,\varphi_i]}, \qquad
  \mathcal{O}:= \normord{\mathcal{O}[\Pi,\varphi_i]}
\end{equation}
In a sense, the gauge-fixed observable $\mathcal{O}[\Pi,\varphi_i]$ contains the same amount of information as the gauge-invariant one $\tilde{O}[\Pi,V,\varphi_i]$. Moreover, the gauge-fixed operator $\mathcal{O}$ is in general significantly simpler to construct than $\Cov{\tilde{O}}$.
It is therefore natural to ask whether we can get away with just using the gauge-fixed operators $\mathcal{O}$ to describe the gauge-invariant physics. In other words, is (i) gauge-fixing and then quantizing `the same as' (ii) quantizing the gauge-invariant operator? Are these two procedures physically equivalent -- do we actually gain anything by using $\Cov{\tilde{O}}$ instead of $\mathcal{O}$?

The answer is that they are \emph{different}, and the differences lie in the algebraic structure and fluctuations of the gauge-fixed and gauge-invariant operators. As we have seen, the map between the gauge-fixed operators and the gauge-invariant ones is the dressing map $\Dmap$. If this map were an isomorphism, then one would be justified in saying that (i) is physically equivalent to (ii). But, as explained above, the dressing map is a \emph{deformation} -- it changes the operator product of the gauge-fixed observables from ordinary operator composition to $\Dstar$.

A similar thing happens in the classical theory. The Poisson bracket of classically dressed observables is the classically dressed \emph{Dirac} bracket of the corresponding gauge-fixed observables. Thus, classical dressing is a \emph{deformation} of the Poisson bracket to the Dirac bracket.  The deformation arising from quantum dressing can be understood as the quantum analogue of this classical deformation. In particular, we show below that
\begin{equation}
  \lim_{\hbar\to 0}\frac1{i\hbar}\comm{\mathcal{O}_1}{\mathcal{O}_2}_{\Dstar} = \db{\mathcal{O}_1}{\mathcal{O}_2},
  \label{Equation: Dstar Dirac}
\end{equation}
where
\begin{equation}
  \comm{\mathcal{O}_1}{\mathcal{O}_2}_{\Dstar}:= \mathcal{O}_1\Dstar\mathcal{O}_2 - \mathcal{O}_2\Dstar\mathcal{O}_1.
  \label{Equation: Dstar commutator}
\end{equation}
denotes the $\Dstar$-commutator, and $\db{\cdot}{\cdot}$ denotes the classical Dirac bracket. Thus, the subscript ${}_\text{D}$ notation serendipitously stands for both `dressing' and `Dirac' at once.

Note that there are non-trivial higher order in $\hbar$ terms that contribute to the $\Dstar$-commutator, relative to the Dirac bracket. For example, we have~\cite{LocalizationAnomalous}
\begin{align}
  \db{\Pi(u)}{\Pi(v)} &= -2\partial_u\delta(u-v) \Pi(u) - \delta(u-v)\partial_u\Pi(u),\\
  \intertext{but (using~\eqref{Equation: reorietnation unintegrated})}
  \frac1{i\hbar}\comm{\Pi(u)}{\Pi(v)}_{\Dstar} &= -2\partial_u\delta(u-v) \Pi(u) - \delta(u-v)\partial_u\Pi(u) + \frac{c_{\tilde\tau}\hbar}{24\pi}\partial_{v}^3\delta(u-v).
\end{align}
These differ by the term involving the central charge. This and similar terms would not be captured by a na\"ive direct quantization of the Dirac bracket with no central charge. One could on the other hand choose to include a central charge when directly quantizing the Dirac bracket; one can then understand the quantization with the dressing map as selecting a particular value $c_{\tilde\tau}$ for this central charge.

Let us now demonstrate~\eqref{Equation: Dstar Dirac}. We start by deriving an $\order{\hbar}$ formula for the $\Dstar$-product. First, note that the classical observables corresponding to $\Dmap(\mathcal{O}_1)$, $\Dmap(\mathcal{O}_2)$ are
\begin{equation}
  \tilde{\mathcal{O}}_{a}[\Pi,V,\varphi] = \mathcal{O}_{a}[\tilde\Pi,\tilde\varphi]
\end{equation}
(with $a=1,2$),
while the observable corresponding to $\Dmap(\mathcal{O}_1\Dstar \mathcal{O}_2)$ is the covariant star product
\begin{equation}
  \qty(\mathcal{O}_{1}[\tilde\Pi,\tilde\varphi]) \star \qty(\mathcal{O}_{2}[\tilde\Pi,\tilde\varphi]),
\end{equation}
and that we can recover $\mathcal{O}_1\Dstar\mathcal{O}_2 = \GFmap(\Dmap(\mathcal{O}_1\Dstar\mathcal{O}_2))$ by evaluating this observable at $V(v)=v$:
\begin{equation}
  (\mathcal{O}_1\Dstar\mathcal{O}_2)[\Pi,\varphi] = \left.\qty(\mathcal{O}_{1}[\tilde\Pi,\tilde\varphi]) \star \qty(\mathcal{O}_{2}[\tilde\Pi,\tilde\varphi])\right|_{V(v)=v}.
\end{equation}
We use~\eqref{Equation: other covariant Wick's theorem to order hbar}. In particular, we have
\begin{align}
  \left.\fdv{P_\pm\tilde\Pi}\mathcal{O}_{a}[\tilde\Pi,\tilde\varphi]\right|_{V(v)=v} &= \fdv{P_\pm\Pi} \mathcal{O}_{a}[\Pi,\varphi],\\
  \left.\fdv{\tilde\varphi_i}\mathcal{O}_{a}[\tilde\Pi,\tilde\varphi_i]\right|_{V(v)=v} &= \fdv{\varphi_i} \mathcal{O}_{a}[\Pi,\varphi],
\end{align}
and
\begin{equation}
  \left.\tilde{\mathcal{L}}\qty(\mathcal{O}_{a}[\tilde\Pi,\tilde\varphi])\right|_{V(v)=v} =\mathcal{L}\qty(\mathcal{O}_{a}[\Pi,\varphi]),
\end{equation}
where
\begin{equation}
  \mathcal{L} = \sum_i\partial_v\varphi_i\fdv{\varphi_i} - \Pi \partial_\mv\fdv{\Pi},
\end{equation}
which gives
\begin{multline}
  (\mathcal{O}_1\Dstar \mathcal{O}_2)[\Pi,\varphi]
  = \mathcal{O}_1[\Pi,\varphi]\mathcal{O}_2[\Pi,\varphi] - \frac{i\hbar}{2}\int\dd{u}\dd{v}\ln(u-v-i\epsilon)\sum_i\fdv{O_1[\Pi,\varphi]}{\varphi_i(u)}\fdv{O_2[\Pi,\varphi]}{\varphi_i(v)} \\
  + i\hbar{\int\dd{v}\qty(\fdv{\mathcal{O}_1[\Pi,\varphi]}{P_+\Pi}{\mathcal{L}}\qty(\mathcal{O}_2[\Pi,\varphi]) -{\mathcal{L}}\qty(\mathcal{O}_1[\Pi,\varphi])\fdv{\mathcal{O}_2[\Pi,\varphi]}{P_-\Pi})}
  + \order{\hbar^2}.
  \label{Equation: classical star product to order hbar}
\end{multline}
The terms on the first line are simply the usual star product which follows  from  Wick's theorem for $\normord{\mathcal{O}_1}\normord{\mathcal{O}_2}=\normord{O_1 * O_2}$. For the terms on the second line, we note that
\begin{align}
  \normord{\fdv{P_\pm\Pi} \mathcal{O}[\Pi,\varphi]} &= \frac1{i\hbar}[P_\mp V,\normord{\mathcal{O}[\Pi,\varphi]}],\\
  \normord{\mathcal{L}(\mathcal{O}[\Pi,\varphi])}&= -\frac1{i\hbar}[T,\normord{\mathcal{O}[\Pi,\varphi]}]+\order{\hbar}.
\end{align}
Any higher order Wick contractions are suppressed in powers of $\hbar$, and one finds after normal ordering
\begin{equation}
  \mathcal{O}_1\Dstar \mathcal{O}_2 = \mathcal{O}_1 \mathcal{O}_2
  - \frac1{i\hbar}\int\dd{v}
  \qty([\mathcal{O}_1,P_-V][T,\mathcal{O}_2] - [\mathcal{O}_1,T][P_+V,\mathcal{O}_2]) + \order{\hbar^2}.
  \label{Equation: star product to order hbar}
\end{equation}

Let's now consider the $\Dstar$-commutator~\eqref{Equation: Dstar commutator}.
Using~\eqref{Equation: star product to order hbar}, and assuming any nested commutators introduce extra powers of $\hbar$, we have the formula
\begin{equation}
  \comm{\mathcal{O}_1}{\mathcal{O}_2}_{\Dstar} = \comm{\mathcal{O}_1}{\mathcal{O}_2}
  - \frac1{i\hbar}\int\dd{v}\qty([\mathcal{O}_1,V][T,\mathcal{O}_2] - [\mathcal{O}_1,T][V,\mathcal{O}_2]) + \order{\hbar^2}.
  \label{Equation: star commutator to order hbar}
\end{equation}
If we identify $\pb{\cdot}{\cdot} = \lim_{\hbar\to 0}\frac1{i\hbar}\comm{\cdot}{\cdot}$ with the classical Poisson bracket,\footnote{To be more precise, this holds in the sense $\normord{\pb{\cdot}{\cdot}} + \order{\hbar} = \frac1{i\hbar}\comm{\cdot}{\cdot}$. One can also derive~\eqref{Equation: Dirac bracket deformation quantization} directly from~\eqref{Equation: classical star product to order hbar}.} then~\eqref{Equation: star commutator to order hbar} implies
\begin{equation}
  \lim_{\hbar\to 0}\frac1{i\hbar}\comm{\mathcal{O}_1}{\mathcal{O}_2}_{\Dstar}= \pb{\mathcal{O}_1}{\mathcal{O}_2}
  - \int\dd{v}\qty\big(\pb{\mathcal{O}_1}{V}\pb{T}{\mathcal{O}_2} - \pb{\mathcal{O}_1}{T}\pb{V}{\mathcal{O}_2})= \db{\mathcal{O}_1}{\mathcal{O}_2}.
  \label{Equation: Dirac bracket deformation quantization}
\end{equation}
This matches exactly with the Dirac bracket obtained in~\cite{LocalizationAnomalous}.

\subsection{Fluctuations of dressed operators vs.\ bare operators}
\label{Subsection: fluctuations}

A key way in which dressing modifies the physical properties of the quantum theory is through the \emph{fluctuations} of operators.
Let us consider a gauge-invariant quantum state
\begin{equation}
  \tilde\Psi: \mathcal{A}_{\text{kin}}^{\Diff^+(\RR)} \to \CC,
\end{equation}
i.e.\ a normalized ($\tilde\Psi(1)=1$) linear functional on the algebra $ \mathcal{A}_{\text{kin}}^{\Diff^+(\RR)}$ of gauge-invariant operators.\footnote{Concretely we might have $\tilde\Psi(\cdot) = \bra{\tilde\Psi}\cdot\ket{\tilde\Psi}$ for some Hilbert space state $\ket{\tilde\Psi}$.} By pulling back $\tilde\Psi$ through the dressing map $\Dmap:\gf{\mathcal{A}}\to \mathcal{A}_{\text{kin}}^{\Diff^+(\RR)}$ we get the corresponding gauge-fixed state
\begin{equation}
  \Psi = \tilde\Psi\circ \Dmap: \gf{\mathcal{A}}\to\CC,
\end{equation}
satisfying $\Psi(\mathcal{O}) = \tilde\Psi(\Dmap(\mathcal{O}))$.

Consider a gauge-invariant observable $\tilde O[\Phi]$, and its covariant quantization $\Cov{\tilde O}$. The expectation value and fluctuations of this operator in the state $\tilde\Psi$ are
\begin{equation}
  \expval*{\Cov{\tilde O}}_{\tilde\Psi} = \tilde\Psi(\Cov{\tilde O}), \qquad \Delta_{\tilde\Psi}^2(\Cov{\tilde O}) = \tilde\Psi\qty\big((\Cov{\tilde O})^2) - [\tilde\Psi(\Cov{\tilde O})]^2.
\end{equation}
On the other hand, the expectation value and fluctuations of the gauge-fixed operator $\mathcal{O}$ (where $\mathcal{O}[\Pi,\varphi_i] = \tilde O[\Pi,V,\varphi_i]|_{V(v)=v}$) in the corresponding gauge-fixed state $\Psi$ are
\begin{equation}
  \expval*{\mathcal{O}}_{\Psi} = \Psi(\mathcal{O}), \qquad \Delta_{\Psi}^2(\mathcal{O}) = \Psi\qty\big(\mathcal{O}^2) - \Psi(\mathcal{O})^2.
\end{equation}
The expectation values agree within the gauge-invariant and gauge-fixed quantizations:
\begin{equation}
  \expval*{\Cov{\tilde O}}_{\tilde\Psi}
  = \tilde\Psi(\Cov{\tilde O})
  = \tilde\Psi(\Dmap(\mathcal{O}))
  = \Psi(\mathcal{O})
  = \expval*{\mathcal{O}}_\Psi.
\end{equation}
But the fluctuations, whose difference can be written
\begin{align}
  \Delta_{\tilde\Psi}^2(\Cov{\tilde O}) - \Delta_{\Psi}^2(\mathcal{O})
  &= \tilde\Psi(\Dmap(\mathcal{O})^2) - \Psi(\mathcal{O}^2)\\
  &= \tilde\Psi(\Dmap(\mathcal{O})^2 - \Dmap(\mathcal{O}^2)) \\
  &=\Psi(\mathcal{O}\Dstar\mathcal{O}-\mathcal{O}^2),
\end{align}
generally \emph{do not agree}.\footnote{Note that the fluctuations are the same when $\mathcal{O}$ acts purely on the radiative degrees of freedom, but we get a non-trivial difference when $\mathcal{O}$ depends on $\Pi$. } A similar conclusion was reached by Perez and Sudarsky in \cite{Perez:2025tvg}, which describes, from a different perspective, the fluctuations of covariantly renormalized operators. The same is true of higher order correlation functions of the observables. This captures quantum fluctuations inherent to the frame.

For example, consider the spin 0 stress tensor, which has the gauge-fixed and gauge-invariant quantizations
\begin{equation}
  \mathcal{O}=\int\dd{v} g\Pi, \qquad \Cov{\tilde{O}} = \int\dd{\mv}g\Cov{\tilde\tau}.
\end{equation}
Using~\eqref{Equation: reorientation Wick}, we have
\begin{equation}
  \mathcal{O}\Dstar\mathcal{O} = \int\dd{v}\dd{u}\qty(g(v)g(u) \Pi(v)\Pi(u) + \frac\hbar{\pi}\frac{g'(v)g(u)}{[v-u]}\Pi(v)+\frac{c_{\tilde\tau}}{2}\qty(\frac{\hbar}{2\pi})^2\frac{g(v)g(u)}{[(v-u)^4]}),
\end{equation}
where the square bracket indicates that we are taking the Cauchy principal value, i.e.\ $\frac{1}{[u-v]}= \mathrm{P}\left(\frac1{u-v}\right)$ (note that all the terms that would come from delta functions in the contractions $\frac1{u-v-i\epsilon}=\frac1{[u-v]}+i\pi\delta(u-v)$ vanish due to symmetry).
Therefore, one finds that the fluctuations differ by
\begin{equation}
  \Delta_{\tilde\Psi}^2(\Cov{\tilde O}) - \Delta_{\Psi}^2(\mathcal{O}) = \frac\hbar{\pi}\int\dd{v}\dd{u}\qty( \frac{g'(v)g(u)}{[v-u]}\expval{\Pi(v)}_\Psi+\frac{c_{\tilde\tau}\hbar}{4\pi}\frac{g(v)g(u)}{[(v-u)^4]}).
\end{equation}

For more general observables, a formula to $\order{\hbar}$ can be obtained by using~\eqref{Equation: star product to order hbar}; one finds
\begin{align}
  \Delta_{\tilde\Psi}^2(\Cov{\tilde O}) - \Delta_{\Psi}^2(\mathcal{O}) = \frac1{\pi\hbar}\int\dd{u}\dd{v}\frac1{[v-u]}\expval{\comm{\mathcal{O}}{T(v)}\comm{V(u)}{\mathcal{O}}}_\Psi + \order{\hbar^2}.
\end{align}

\section{Three kinds of quantum diffeomorphisms and their central charges}
\label{Section: three quantum diffeomorphisms}

\begin{table}
  \centering
  \begin{tabular}{lcc}\toprule
    & Central charge  & \small Generator \\  \midrule
    {\small Reparametrization} & $c_T=2+M$ &$T_f = \int\dd{v}fT$ \\\addlinespace
    {\small Reorientation} & $c_{\tilde\tau}=24$ & $\Cov{\tilde\tau}_g = \int\dd{\mv}g\Cov{\tilde\tau}$ \\\addlinespace
    {\small Dressed reparametrization} & $c_{\tilde T}=24-M$ &$\Cov{\tilde T_{\tilde f}} = \int\dd{\mv}\tilde f \Cov{\tilde T}$
  \end{tabular}
  \caption{The three diffeomorphism actions on the null ray and their key properties.}
  \label{Table: diffeomorphisms}
\end{table}

There are three diffeomorphism actions relevant to the null ray~\cite{LocalizationAnomalous}:
\begin{itemize}
  \item \textbf{Reparametrizations}: Gauge transformations of the null ray, as described previously.
  \item \textbf{Reorientations}: Gauge-invariant transformations which act on the spin 0 fields $\Pi,V$ without affecting the \emph{bare} radiative fields $\varphi_i$. They are important because they underlie the crossed product structure of the algebra of gauge-invariant operators, as already described.
  \item \textbf{Dressed reparametrizations}: Gauge-invariant transformations which act on the spin 0 fields $\Pi,V$, but without affecting the \emph{dressed} radiative fields $\tilde\varphi_i$. They are important because they determine properties of the physical Hilbert space, as detailed in Section~\ref{Section: physical representation}.
\end{itemize}

As explained in Section~\ref{Section: kinematical quantization}, the generator of a reparametrization is the Raychaudhuri constraint
\begin{equation}
  T_f = \int\dd{v}f T.
\end{equation}
In Appendix~\ref{Subsection: reparametrization algebra}, we show by a standard computation that these operators obey
\begin{equation}
  \comm{T_{f_1}}{T_{f_2}} = i\hbar \qty(T_{[f_1,f_2]}+\frac{c_T\hbar}{48\pi}\int\dd{v}(f_1'f_2'' - f_2'f_1'')),
  \label{Equation: reparametrization algebra}
\end{equation}
with the central charge $c_T=2+M$.

The reorientations and dressed reparametrizations are generated similarly by
\begin{equation}
  \Cov{\tilde\tau}_g = \int\dd{v}g \Cov{\tilde\tau},
  \qquad
  \Cov{\tilde T}_{\tilde f} = \int\dd{v}{\tilde f} \Cov{\tilde T},
\end{equation}
and we show in Subsections~\ref{Subsection: reorientation algebra} and~\ref{Subsection: dressed reparametrization algebra} respectively that these obey
\begin{align}
  \comm{\Cov{\tilde\tau}_{g_1}}{\Cov{\tilde\tau}_{g_2}} &= i\hbar\qty(-\Cov{\tilde\tau}_{[g_1,g_2]} + \frac{c_{\tilde\tau}\hbar}{48\pi}\int\dd{\mv}(g_1'g_2''-g_2'g_1'')),
  \label{Equation: reorientation algebra}\\
  \comm{\Cov{\tilde T}_{\tilde{f}_1}}{\Cov{\tilde T}_{\tilde{f}_2}} &= i\hbar\qty(-\Cov{\tilde T}_{[\tilde{f}_1,\tilde{f}_2]} + \frac{c_{\tilde T}\hbar}{48\pi}\int\dd{\mv}(\tilde{f}_1'\tilde{f}_2''-\tilde{f}_2'\tilde{f}_1'')),
  \label{Equation: dressed reparametrization algebra}
\end{align}
where the central charges are given by $c_{\tilde \tau}=24$ and $c_{\tilde T}=24-M$. Thus, all three classical diffeomorphism actions are promoted to Virasoro actions in the quantum theory, with different central charges. Note the minus signs in~\eqref{Equation: reorientation algebra} and~\eqref{Equation: dressed reparametrization algebra} relative to~\eqref{Equation: reparametrization algebra}; these reflect the fact that reparametrizations form an antimorphism of diffeomorphisms, while reorientations / dressed reparametrizations form morphisms (see~\cite{LocalizationAnomalous}). These properties are summarized in Table~\ref{Table: diffeomorphisms}.

In Subsection~\ref{Section: anomaly shift}, we explain how one may simultaneously shift the central charges of each of these Virasoro representations by introducing a parameter $c_{\text{cl}}$ at the level of the \emph{classical} theory, before quantization. The result of this procedure is that each of the central charges take certain deformed values~\eqref{Equation: deformed central charges}. The consequences of the deformation are summarized in Table~\ref{Table: diffeomorphisms shifted}. As explained in Subsection~\ref{Section: anomaly cancel}, such a deformation is required for a cancellation of gauge anomalies.

\subsection{Reorientations}
\label{Subsection: reorientation algebra}

The generator of reorientations is the covariantly normal ordered dressed spin 0 stress tensor $\Cov{\tilde\tau}$. Let us define\footnote{Note that in~\cite{LocalizationAnomalous} we used the alternative notation $Q_g$ to refer to this object.}
\begin{equation}
  \tilde{\tau}_g := \int\dd{\mv}g\tilde\tau = \int\dd{v}\normord{\Pi g(V)},
\end{equation}
so that
\begin{equation}
  \Cov{\tilde\tau_g} = \tilde{\tau}_g -\frac\hbar{4\pi}\int\dd{\mv}g'\frac{\partial_\mv^2X}{\partial_\mv X} .
  \label{Equation: reorientation generator}
\end{equation}
To compute the bracket of these operators one can first use Wick's theorem to find
\begin{align}
  \label{Equation: reorientation Wick}
  \MoveEqLeft\normord{[\Pi g_1(V)](v)} \normord{[\Pi g_2(V)](u)}= \normord{[\Pi g_1(V)](v) [\Pi g_2(V)](u)} \\
  &+ \frac\hbar{2\pi}\frac1{v-u-i\epsilon}\normord{(\Pi(v)g_1'(V(v))g_2(V(u))-\Pi(u)g_2'(V(u))g_1(V(v)))}
  -\qty(\frac\hbar{2\pi})^2\frac{g_1'(V(v))g_2'(V(u))}{(v-u-i\epsilon)^2},\nonumber
\end{align}
so that the commutator takes the form
\begin{multline}
  \comm{\normord{\Pi g_1(V)(v)}}{\normord{\Pi g_2(V)(u)}}\\
  =-i\hbar\left(\delta(u-v) \normord{\Pi(u) [g_1,g_2](V)(u)}
    +\frac{\hbar}{2\pi}\partial_u\delta(v-u)g_1'(V(v))g_2'(V(u))
  \right),
  \label{Equation: reorientation bracket without correction}
\end{multline}
This can be written after integration as
\begin{align}
  \frac1{i\hbar}\left[
    \tilde\tau_{g_1},
  \tilde\tau_{g_2}\right] &= -
  \tilde\tau_{[g_1,g_2]}
  + \frac{\hbar}{4\pi}\int\dd{\mv}(g_1'g_2''-g_2'g_1'')
\end{align}
One also has, using $\frac1{i\hbar}[\tilde\tau_g,X] = g\partial_\mv X$,
\begin{equation}
  \comm{\tilde\tau_{g_1}}{\int\dd{\mv} g_2'\frac{\partial_{\mv}^2 X}{\partial_{\mv} X}}
  = i\hbar\int\dd{\mv}g_2'\partial_{\mv}\qty(\frac{\partial_{\mv}^2 X}{\partial_{\mv} X}g_1 + g_1').
\end{equation}
Combining these one obtains
\begin{align}
  \comm{\Cov{\tilde\tau}_{g_1}}{\Cov{\tilde\tau}_{g_2}} &= i\hbar\qty(-\tilde\tau_{[g_1,g_2]}+\frac\hbar{4\pi}\int\dd{\mv}[g_1,g_2]'\frac{\partial_\mv^2X}{\partial_\mv X} + \frac\hbar{2\pi}\int\dd{\mv}(g_1'g_2''-g_2'g_1'')) \\
  &= i\hbar\qty(-\Cov{\tilde\tau}_{[g_1,g_2]} + \frac{c_{\tilde\tau}\hbar}{48\pi}\int\dd{\mv}(g_1'g_2''-g_2'g_1'')),
\end{align}
where the central charge is $c_{\tilde\tau}=24$. One finds also the unintegrated version of this formula
\begin{equation}
  \comm{\Cov{\tilde\tau}(\mv)}{\Cov{\tilde\tau}(\rm u)}= i\hbar\qty(-\delta(\mv-\rm u) \partial_{\rm u}\Cov{\tilde\tau}({\rm u}) + 2\partial_{\rm u}(\delta(\mv-{\rm u})\Cov{\tilde\tau}({\rm u})) + \frac{c_{\tilde\tau}\hbar}{24\pi}\partial_{\rm u}^3\delta(\mv-{\rm u})).
  \label{Equation: reorietnation unintegrated}
\end{equation}
It is worth noting that if one did not include the correction due to covariant normal ordering, i.e.\ the final term in~\eqref{Equation: reorientation generator}, one would find from~\eqref{Equation: reorientation bracket without correction} a central charge of $12$ instead. But the requirement of gauge-invariance for reorientations shifts their central charge from $12$ to $24$.

Let's now compute the action of $\Cov{\tilde{\tau}}$ on dressed radiative observables. We use that
\begin{equation}
  e^{-\frac1{i\hbar}\Cov{\tilde\tau}_{g}}U_{\rad}[V]e^{\frac1{i\hbar}\Cov{\tilde\tau}_{g}}= U_{\rad}[G\circ V] = U_{\rad}[G] U_{\rad}[V] \exp(\frac{ic^{\text{rad}}}{24}C(G,V)),
  \label{Equation: tau U rad V}
\end{equation}
where the cocycle $C(G,V)$ is given in Appendix~\ref{Appendix: cocycle}.
Suppose that $\mathcal{O}_{\text{rad}}$ is a radiative operator that does not act on the spin 0 variables, and consider its dressing
\begin{equation}
  \Dmap(\mathcal{O}_{\text{rad}}) = U_{\rad}[V]^\dagger \mathcal{O}_{\text{rad}} U_{\rad}[V].
\end{equation}
Using the above with $[\mathcal{O}_{\text{rad}},V]=0$, one has
\begin{align}
  e^{-\frac1{i\hbar}\Cov{\tilde\tau}_{g}}\Dmap(\mathcal{O}_{\text{rad}}) e^{\frac1{i\hbar}\Cov{\tilde\tau}_{g}} &= e^{-\frac1{i\hbar}\Cov{\tilde\tau}_{g}}U_{\rad}[V]^\dagger \mathcal{O}_{\text{rad}} U_{\rad}[V] e^{\frac1{i\hbar}\Cov{\tilde\tau}_{g}}\\
  &= U_{\rad}[V]^\dagger U_{\rad}[G]^\dagger \mathcal{O}_{\text{rad}} U_{\rad}[G] U_{\rad}[V]\\
  &= \Dmap(U_{\rad}[G]^\dagger \mathcal{O}_{\text{rad}}U_{\rad}[G]),
\end{align}
or, infinitesimally,
\begin{equation}
  [\Cov{\tilde\tau},\Dmap(\mathcal{O}_{\text{rad}})] = -\Dmap([T^\rad,\mathcal{O}_{\text{rad}}]).
  \label{Equation: reorientation on D(O)}
\end{equation}
Therefore, reorientations act on such dressed operators $\Dmap(\mathcal{O}_{\text{rad}})$ as inverse reparametrizations on the bare operator $\mathcal{O}_{\text{rad}}$. This matches the classical setup.

Expanding~\eqref{Equation: tau U rad V} in small $g$ gives
\begin{equation}
  [U_{\text{rad}}[V],\Cov{\tilde\tau}] = -\qty(T^\text{rad}+\frac{c^\text{rad}\hbar}{48\pi}\partial_\mv\qty(\frac{\partial_\mv^2X}{\partial_\mv X}))U_{\text{rad}}[V],
\end{equation}
which can be rearranged to
\begin{equation}
  U_{\text{rad}}[V]\Cov{\tilde\tau}U_{\text{rad}}[V]^\dagger = \Cov{\tilde\tau} - T^\text{rad}-\frac{c^\text{rad}\hbar}{48\pi}\partial_\mv\qty(\frac{\partial_\mv^2X}{\partial_\mv X})
  \label{Equation: U rad V tilde tau}
\end{equation}
Recall that the algebra of gauge-invariant operators has the form of a crossed product
\begin{equation}
  \mathcal{A}_{\text{kin}}^{\Diff^+(\RR)} = \{\Cov{\tilde\tau},\, U_\text{rad}[V]^\dagger a U_\text{rad}[V]\,\mid\,a\in\mathcal{B}(\mathcal{H}^\text{rad})\}''.
\end{equation}
We can now write this equivalently as $\mathcal{A}_{\text{kin}}^{\Diff^+(\RR)} = U_\text{rad}[V]^\dagger \hat{\mathcal{A}} U_\text{rad}$, where
\begin{equation}
  \hat{\mathcal{A}} = \{\Cov{\tilde\tau} - T^\text{rad}-\tfrac{c^\text{rad}\hbar}{48\pi}\partial_\mv\qty(\tfrac{\partial_\mv^2X}{\partial_\mv X}),\,  a\,\mid\,a\in\mathcal{B}(\mathcal{H}^\text{rad})\}''.
  \label{Equation: crossed product 3}
\end{equation}
The expression on the right-hand side of~\eqref{Equation: crossed product 3} reproduces another common way in which the crossed product is written~\cite{Connes1994,Takesaki2003II,Takesaki2003III,Witten2022,CLPW}; it is the Virasoro version of~\eqref{Equation: undressed crossed product G}, with the $\tilde\pi-P$ appearing in that equation analogous to the $\Cov{\tilde\tau} - T^\text{rad}-\tfrac{c^\text{rad}\hbar}{48\pi}\partial_\mv\qty(\tfrac{\partial_\mv^2X}{\partial_\mv X})$ appearing in~\eqref{Equation: crossed product 3}. It is interesting that here there is a term depending on the reference frame configuration $X$, but there is no analogous term depending on $\hat g$ appearing in the discussion of Subsubsection~\ref{Subsubsection: Lie group crossed product}. This term is ultimately due to the fact that there is a cocycle in the representation of the gauge group, which was not considered in that discussion.

\subsection{Dressed reparametrizations}
\label{Subsection: dressed reparametrization algebra}

The generator of dressed reparametrizations is the covariantly normal ordered dressed Raychaudhuri constraint
\begin{align}
  \Cov{\tilde T} &= \Cov{\tilde\tau}+\Covrad{\tilde T} = \Cov{\tilde\tau} + \Dmap(T^\rad).
\end{align}
Now with~\eqref{Equation: reorientation on D(O)} we have
\begin{align}
  \comm{\Cov{\tilde\tau}(\mv)}{\Dmap(T^\rad)({\rm u})}
  &= - D\qty(\comm{T^\rad(\mv)}{T^\rad({\rm u})}),
\end{align}
while
\begin{equation}
  \comm{\Dmap(T^\rad)(\mv)}{\Dmap(T^\rad)({\rm u})}
  = D\qty(\comm{T^\rad(\mv)}{T^\rad({\rm u})}),
\end{equation}
Therefore
\begin{align}
  \comm{\Cov{\tilde T}(\mv)}{\Cov{\tilde T}({\rm u})}
  &= \comm{\Cov{\tilde \tau}(\mv)}{\Cov{\tilde \tau}({\rm u})} -D\qty(\comm{T^{\text{rad}}(\mv)}{T^\text{rad}({\rm u})})
\end{align}
and integrating this against functions $g_1,g_2$ gives
\begin{align}
  \comm{\Cov{\tilde T}_{g_1}}{\Cov{\tilde T}_{g_2}} &= \comm{\Cov{\tilde \tau}_{g_1}}{\Cov{\tilde \tau}_{g_2}} -D\qty(\comm{T^{\text{rad}}_{g_1}}{T^\text{rad}_{g_2}})\\
  &= i\hbar\qty(-\Cov{\tilde\tau}_{[g_1,g_2]} + \frac{c_{\tilde\tau}\hbar}{48\pi}\int\dd{\mv}(g_1'g_2''-g_2'g_1'')) \\
  &\hspace*{2em}-i\hbar D\qty(T^{\text{rad}}_{[g_1,g_2]} + \frac{c_{\rad}\hbar}{48\pi}\int\dd{\mv}(g_1'g_2''-g_2'g_1''))\\
  &= i\hbar\qty(-\Cov{\tilde T}_{[g_1,g_2]} + \frac{c_{\tilde T}\hbar}{48\pi}\int\dd{\mv}(g_1'g_2''-g_2'g_1'')),
\end{align}
where the central charge is $c_{\tilde T}=c_{\tilde\tau} - c_{\text{rad}} = 24-M$.

Combining~\eqref{Equation: reorientation on D(O)} with~\eqref{Equation: Dstar rad}, one may observe that $\Cov{\tilde T}$ commutes with any dressed radiative operator:
\begin{align}
  [\Cov{\tilde T},\Dmap(\mathcal{O}_{\text{rad}})] &= [\Cov{\tilde\tau},\Dmap(\mathcal{O}_{\text{rad}})] + [\Dmap(T_\rad),\Dmap(\mathcal{O}_{\text{rad}})] \\
  &= -\Dmap([T_\rad,\mathcal{O}_{\text{rad}}]) + \Dmap([T_\rad,\mathcal{O}_{\text{rad}}])=0.
  \label{Equation: T tilde commutes with D(rad)}
\end{align}

\subsection{Shifting the diffeomorphism anomalies}
\label{Section: anomaly shift}

\begin{table}
  \centering
  \begin{tabular}{lcc}\toprule
    & Central charge  & \small Generator \\  \midrule
    {\small Reparametrization} & $c_T=2+M+c_{\text{cl}}$ &$T^{c}_f = \int\dd{v}fT^{c}$ \\\addlinespace
    {\small Reorientation} & $c_{\tilde\tau}=24-c_{\text{cl}}$ & $\Cov{\tilde\tau_g} = \int\dd{\mv}g\Cov{\tilde\tau}$ \\\addlinespace
    {\small Dressed reparametrization} & $c_{\tilde T}=24-M-c_{\text{cl}}$ &$\Cov{\tilde T_{\tilde f}} = \int\dd{\mv}\tilde f \Cov{\tilde T}$
  \end{tabular}
  \caption{Under a deformation of the theory parametrized by a classical central charge $c_{\text{cl}}$, the properties of the three diffeomorphism actions are modified as shown here.}
  \label{Table: diffeomorphisms shifted}
\end{table}

As described in~\cite{LocalizationAnomalous}, one may introduce a central charge $c_{\text{cl}}$ into the diffeomorphisms of the classical theory by including specific counterterms into the symplectic form and Raychaudhuri constraint. These counterterms do not affect the radiative modes, only modifying the spin 0 contribution. As explained in Subsection~\ref{Section: anomaly cancel} and Section~\ref{Section: physical representation}, such a modification is required for a consistent quantization of the gauge-invariant degrees of freedom.

One way to summarize the results obtained in \cite{LocalizationAnomalous} is to start with the same kinematical phase space labelled by $(\Pi,V,\varphi_i)$ and to modify the constraints.\footnote{Note that this is a different perspective to the one used in \cite{LocalizationAnomalous}, where we labelled the phase space by $(\Omega,\beta,\varphi_i)$ and showed that the classical central charge leads to a deformation of the phase space structure and of the constraint. If we work in terms of the free fields $(\Pi,V)$ instead the phase space is not deformed, only the constraint.} After inclusion of the classical central charge the constraint is deformed into
\begin{equation}
  T\to T^c= \tau^c + T^{\text{rad}}=0,
\end{equation}
where $\tau^c$ is constructed in terms of $V$ and its conjugate momentum $\Pi$ as
\begin{equation}
  \tau^{c}:= \Pi\partial_vV - \frac{c_{\text{cl}}\hbar}{48\pi}\partial_v\qty(\frac{\partial_v^2 V}{\partial_vV}).
\end{equation}
At the classical level we can identify $\tau$ as an operator which transforms as a rank 2 tensor with respect to the reparametrization generated by $T^c_f:= \int fT^c$. This functional is given by
\begin{equation}
  \tau =   \tau^{c}+\frac{c_{\text{cl}}\hbar}{24\pi} \Schwarzian{V}{v},
\end{equation}
which shows that the classical deformation simply adds the dressing time Schwarzian stress tensor to the spin zero stress tensor.
Note that, after dressing to $V$, we have $V(v)\to \tilde V(v)= v$. Therefore,
\begin{equation}
  \tilde\tau^c=\tilde\Pi^c=\tilde\Pi=\tilde\tau,
\end{equation}
so these deformations do not affect the forms of classical dressed observables.

Including the radiative fields and employing normal ordering, the full quantum stress tensor / Raychaudhuri constraint is now given by
\begin{equation}
  T^c = \normord{\Pi\partial_vV} - \frac{c_{\text{cl}}\hbar}{48\pi}\partial_v\qty(\frac{\partial_v^2 V}{\partial_vV}) + \sum_i\normord{(\partial_v\varphi_i)^2}.
\end{equation}

In the classical theory, this deformation causes reparametrization, reorientations and dressed reparametrizations to have central charges $c_{\text{cl}},-c_{\text{cl}},-c_{\text{cl}}$ respectively. In this section we see what changes after a canonical quantization of the deformed theory. The result is a theory whose central charges have contributions from both the classical deformation and quantum anomalies:
\begin{equation}
  c_T = 2+M+c_{\text{cl}}, \qquad c_{\tilde\tau} = 24-c_{\text{cl}}, \qquad c_{\tilde T} = 24-M-c_{\text{cl}}.
  \label{Equation: deformed central charges}
\end{equation}
Because the stress tensor is deformed, the action of a gauge transformation is deformed, and therefore so is the definition of covariant normal ordering. However, all the formulas in Section~\ref{Section: covariant normal ordering} still apply, with the following caveat: one must use the deformed action of reparametrizations on $\Pi$ implied by the deformed stress tensor. We give precise formulas and their derivations in Appendix~\ref{Subsection: anomaly covnormord} for the covariant normal ordering of arbitrary observables. The formula~\eqref{Equation: covnormord to order hbar} suffices for the main text of the paper and applies as before, i.e.\ it is exact at $\order{\hbar}$ for linear in $\Pi$ observables.

\subsubsection{Reparametrizations}
Let us decompose $
T^c = T - \frac{c_{\text{cl}}\hbar}{48\pi}\partial_v\qty(\frac{\partial_v^2 V}{\partial_vV}),
$
where
\begin{equation}
  T = \normord{\Pi\partial_vV} + \sum_i\normord{(\partial_v\varphi_i)^2}
\end{equation}
is the stress tensor of the undeformed theory, and define
\begin{equation}
  T^c_f = \int\dd{v}fT^c = T_f + \frac{c_{\text{cl}}\hbar}{48\pi}\int\dd{v}\partial_vf \frac{\partial_v^2V}{\partial_vV},
\end{equation}
where $T_f=\int\dd{v}f T$.

We already know from Appendix~\ref{Subsection: reparametrization algebra} that
\begin{equation}
  \comm{T_{f_1}}{T_{f_2}} = i\hbar \qty(T_{[f_1,f_2]}+\frac{(2+M)\hbar}{48\pi}\int\dd{v}(\pa_vf_1\partial_v^2f_2 - \pa_vf_2\partial_v^2f_1)),
\end{equation}
The only new contributions to the full stress tensor commutator arising from the deformation comes from terms of the form
\begin{align}
  \comm{T_{f_1}}{\int\dd{v}\partial_vf_2 \frac{\partial_v^2V}{\partial_vV}} = -i\hbar\int\dd{v}\partial_vf_2 \partial_v\qty(f_1\frac{\partial_v^2V}{\partial_vV} + \partial_vf_1).
\end{align}
Overall, one finds
\begin{align}
  \comm{T_{f_1}^c}{T_{f_2}^c} &= i\hbar \qty(T_{[f_1,f_2]}+\frac{(2+M)\hbar}{48\pi}\int\dd{v}(\pa_vf_1\partial_v^2f_2 - \pa_vf_2\partial_v^2f_1)) \\
  &\qquad -i\hbar\frac{c_{\text{cl}}\hbar}{48\pi}\int\dd{v}\qty[\partial_vf_2 \partial_v\qty(f_1\frac{\partial_v^2V}{\partial_vV} + \partial_vf_1)-\partial_vf_1 \partial_v\qty(f_2\frac{\partial_v^2V}{\partial_vV} + \partial_vf_2)]\\
  &= i\hbar\qty(T^c_{[f_1,f_2]}+\frac{c_T\hbar}{48\pi}\int\dd{v}(\pa_vf_1\partial_v^2f_2 - \pa_vf_2\partial_v^2f_1))
\end{align}
where the central charge is now $c_T=2+M+c_{\text{cl}}$.

\subsubsection{Reorientations}
\label{Subsection: shifted reorientations}

In the deformed classical theory, $\tilde\tau=\tilde\tau^c$ is still the generator of reorientations. But it decomposes into $X$ and the conjugate momentum $\Pi$ differently:
\begin{equation}
  \tilde\tau = (\partial_\mv X)^2\tau\circ X =
  \partial_v X\qty(\Pi \circ X + \frac{c_{\text{cl}}\hbar}{48\pi}\partial_\mv\qty(\frac{\partial_\mv^2X}{\partial_\mv X})).
\end{equation}
The covariant normal ordering of this observable proceeds as in the undeformed theory, and one finds in exactly the same way a correction term to the ordinary normal ordering:
\begin{equation}
  \Cov{\tilde\tau} = \tilde\tau -\frac\hbar{4\pi}\partial_\mv\qty(\frac{\partial_\mv^2 X}{\partial_\mv X}) = \normord{\partial_v X\qty(\Pi \circ X + \frac{(c_{\text{cl}}-12)\hbar}{48\pi}\partial_\mv\qty(\frac{\partial_\mv^2X}{\partial_\mv X}))},
\end{equation}
and we then have
\begin{equation}
  \Cov{\tilde\tau_g} = \int\dd{\mv}g\Cov{\tilde\tau} = \int\dd{v}\normord{\Pi g(V)}+\frac{(c_{\text{cl}}-12)\hbar}{48\pi}\int\dd{\mv}\partial_\mv g\frac{\partial_\mv^2X}{\partial_\mv X}.
\end{equation}
This is very similar to the undeformed theory~\eqref{Equation: reorientation generator}, with the only difference being a shift in the coefficient of the final term. Following the same steps as in Section~\ref{Subsection: reorientation algebra}, one obtains
\begin{equation}
  \comm{\Cov{\tilde\tau}_{g_1}}{\Cov{\tilde\tau}_{g_2}} = -i\hbar\qty(\Cov{\tilde\tau}_{[g_1,g_2]} - \frac{c_{\tilde\tau}\hbar}{48\pi}\int\dd{\mv}(\partial_\mv g_1\partial_\mv^2 g_2-\partial_\mv g_2\partial_\mv^2 g_1)),
\end{equation}
where the central charge $c_{\tilde\tau} = 24-c_{\text{cl}}$ now contains the contribution of the classical anomaly term.

\subsubsection{Dressed reparametrizations}

The generator of a dressed reparametrization is still $\Cov{\tilde T}$, and the analysis proceeds exactly as in Section~\ref{Subsection: dressed reparametrization algebra}, with the only change being in the value of the central charge, which is now given by $c_{\tilde T} = c_{\tilde \tau} - c_{\text{rad}} = 24-M-c_{\text{cl}}$.

\subsection{Anomaly cancellation, \texorpdfstring{$c_{\textnormal{cl}}=24-M$}{c cl = 24-M}}
\label{Section: anomaly cancel}

At this stage $c_{\text{cl}}$ is a parameter which we can freely specify in order to make the central charges take whatever values we would like.
In the purely classical theory, one should set $c_{\text{cl}}=0$ in order for the Raychaudhuri constraint to be non-anomalous, such that diffeomorphisms preserve the constraint surface in phase space.

But in the quantum theory this argument does not apply. Instead, we must pick some non-zero value of $c_{\text{cl}}$ such that the Raychaudhuri constraint is non-anomalous in the quantum theory. The aim is to set a central charge to zero, but since there are various central charges $c_T,c_{\tilde\tau},c_{\tilde T}$ that play a role, it may not be immediately obvious which one we should cancel. Note that we can cancel at most one of them in this way.

It turns out that the correct thing to do is to pick $c_{\text{cl}}$ such that the \emph{dressed} Raychaudhuri constraint central charge vanishes, $c_{\tilde T}=0$. This means we require $c_{\text{cl}}=24-M$, which leads to
\begin{equation}
  c_T = 26, \qquad c_{\tilde \tau} = M, \qquad c_{\tilde{T}}=0.
\end{equation}
As explained in Section~\ref{Section: physical representation}, it is only at this specific value of $c_{\text{cl}}$ that spurious extra degrees of freedom are eliminated from the physical Hilbert space of the theory.
The reorientations then have a final central charge $c_{\tilde \tau}=M$. This is the central charge of the undressed radiative fields and this choice is compatible with~\eqref{Equation: reorientation on D(O)} and~\eqref{Equation: physical reorientation is inverse reparametrization}, which equate the action of reorientations on dressed radiative operators to the radiative action on the undressed ones.  Note that reorientations are a physical symmetry, so from this point of view there is no issue with a non-zero value for their central charge.

Since this leaves the \emph{undressed} Raychaudhuri constraint central charge at $c_T=26$, one may be concerned that the gauge transformations are still anomalous. However, since the undressed Raychaudhuri constraint is not a gauge-invariant quantity, this non-vanishing central charge does not affect the gauge-invariant physics.
What is remarkable in our analysis is that we manage to construct the gauge-invariant observables and the gauge-invariant Hilbert space without the use of ghost fields.
It would be however  interesting to understand how the dressing picture described here connects to a BRST analysis where ghosts are introduced.

The construction given here of the anomaly cancellation bears some resemblance to the old school proof of the no-ghost theorems in string theory involving DDF operators; see \cite{Goddard:1972, GoddardThorn:1972, Brower:1972, Thorn:1987}.
In string theory the no-ghost theorem has also been proven in a covariant fashion using BRST techniques \cite{KatoOgawa:1983,KugoOjima:1979,Furuuchi:2006wv, Asano:2003jn} and the connections between BRST techniques and DDF constructions have been investigated in \cite{Schreiber_2004, Thorn:2011, Green_Schwarz_Witten_2012, Polchinski}.

In our case, it is not clear yet what the connection is between dressing and the BRST construction. Although we do not so in this paper, we expect that we could cancel the $c_T$ charge as well by additionally including a BRST ghost sector with central charge $c_{\text{gh}}=-26$.
However, in the gauge fixed BRST picture there is no longer a difference between $T$ and $\tilde{T}$. A deeper analysis of this question is needed, and we leave this for future work.

\section{Physical Hilbert space, and the dressing time QRF}
\label{Section: physical representation}

As explained above, the complete algebra of gauge-invariant operators is given by the crossed product
\begin{equation}
  \mathcal{A}_{\text{kin}}^{\Diff^+(\RR)} = \mathcal{B}(\mathcal{H}^\text{rad})\rtimes\Diff^+(\RR) = \{\Cov{\tilde\tau},\, \Dmap(a)\,\mid\,a\in\mathcal{B}(\mathcal{H}^\text{rad})\}'',
\end{equation}
where $\Cov{\tilde\tau}$ is the reorientation generator, and $\Dmap(a)=U_{\rad}[V]^\dagger a U_{\rad}[V]$ is a dressed system operator.

To complete the physical quantization of the theory, it remains to find an appropriate Hilbert space representation of this algebra. We use the GNS Hilbert space of the vacuum, which is equivalent to the span of the vacuum state $\ket{0}$ under the action of operators in $\mathcal{A}_{\text{kin}}^{\Diff^+(\RR)}$, with the null states quotiented out:
\begin{equation}
  \mathcal{H}_{\text{phys}} = \mathcal{A}_{\text{kin}}^{\Diff^+(\RR)}\ket{0}/\mathcal{N}.
\end{equation}
As explained in Section~\ref{Section: kinematical quantization}, the kinematical state space $\mathcal{K}_{\text{kin}}$ has indefinite inner product and is therefore not a Hilbert space. However, for $c_{\tilde T}\ge 0$, the inner product restricted to $\mathcal{A}_{\text{kin}}^{\Diff^+(\RR)}\ket{0}$ is positive semi-definite, so we can quotient this space by the subspace of null states (denoted $\mathcal{N}$) to obtain a space with positive definite inner product. Therefore, $\mathcal{H}_{\text{phys}}$ forms a genuine Hilbert space. This is an instantiation of  the no-ghost theorem, as we show explicitly below.

To understand the structure of $\mathcal{H}_{\text{phys}}$, it is more useful to write
\begin{equation}
  \mathcal{A}_{\text{kin}}^{\Diff^+(\RR)} = \{\Cov{\tilde T},\, \Dmap(a)\,\mid\,a\in\mathcal{B}(\mathcal{H}^\text{rad})\}''.
\end{equation}
Because $\Cov{\tilde T}=\Cov{\tilde\tau}+\Dmap(T_\text{rad})$, this is indeed the same algebra. This change of variables is useful because $\Cov{\tilde T}$ commutes with $\Dmap(a)$ for $a\in\mathcal{B}(\mathcal{H}^\text{rad})$ (see~\eqref{Equation: T tilde commutes with D(rad)}).
Hence, $\{\Cov{\tilde T}\}''$ and $\Dmap(\mathcal{B}(\mathcal{H}^\text{rad}))$ form two commuting subalgebras of $\mathcal{A}_{\text{kin}}^{\Diff^+(\RR)}$. The former is a Virasoro representation with central charge $c_{\tilde T}$, while the latter is unitarily equivalent to $\mathcal{B}(\mathcal{H}^\text{rad})$. Appropriately, the physical Hilbert space we construct has tensor factors corresponding to these two subalgebras. Indeed, it is isomorphic to the tensor product of a Virasoro Verma module with the radiative Hilbert space $\mathcal{H}^{\text{rad}}$ (the former factor becomes trivial if $c_{\tilde T}=0$).

It is useful to explicitly exhibit this isomorphism. To that end, using $\Dmap(a)=U_{\text{rad}}[V]^\dagger a U_{\text{rad}}[V]$ for $a\in\mathcal{B}(\mathcal{H}^\text{rad})$, and
\begin{equation}
  U_\text{rad}[V]\Cov{\tilde T}U_\text{rad}[V]^\dagger = \Cov{\tilde\tau} -\frac{c^\text{rad}\hbar}{48\pi}\partial_\mv\qty(\frac{\partial_\mv^2X}{\partial_\mv X}),
  \label{Equation: U rad V tilde T}
\end{equation}
(which follows from~\eqref{Equation: U rad V tilde tau}) we can write
\begin{equation}
  \mathcal{A}_{\text{kin}}^{\Diff^+(\RR)} = U_\text{rad}[V]^\dagger\qty(\qty{\Cov{\tilde\tau} -\tfrac{c^\text{rad}\hbar}{48\pi}\partial_\mv\qty(\tfrac{\partial_\mv^2X}{\partial_\mv X})}'' \otimes\mathcal{B}(\mathcal{H}^\text{rad}))U_{\text{rad}}[V].
\end{equation}
Note that $U_\text{rad}[V]$ is a unitary operator, which implies that~\eqref{Equation: U rad V tilde T} generates a Virasoro representation at the same central charge $c_{\tilde T}$ as $\Cov{\tilde T}$. This can also be directly checked using the right-hand side of~\eqref{Equation: U rad V tilde T}; the $\Cov{\tilde \tau}$ term gives a representation with central charge $c_{\tilde\tau}$, which the latter term corrects to $c_{\tilde\tau}-c^\text{rad}=c_{\tilde T}$ in a manner similar to that described in Subsection~\ref{Subsection: shifted reorientations}.

Note also that the vacuum is fixed by $U_\text{rad}[V]$, i.e.\ $U_\text{rad}[V] \ket{0} =\ket{0}$. Indeed, when $V$ acts on the spin 0 vacuum $\ket{0}^0$ it can be treated as a negative frequency operator, and negative frequency reparametrizations fix the radiative vacuum $\ket{0}^\text{rad}$.
We therefore have
\begin{align}
  \mathcal{A}_{\text{kin}}^{\Diff^+(\RR)}\ket{0} &= U_\text{rad}[V]^\dagger\qty(\qty{\Cov{\tilde\tau} -\tfrac{c^\text{rad}\hbar}{48\pi}\partial_\mv\qty(\tfrac{\partial_\mv^2X}{\partial_\mv X})}'' \otimes\mathcal{B}(\mathcal{H}^\text{rad}))\ket{0}\\
  &= U_{\text{rad}}[V]^\dagger\qty(\qty{\Cov{\tilde\tau} -\tfrac{c^\text{rad}\hbar}{48\pi}\partial_\mv\qty(\tfrac{\partial_\mv^2X}{\partial_\mv X})}''\ket{0}^0 \otimes\mathcal{B}(\mathcal{H}^\text{rad})\ket{0}^\text{rad}),
\end{align}
where $\ket{0}^0$ and $\ket{0}^\text{rad}$ are the spin 0 and radiative vacuums respectively.
This implies
\begin{equation}
  \mathcal{A}_{\text{kin}}^{\Diff^+(\RR)}\ket{0} = U_{\text{rad}}[V]^\dagger \qty(\mathcal{V}_{c_{\tilde T}}\otimes \mathcal{H}^{\text{rad}}).
\end{equation}
Here, $\mathcal{H}^{\text{rad}}$ is the usual radiative Hilbert space, while $\mathcal{V}_{c_{\tilde T}}$ is a \emph{vacuum Verma module}\footnote{This is a Virasoro representation generated by acting with a stress tensor on an $\SLTR$-invariant, positive-frequency-annihilated vacuum.} for a stress tensor of central charge $c_{\tilde T}$. This Verma module contains extra, spurious degrees of freedom that result from the gauge anomaly in the quantization, leading to the existence of states in the physical Hilbert space $\mathcal{H}_{\text{phys}}$ for which the dressed stress tensor $\Cov{\tilde T}$ has non-vanishing expectation value, i.e.\ the Raychaudhuri constraint is violated. For example, consider the state $e^{i\Cov{\tilde T}_g/\hbar}\ket{0}$, where $\Cov{\tilde T}_g = \int\dd{\mv}g\Cov{\tilde T}$. By the anomalous transformation law of the stress tensor, we have
\begin{equation}
  \bra{0}e^{-i\Cov{\tilde T}_g/\hbar}\Cov{\tilde T}e^{i\Cov{\tilde T}_g/\hbar}\ket{0}
  = \bra{0}\qty((\partial_\mv G)^2\Cov{\tilde T}\circ G + \frac{c_{\tilde T}\hbar}{24\pi}\Schwarzian{G}{\mv})\ket{0}
  = \frac{c_{\tilde T}\hbar}{24\pi}\Schwarzian{G}{\mv}.
  \label{Equation: dressed Raychaudhuri violation}
\end{equation}
where $G=\exp(g\partial_\mv)$.
The purpose of anomaly cancellation is to eliminate this possibility, and thus to enforce that the Raychaudhuri constraint is satisfied by all states.
It is clear that~\eqref{Equation: dressed Raychaudhuri violation} vanishes if we set $c_{\tilde T}=0$ by introducing a classical central charge $c_{\text{cl}}=c_{\tilde T}$ as described in Section~\ref{Section: anomaly shift} -- so that the state $\ket{G}$ is no longer a problem. In fact, setting $c_{\tilde T}=0$ eliminates all such problematic states.

To see this, we need to compute the space of null states $\mathcal{N}$, which depends on the value of the central charge, via the properties of the Verma module $\mathcal{V}_{c_{\tilde T}}$:
\begin{itemize}
  \item $\bm{c_{\tilde T}>0}$ gives a Verma module $\mathcal{V}_{c_{\tilde T}}$ with a positive-definite inner product. There are then no null states, and we have
    \begin{equation}
      \mathcal{H}_{\text{phys}}=U_\text{rad}[V]^\dagger(\mathcal{V}_{c_{\tilde T}}\otimes \mathcal{H}^\text{rad}).
    \end{equation}
  \item $\bm{c_{\tilde T}<0}$ gives a Verma module $\mathcal{V}_{c_{\tilde T}}$ with an indefinite inner product, so we cannot form a physical Hilbert space when this is the case.\footnote{Although, note that if we had defined the spin 0 vacuum $\ket{0}^0$ to be annihilated by \emph{negative}-frequency modes (rather than positive-frequency ones), we would get a positive-definite inner product at $c_{\tilde T}<0$.}
  \item At $\bm{c_{\tilde T}=0}$, the Verma module becomes trivial: $\mathcal{V}_{c_{\tilde T}}\simeq\CC$. In particular, the only non-null state in $\mathcal{V}_{c_{\tilde T}}$ is the vacuum $\ket{0}^0$. As a consequence we have
    \begin{equation}
      \mathcal{N} = U_{\text{rad}}[V]^\dagger(\{\ket{0}^0\}^\perp \otimes \mathcal{H}^{\text{rad}}),
    \end{equation}
    and it is simplest to implement the quotient by $\mathcal{N}$ by just setting
    \begin{equation}
      \mathcal{H}_{\text{phys}}=U_\text{rad}[V]^\dagger\qty(\ket{0}^0\otimes \mathcal{H}^\text{rad}).
      \label{Equation: perspective-neutral H phys}
    \end{equation}
\end{itemize}

Let us from now on assume we have picked $c_{\text{cl}}=24-M$ to cancel the anomaly, so that $c_{\tilde T}=0$.\footnote{Here we have only considered bosonic radiation, but it is simple to extend to fermionic fields too. In a theory with $N_b$ bosons and $N_f$ Weyl fermions in the radiation, one would have $M=N_b+\frac12N_f$ (the fermions each contribute $\frac12$ to the central charge, relative to the bosons).

A back-of-the-envelope calculation with the Standard Model (after electroweak symmetry breaking, without right-handed neutrinos, and additionally including the two polarizations of spin 2 gravitational radiation) gives $M=52.5$ ($2\times 1$ for the gravitons, $2\times 8$ for the gluons, $2\times 1$ for the photon, $3\times 3$ for the $\mathrm{SU}(2)$ massive vectors, $1$ for the Higgs, and $\frac12\times 45$ for the fermions), so one would take $c_{\text{cl}}=-28.5$.} Then quotienting by the null states causes all spurious degrees of freedom to be eliminated. Indeed, gauge-invariant operators $A\in\mathcal{A}_{\text{kin}}^{\Diff^+(\RR)}$ are then represented on this physical Hilbert space in the following way:
\begin{equation}
  r(A)\ket{\psi} = \Pi_{\text{phys}} A\ket{\psi},
\end{equation}
where
\begin{equation}
  \Pi_{\text{phys}} = U_{\text{rad}}[V]^\dagger\qty(\ket{0}^0\bra{0}^0\otimes\mathds{1}_{\text{rad}})U_{\text{rad}}[V]
\end{equation}
is the projection onto $\mathcal{H}_{\text{phys}}\subset \mathcal{A}_{\text{kin}}^{\Diff^+(\RR)}\ket{0}$. This defines the representation
\begin{equation}
  r:\mathcal{A}_\text{kin}^{\Diff^+(\RR)}\to \mathcal{B}(\mathcal{H}_{\text{phys}}).
\end{equation}
Since $\Dmap(a)=U_{\text{rad}}[V]^\dagger a U_\text{rad}[V]$ (with $a\in\mathcal{B}(\mathcal{H}^\text{rad})$) commutes with $\Pi_{\text{phys}}$, we have $r(\Dmap(a))=\Dmap(a)$. On the other hand, the dressed Raychaudhuri constraint satisfies $r(\Cov{\tilde T})=0$, since
\begin{align}
  r(\Cov{\tilde T}) \ket{\psi} &= \Pi_{\text{phys}}\Cov{\tilde T} \Pi_{\text{phys}}\ket{\psi} \\
  &= U_{\text{rad}}[V]^\dagger\qty(\ket{0}^0\bra{0}^0\otimes\mathds{1}_{\text{rad}})U_{\text{rad}}[V]\Cov{\tilde T} U_{\text{rad}}[V]^\dagger\qty(\ket{0}^0\bra{0}^0\otimes\mathds{1}_{\text{rad}})U_{\text{rad}}[V] \ket{\psi}\\
  &= \underbrace{\bra{0}^0\qty(\Cov{\tilde\tau} -\tfrac{c^\text{rad}\hbar}{48\pi}\partial_\mv\qty(\tfrac{\partial_\mv^2X}{\partial_\mv X}))\ket{0}^0}_{=0} \, U_{\text{rad}}[V]^\dagger\qty(\ket{0}^0\bra{0}^0\otimes\mathds{1}_{\text{rad}})U_{\text{rad}}[V] \ket{\psi}
\end{align}
Thus, the triviality of the Verma module at $c_{\tilde T}=0$ leads to a physical Hilbert space for which $\Cov{\tilde T}=0$ is satisfied as an operator. The reorientation operator can then be identified with minus the dressed radiative stress tensor:
\begin{equation}
  r(\Cov{\tilde \tau}) = -r(\Covrad{\tilde T}) = -U_\text{rad}[V]^\dagger T^\text{rad} U_\text{rad}[V].
  \label{Equation: physical reorientation is inverse reparametrization}
\end{equation}
In what follows we sometimes leave the representation map $r$ implicit.

\subsection{The perspective of the dressing time frame}

It is clear from~\eqref{Equation: perspective-neutral H phys} that the physical Hilbert space is unitarily equivalent to $\mathcal{H}^{\text{rad}}$. We can explicitly write down the map relating these spaces as
\begin{equation}
  \mathcal{R} = (\bra{0}^0\otimes\mathds{1}_\text{rad}) U_\text{rad}[V].
  \label{Equation: reduction map}
\end{equation}
We have
\begin{equation}
  \mathcal{R}\mathcal{H}_{\text{phys}} = \mathcal{H}^{\text{rad}},
\end{equation}
and with
\begin{equation}
  \mathcal{R}^\dagger = U_\text{rad}[V]^\dagger(\ket{0}^0\otimes\mathds{1}_\text{rad})
\end{equation}
we have
\begin{equation}
  \mathcal{R}^\dagger \mathcal{R} = \Pi_{\text{phys}}, \qquad \mathcal{R}\mathcal{R}^\dagger = \mathds{1}_{\text{rad}}.
\end{equation}

In the language of the perspective-neutral formalism for quantum reference frames~\cite{delaHamette:2021oex,Hoehn2023,Hoehn2021a,Hoehn2021,AliAhmad2022,CastroRuiz2020,Vanrietvelde2020,Hoehn2022,Suleymanov:2023wio,DeVuyst:2024pop,DeVuyst:2024uvd,kirklin2024generalisedsecondlawsemiclassical}, $\mathcal{H}_{\text{phys}}$ is the `perspective-neutral' Hilbert space, the image $\mathcal{H}^{\text{rad}}$ of $\mathcal{R}$ is the reduced Hilbert space `in the perspective of' the dressing time QRF $V$, and $\mathcal{R}$ itself is the Page--Wootters `reduction' map relating these two Hilbert spaces. Given a perspective-neutral state $\ket{\psi}\in\mathcal{H}_{\text{phys}}$, the state `in the perspective of' $V$ is defined by $\mathcal{R}\ket{\psi}$.
To be clear, these are two \emph{isomorphic} descriptions of the degrees of freedom in the physical theory, just with different interpretations. The former $\mathcal{H}_{\text{phys}}$ gives a gauge-invariant description, while the latter $\mathcal{H}^{\text{rad}}$ gives the description appropriate after gauge-reducing with $V(v)=v$.

The algebra of operators acting in the reduced Hilbert space is obtained by conjugating $\mathcal{A}_{\text{phys}}$ with $\mathcal{R}$:
\begin{equation}
  \mathcal{R}\mathcal{A}_{\text{phys}}\mathcal{R}^\dagger.
\end{equation}
For example, the dressed radiative operators $\Dmap(a)$ are mapped to their undressed counterparts:
\begin{equation}
  \mathcal{R}\Dmap(a)\mathcal{R}^\dagger = a.
\end{equation}
At the level of the reduced Hilbert space, the reorientation generator is simply minus the radiative stress tensor:
\begin{equation}
  \mathcal{R}\Cov{\tilde\tau}\mathcal{R} = -U_\text{rad}[V]\Covrad{\tilde T}U_{\text{rad}}[V]^\dagger = -T^\text{rad}.
\end{equation}

\subsection{The dressing time frame is `Heisenberg ideal' but `Schr\"odinger non-ideal'}
\label{Subsection: nonideal}

An important classification of quantum reference frames involves the distinction between \emph{ideal} QRFs and \emph{non-ideal} QRFs. Actually, it is useful to distinguish two notions of ideality/non-ideality. Let us use the following terminology:
\begin{itemize}
  \item A QRF is \textbf{Heisenberg ideal} if the dressed system operators are isomorphic to the underlying system algebra, i.e.\ $\Dmap(a)\Dmap(b)=\Dmap(ab)$, for $a,b$ acting on the system.
  \item A QRF is \textbf{Schr\"odinger non-ideal} if its distinct coherent states have non-trivial overlaps -- in other words, its quantum orientation is fuzzy.
\end{itemize}
We have already seen in Subsection~\ref{Subsection: radiative endomorphism} and elsewhere that the dressing time frame is Heisenberg ideal. On the other hand, it is Schr\"odinger non-ideal, as described below.

Schr\"odinger ideal frames are characterized by permitting states of arbitrarily sharply peaked orientation. For example, the prototypical QRF for a locally compact gauge group $G$ has Hilbert space $L^2(G)$. The group $G$ is the space of orientations for the QRF, and $L^2(G)$ contains arbitrarily sharply peaked wavefunctions around any group element, so this QRF is ideal. It has an orientation operator $\hat g$, with eigenstates $\ket{g}$ with delta-function-like overlap:
\begin{equation}
  \braket{g}{g'} = \delta(g,g').
\end{equation}

On the other hand, Schr\"odinger non-ideal QRFs have some necessary minimum amount of `fuzziness' in their orientation, and are typically more physically realistic than ideal ones. For example, consider the simple case that the group $G=\RR$ consists of time evolution, so that the QRF is a clock. Physically realistic clocks should have a lower bound on their energy, which would imply that the conjugate variable (the clock time, which is its orientation in this case) must necessarily be non-compactly smeared. More generally, the reorientation operators have some restricted spectrum, and the conjugate orientations are then necessarily smeared. In this case, by Pauli's argument (see e.g.~\cite{Hoehn2021a}) there is no QRF orientation operator, but this can be replaced by a notion of covariant POVM. In many convenient situations the covariant POVM is associated with a set of coherent states $\ket{g}$, and the mark of Schr\"odinger non-idealness is that these coherent states have some non-trivial overlaps:
\begin{equation}
  \abs{\braket{g}{g'}} = e^{-k(g,g')}.
\end{equation}
These coherent states are generated by acting with reorientations on some base state $\ket{0}$.

In the case of the dressing time QRF, the group is $G=\operatorname{Diff}^+(\RR)$, and the orientations are the configurations of $V$. Let us demonstrate in two ways that this QRF is Schr\"odinger non-ideal.
First, note that the reorientation generator $\Cov{\tilde\tau}=-\Covrad{\tilde T}$ has a restricted spectrum. For example, the averaged null energy $\int\dd{v}\Covrad{\tilde T}$ is strictly positive by the ANEC. Thus, there cannot be a state of arbitrarily sharply peaked conjugate orientation to $\int\dd{v}\Cov{\tilde T}$. There are similar constraints on other modes of the stress tensor (e.g.~\cite{Flanagan:1997gn}) which also limit how peaked we can make the dressing time orientation.

Second, we can consider the overlaps of the coherent states of the dressing time QRF. Starting with the vacuum state $\ket{0}$, we produce a coherent state by acting with a reorientation. Defining $\tilde U[G]=\exp(-i\Cov{\tilde{\tau}}_g/\hbar)$ for $G=\exp(g\partial_\mv)$, the coherent state is therefore given by
\begin{equation}
  \ket{G} = \tilde U[G^{-1}] \ket{0},
\end{equation}
for some diffeomorphism $G$. The overlap of two such coherent states is then given by
\begin{equation}
  \braket{G_1}{G_2} = \bra{0}\tilde U[G_1\circ G_2^{-1}]\ket{0} \exp(-i\frac{c_{\tilde\tau}}{24}C(G_1,G_2^{-1})),
\end{equation}
where the cocycle $C(G_1,G_2^{-1})$ is described in Appendix~\ref{Appendix: cocycle}. Focusing on the absolute value of this overlap, we have
\begin{equation}
  \abs{\braket{G_1}{G_2}} = \abs{\bra{0}\tilde U[G_1\circ G_2^{-1}]\ket{0}} = \exp(-\frac{c_{\tilde \tau}}{24}K(G_1\circ G_2^{-1})),
  \label{Equation: dressing time coherent state overlap}
\end{equation}
where, as shown in Appendix~\ref{Appendix: Kahler},
\begin{equation}
  K(F) = \frac{1}{2\pi}\qty(\int_{\bar{\mathbb{H}}}\dd[2]{z}\abs{\frac{\partial_z^2F_+}{\partial_zF_+}}^2+\int_{\mathbb{H}}\dd[2]{z}\abs{\frac{\partial_z^2F_-}{\partial_zF_-}}^2).
\end{equation}
Here, we are applying a Birkhoff decomposition to write a given diffeomorphism $F=F_-^{-1}\circ F_+$ as a composition of maps $F_\pm$ which are univalent in the lower/upper half planes $\mathbb{H},\bar{\mathbb{H}}$ respectively (see Appendix~\ref{Appendix: Kahler}). This $K(F)$ is the K\"ahler potential for coadjoint orbits of the Virasoro group, also known as the Teo-Takhtajan energy \cite{Takhtajan:2003hm, Alekseev:2022efp} (see also~\cite{nair2024notecoherentstatesvirasoro}). From~\eqref{Equation: dressing time coherent state overlap} we see that different coherent states of the dressing time QRF have non-trivial overlap, and therefore this frame is Schr\"odinger non-ideal. The strength of this overlap is controlled by the central charge $c_{\tilde\tau}$. Having cancelled the anomaly, this takes the value $c_{\tilde\tau}=M$. Thus, with more radiative fields, the coherent states are more sharply peaked. The dressing time frame would only become ideal in the limit $M\to\infty$ of infinitely many radiative fields.

\section{Conclusion}
\label{Section: conclusion}

In this paper, we have carried out the covariant quantization of gravitational null rays in terms of observables dressed to the dressing time $V$. Before ending, let us comment on possible future directions.

As already mentioned in the main text, the methods we have used bear a resemblance to similar approaches in the foundations of string theory \cite{DelGiudice:1971yjh, Brower:1972}. It could potentially be very useful to import other ideas from string theory to apply to the physics of gravitational null surfaces.

We were able to avoid using BRST techniques in our covariant quantization of the null ray, by focusing directly on the properties of dressed operators. This is strongly suggestive that a general enough formulation of dressing in quantum theories may more broadly be used as an alternative to BRST. It has been suggested by Grassi and Porrati that quantum dressing should be understood as an intertwiner that trivializes the BRST charge~\cite{Grassi:2024vkb}, and we would like to better understand how that perspective relates to the setup described here.

Our main technical tool in this paper was covariant normal ordering. It would be very interesting to understand the mathematical properties of this prescription more generally, and to understand how the covariant star product fits into the general framework of deformation quantization.

We have worked in a linearized perturbative regime in which all interactions are negligible. It is important to understand what changes once we turn the interactions back on. Such interactions would affect renormalization, including covariant normal ordering. Even in pure gravity, the spin 2 gravitons interact with each other and the spin 0 area degrees of freedom. The $\RR^2$ target space (one direction for each polarization) of the linearized gravitons considered here (in string theory language, the transverse degrees of freedom) would be modified to a coset target space $H_2=\mathrm{SL}(2,\RR)/\mathrm{SO}(2)$ consisting of unimodular 2-metrics. QRFs interacting with the system they are used to observe are much less well understood than those which do not. Such interactions can lead to interesting phenomena such as non-unitary evolution of the state from the perspective of the frame~\cite{Gambini:2004pe,Gambini:2006ph,Smith2019,Dittrich_2017,dittrich2015chaosdiracobservablesconstraint,Paiva_2022, Rijavec:2025vti}. These phenomena are presumably all highly relevant to the complete non-perturbative physics of null rays.

As noted in Subsection~\ref{Subsection: fluctuations}, dressed operators fluctuate differently from their undressed counterparts. It would be interesting to understand whether such fluctuations can actually be measured, i.e.\ whether there is a phenomenology of dressing (see~\cite{Freidel:2026hed}). A related question (and one perhaps more well-defined, given the relational nature of gravity) involves understanding the fluctuations of operators dressed to \emph{different} reference frames. Indeed, we have used dressing time as a QRF in this paper, but there are of course other choices of time (such as affine time, area time, conformal time, etc.~\cite{Ciambelli:2023mir,Ciambelli_2024}) that one could use to dress observables. We then may be able to phenomenologically distinguish the different dressings by accounting for the different ways in which they fluctuate relative to each other. Alternatively, one can always write down a gauge-invariant field-dependent diffeomorphism that maps between different choices of time, e.g. dressing time to affine time, $F=V_{\text{aff}}\circ V^{-1}$. In the quantum theory this becomes a gauge-invariant operator (if one uses covariant normal ordering) whose fluctuations can in principle be measured. With this in mind, it is also worth asking whether there is an `optimal' reference frame, which could be defined as one which minimizes in some way the magnitude of the fluctuations of operators dressed to it. A natural candidate seems to be the dressing time $V$, but we leave exploration of this to future work.

For much of the paper, we have ignored the edge modes $\omega_a,q_a$. They are simple to include back in the analysis, because they are already gauge-invariant, so they contribute the same Hilbert space tensor factor $\mathcal{H}^\text{edge}= L^2(\RR^2)$ and operators $\mathcal{B}(\mathcal{H}^{\text{edge}})$ at the kinematical and physical levels. Including the edge modes, and assuming the anomaly has been cancelled so that $c_{\tilde T}=0$, the complete physical Hilbert space is given by
\begin{equation}
  \mathcal{H}_{\text{phys}} = U_{\text{rad}}[V]^\dagger(\ket{0}^0\otimes \mathcal{H}^\text{rad}) \otimes \mathcal{H}^\text{edge}.
\end{equation}
The edge modes are necessary for including the dressed area $\Cov{\tilde\Omega}$ as an operator acting on the physical Hilbert space, since they provide boundary conditions in~\eqref{area}. Moreover, they are required if one wants to include boosts that act non-trivially at the corners of the null ray segment. They are therefore very important for a correct understanding of gravitational entropies on null ray segments, since boosts correspond to modular flow, and the area is the dominant term in the generalized entropy. We would like to understand the consequences of our construction for entropies, including the identification of the generalized entropy as the Von Neumann entropy of a modular crossed product algebra~\cite{CLPW,Jensen2023}. A key future direction is to further develop the approach of~\cite{kirklin2024generalisedsecondlawsemiclassical} in order to understand the ways in which taking QRFs into account affects the status of the generalized second law. Note that in~\cite{CLPW,Jensen2023} a single boost gauge symmetry was accounted for, and in~\cite{kirklin2024generalisedsecondlawsemiclassical} a two-dimensional affine gauge group was accounted for -- but here we go much further by accounting for all diffeomorphisms of the null ray.

Finally, we have considered only a single null ray in this paper. An essential next step is to promote this ray to a full null surface, which introduces many additional complications. For example, the central charge needs to be regularized~\cite{Ciambelli_2024}, and moreover there are many additional diffeomorphisms and constraints that need to be accounted for~\cite{CiambelliFreidelDamour}. We believe that it will be fruitful to apply the methods developed in this paper to that more complete setup, and look forward to doing so.

\phantomsection
\addcontentsline{toc}{section}{\numberline{}Acknowledgements}
\section*{Acknowledgements}
We thank Shadi Ali Ahmad, Rodrigo Andrade e Silva, Luca Ciambelli, Stefan Eccles, Eanna Flanagan, Philipp H\"ohn, Marc Klinger, Fabio Mele, Eirini Telali, and Tom Wetzstein for helpful and stimulating discussions. This work was
supported by the Simons Collaboration on Celestial Holography.
This project was also made possible through the support of the ID\# 62312 grant from the John Templeton Foundation, as part of the \href{https://www.templeton.org/grant/the-quantum-information-structure-of-spacetime-qiss-second-phase}{\textit{`The Quantum Information Structure of Spacetime'} Project (QISS)}.~The opinions expressed in this project/publication are those of the author(s) and do not necessarily reflect the views of the John Templeton Foundation. Research at Perimeter Institute is supported in part by the Government of Canada through the Department of Innovation, Science and Economic Development and by the Province of Ontario through the Ministry of Colleges and Universities.

\appendix

\section{Reparametrization algebra}
\label{Subsection: reparametrization algebra}
The generator of reparametrizations is the normal ordered Raychaudhuri constraint / stress tensor
\begin{equation}
  T = \normord{\Pi\partial_vV+\sum_i(\partial_v\varphi_i)^2}.
\end{equation}
Wick's theorem, using (\ref{vacexp0}, \ref{vacexp1}), yields
\begin{align}
  T(v)T(u)& = \normord{ T(v)T(u)}
  - \frac\hbar{2\pi}\frac{\normord{\Pi(u)\partial_vV(v)+\Pi(v)\partial_uV(u)+2\sum_i(\partial_u\varphi_i(u)\partial_v\varphi_i(v))}}{(v-u-i\epsilon)^2}\\
  & + \qty(\frac{\hbar}{2\pi}\frac1{(v-u-i\epsilon)^2})^2 \frac{(2+M)}{2}
\end{align}
where $M$ is the number of radiative fields $\varphi_i$, and the central charge is $c_T=2+M$.
Keeping only the singular terms gives
\begin{align}
  T(v)T(u)&  \sim\frac\hbar{2\pi} \left( \frac\hbar{4\pi} \frac{c_T}{(v-u-i\epsilon)^4} - \frac{2 T(u)}{(v-u-i\epsilon)^2} - \frac{\pa_u T(u)}{(v-u-i\epsilon)}
  \right),\\
  T(u)T(v)&  \sim\frac\hbar{2\pi} \left( \frac\hbar{4\pi} \frac{c_T}{(v-u-i\epsilon)^4} - \frac{2 T(u)}{(v-u-i\epsilon)^2} + \frac{\pa_u T(u)}{(v-u-i\epsilon)}
  \right),
\end{align}
so the commutator may be written using
$\tfrac{1}{(v-i\epsilon)}= \mathrm{P}\left(\tfrac1{v}\right) + i\pi \delta(v)$  as
\begin{align}
  \comm{T(v)}{T(u)}
  &= i\hbar\qty(\delta(v-u) \partial_uT(u) - 2\partial_u[\delta(v-u) T(u)] + \frac{c_T\hbar}{24\pi}\partial_u^3\delta(v-u)).
\end{align}
Defining
$
T_f = \int\dd{v}f T,
$
one can integrate the above to find
\begin{align}
  \comm{T_f}{T_g} &= i\hbar\int\dd{u} \qty(-fg\partial_uT + 2 f \pa_u(g T)+\frac{\hbar c_T}{24\pi}g\partial_u^3f) \\
  &= i\hbar \qty(T_{[f,g]}+\frac{c_T\hbar}{48\pi}\int\dd{v}(\pa_vf\partial_v^2g - \pa_vg\partial_v^2f)).
\end{align}

\section{Virasoro cocycle and K\"ahler potential}
\label{Appendix: cocycle}

We assume there is a stress tensor $T$ satisfying
\begin{equation}
  \frac1{i\hbar}[T_f,T_g] = T_{[f,g]} + \frac{c\hbar}{48\pi}\int\dd{v}(\partial_vf \partial_v^2g - \partial_vg\partial_v^2f), \qq{where}
  T_f = \int\dd{v}fT,
\end{equation}
and a dressing field $V$ satisfying
\begin{equation}
  \frac1{i\hbar}[V,T_f] = f\partial_v V, \qquad [V(v),V(u)]=0.
\end{equation}
This covers all cases of interest in the paper, for different values of the central charge $c$. For dressed stress tensors like $\tilde\tau,\tilde T$, we can use $X$ instead of $V$.

We are interested in the cocycle $C(F,G)$ appearing in
\begin{equation}
  U[F]U[G] = U[F\circ G] \exp(-\frac{ic}{24} C(F,G)), \qq{where} U[\exp(f\partial_v)] = \exp(\frac{i}{\hbar}T_f).
\end{equation}
The goal of this section is to show that
\begin{equation}
  C(F,G) = \mathrm{B}(F,G) - b(F\circ G) + b(F) + b(G),
  \label{Equation: cocycle}
\end{equation}
where
\begin{equation}
  \mathrm{B}(F,G) := \frac{1}{2\pi}\int_{\mathbb{R}}\dd{v}\ln(\partial_vF\circ G) \pa_v \ln \pa_v G  = -\frac1{2\pi}\int\dd{v}\ln\partial_v F \partial_v\ln\partial_v G^{-1}\label{BottT}
\end{equation}
is the Bott-Thurston cocycle, and $b(F)$ is the Lagrangian functional
\begin{equation}
  b(F) := \int_0^1 \Theta_s(f) \rd s
\end{equation} obtained by integrating  the Kirillov-Kostant-Souriau symplectic potential~\cite{Guillemin:1990ew,Woodhouse:1980pa} along the path from the identity diffeomorphism to $F_s := \exp(sf\partial_v)$  for $s\in[0,1]$
\begin{equation}
  \Theta_s(f) :=\frac1{2\pi}\int_{\mathbb{R}}\dd{v}\pa_s(\ln\partial_vF_s) \partial_v\ln\partial_v F_s
  \label{Equation: KKS potential}
\end{equation}
This cocycle is such that
\begin{equation}
  b(F^{-1})=-b(F).
\end{equation}
To prove this, one first changes variables $s\to 1-t$:
\begin{align}
  b(F) &= \frac1{2\pi}\int_0^1\dd{s}\int_\RR\dd{v}\partial_s(\ln\partial_vF_s)\partial_v\ln\partial_vF_s\\
  &=  -\frac1{2\pi}\int_0^1\dd{t}\int_\RR\dd{v}\partial_t(\ln\partial_v(F_{-t}\circ F))\partial_v\ln\partial_v(F_{-t}\circ F)
\end{align}
Then one uses
$\partial_t(\ln\partial_v(F_{-t}\circ F)) = \partial_t(\ln\partial_vF_{-t})\circ F$,
and changes variables $v\to F(v)$:
\begin{align}
  b(F)
  &= -\frac1{2\pi}\int_0^1\dd{t}\int_{\RR}\dd{v}\partial_t(\ln\partial_vF_{-t})\qty(\partial_v\ln\partial_vF_{-t} - \partial_v\ln\partial_vF^{-1})\\
  &= -b(F^{-1}) + \frac1{2\pi}\int\dd{v}\ln\partial_vF_{-1}\partial_v\ln\partial_vF_{-1}.
\end{align}
The second term vanishes upon integration, so we have the desired result.

We similarly have the formula from changing variables $v\to G(v)$
\begin{align}
  b(F) &= \frac1{2\pi}\int_0^1\dd{s}\int_\RR\dd{v}\partial_s(\ln\partial_vF_s)\partial_v\ln\partial_vF_s \\
  &= \frac1{2\pi}\int_0^1\dd{s}\int_\RR\dd{v}\partial_s(\ln\partial_v(F_s\circ G))\qty(\partial_v\ln\partial_v(F_s\circ G)-\partial_v\ln\partial_v G)\\
  &= \frac1{2\pi}\int_0^1\dd{s}\int_\RR\dd{v}\partial_s(\ln\partial_v(F_s\circ G))\partial_v\ln\partial_v(F_s\circ G)-\frac1{2\pi}\int_\RR\dd{v}\ln\partial_vF\circ G\partial_v\ln\partial_v G.
\end{align}
The second term is $-B(F,G)$. The first term is the integral of the KKS potential
\begin{equation}
  \Theta[H] = \frac1{2\pi}\int_\RR\dd{v}\delta(\ln\partial_v H) \partial_v\ln\partial_v H
\end{equation}
along the curve $s\mapsto F_s\circ G$. So we can combine everything to find
\begin{equation}
  C(F,G) = \int_{\Gamma(F,G)}\Theta,
\end{equation}
where the curve $\Gamma(F,G)$ goes from the identity to $G$ along $G_s$, then to $F\circ G$ along $F_s\circ G$, then back to the identity along $(F\circ G)_s$.

This nicely generalizes to path-ordered exponentials.
\begin{equation}
  \mathrm{P}\exp(\frac{i}{\hbar}\int_0^1\dd{s} T_{f_s}) = U[\mathrm{P}\exp(\int_0^1\dd{s}f_s\partial_v)] \exp(-\frac{ic}{48\pi}\int_{\Gamma[f]}\Theta),
\end{equation}
where $\Gamma[f]$ is the path going along $F_s=\mathrm{P}\exp(\int_0^s\dd{s}f_s\partial_v)$ from $s=0$ to $s=1$, and then back to the identity along the exponential path $\exp(tf_1\partial_v)$ from $t=1$ to $t=0$.

Note that for small $f$, $b(F)$ is cubic in $f$. Indeed, one has
\begin{equation}
  \ln\partial_vF_s
  = s\partial_vf + \frac12 s^2 f\partial_v^2f + \order{f^3},
\end{equation}
so that
\begin{align}
  b(F)
  &= \frac1{4\pi}\int_0^1\dd{s} \int\dd{v}\qty(\partial_v\qty(s(\partial_vf)^2+s^2f\partial_vf\partial_v^2f) + s^2f(\partial_v^2f)^2) +\order{f^4}\\
  &= \frac1{12\pi}\int\dd{v}f(\partial_v^2f)^2 + \order{f^4}.
\end{align}

Let's now show how to construct the cocycle. An efficient way to compute it is to define the `corrected' generator
\begin{align}
  \mathcal{T}_f &= T_f + \mathcal{F}_f,
\end{align}
where $\mathcal{F}_f$ is a functional of $V$ defined as
\begin{align}
  \mathcal{F}_f[V] :=
  - \frac{c\hbar}{48\pi}\int_{\mathbb{R}}\dd{v}\partial_v f  \pa_v \ln  \pa_v V,
\end{align}
which satisfies
\begin{align}
  \frac1{i\hbar}[\mathcal{T}_f,\mathcal{T}_g]
  &= \frac1{i\hbar}[T_f,T_g] - \frac{c\hbar}{48\pi} \frac1{i\hbar}\int_{\mathbb{R}}\dd{v}\qty(\partial_v f \comm{\frac{\partial_v^2 V}{\partial_vV}}{T_g}-\partial_v g \comm{\frac{\partial_v^2 V}{\partial_vV}}{T_f})\\
  &= T_{[f,g]} + \frac{c\hbar}{48\pi}\int_{\mathbb{R}}\dd{v}(\partial_vf \partial_v^2g - \partial_vg\partial_v^2f)\\
  &\hspace*{3em}- \frac{c\hbar}{48\pi}\int_{\mathbb{R}}\dd{v}\qty(\partial_v f \partial_v\qty(g\partial_v \ln\pa_v V+\partial_vg)-\partial_v g\partial_v\qty(f\frac{\partial_v^2 V}{\partial_vV}+\partial_vf))\\
  &= \mathcal{T}_{[f,g]},
\end{align}
i.e.\ it gives a zero central charge representation of the diffeomorphism algebra. The Poincar\'e-Birkhoff-Witt theorem implies that $U[F]=\exp(\frac{i}{\hbar}\mathcal{T}_f)$ then provides a representation of the diffeomorphism group:
\begin{equation}
  \mathcal{U}[F]\mathcal{U}[G] = \mathcal{U}[F\circ G], \qq{where} \mathcal{U}[\exp(f\partial_v)] = \exp(\frac{i}{\hbar}\mathcal{T}_f).
\end{equation}
Now, we can establish by taking the $s$ derivative that\footnote{
  The equation is clearly satisfied at $s=0$. Denoting the right-hand side by $A(s)$, its $s$ derivative gives
  \begin{align}
    \pa_sA(s)
    &= \frac{i}{\hbar} T_f\exp(\frac{is}{\hbar}T_f)\exp(\frac{i}{\hbar}\int_0^s\dd{s'}\mathcal{F}_f[V\circ F_{-s'}]) + \frac{i }{\hbar} \exp(\frac{is}{\hbar}T_f) \mathcal{F}_f[V\circ F_{-s}] \exp(\int_0^s\dd{s'}\mathcal{F}_f[V\circ F_{-s'}])\nonumber \\
    &= \frac{i}{\hbar}
    \left(T_f + \mathcal{F}_f[V]\right) A(s),
  \end{align}
  which is what the left-hand side $s$ derivative provides.
}
\begin{equation}
  \exp(\frac{i s}{\hbar}\left(T_f + \mathcal{F}_f[V]\right)) = \exp(\frac{i s}{\hbar}T_f) \exp(\frac{i }{\hbar}\int_0^s\dd{s'}\mathcal{F}_f[V\circ F_{-s'}]);
\end{equation}
Taking $s=1$ provides us with the identity
\begin{equation}
  \mathcal{U}[F]=U[F]
  \exp(-\frac{ic}{48\pi}\int_0^1\dd{s}\int_{\mathbb{R}}\dd{v}\partial_vf \pa_v \ln \pa_v(V\circ F_{-s})).
\end{equation}
Using $\partial_s F_s(v)= \exp(f\pa_v) f(v)= (f\circ F_s)(v) $, hence $\partial_s\ln\partial_v F_s=\pa_vf\circ F_s$, we can rewrite the integral in the exponent as
\begin{align}
  \int_{\mathbb{R}}\dd{v}\partial_vf \frac{\partial_v^2(V\circ F_{-s})}{\partial_v(V\circ F_{-s})} &= \int_{\mathbb{R}}\dd{v} \partial_vf\circ F_s \qty(\frac{\partial_v^2V}{\partial_vV}-\frac{\partial_v^2F_s}{\partial_vF_s}) \\
  &=  \pa_s \int_{\mathbb{R}}\dd{v}\ln\partial_v F_s\frac{\partial_v^2V}{\partial_vV} -\int_{\mathbb{R}}\dd{v}\partial_s(\ln\partial_vF_s) \partial_v\ln\partial_v F_s .
\end{align}
The first term above is a representative of the Bott-Thurston cocycle. The second term can be recognized as the  KKS potential contracted along the diffeomorphism  $ F_s$. Altogether, after taking the $s$ integral and using the definition of the Bott-cocycle \eqref{BottT} we have
\begin{equation}
  U[F] = \mathcal{U}[F] \exp(-\frac{ic}{24} \mathrm{B}(F,X))
  \exp(-\frac{ic}{24}b(F)),
\end{equation}
therefore the composition gives
\begin{align}
  U[F]U[G]
  &= \mathcal{U}[F] e^{-\frac{ic}{24} \mathrm{B}(F,X)}
  e^{-\frac{ic}{24}b(F)}\mathcal{U}[G] e^{-\frac{ic}{24} \mathrm{B}(G,X)}
  e^{-\frac{ic}{24}b(G)}\\
  &= \mathcal{U}[F]\mathcal{U}[G]
  e^{-\frac{ic}{24} \left[\mathrm{B}(F,G X)+ \mathrm{B}(G,X)\right]}
  e^{-\frac{ic}{24}[b(F)+b(G)]}\\
  &= \mathcal{U}[F\circ G] e^{-\frac{ic}{24} \left[\mathrm{B}(F,G X)+ \mathrm{B}(G,X)\right]}
  e^{-\frac{ic}{24}[b(F)+b(G)]}\\
  &= U[F\circ G]
  e^{-\frac{ic}{24} \left[\mathrm{B}(F,G X)+ \mathrm{B}(G,X) - \mathrm{B}(FG,X)\right]}
  e^{-\frac{ic}{24}[b(F)+b(G)-b(FG)]}\\
  &= U[F\circ G]
  e^{-\frac{ic}{24} \left[\mathrm{B}(F,G )\right]}
  e^{-\frac{ic}{24}[b(F)+b(G)-b(FG)]}.
\end{align}
This proves~\eqref{Equation: cocycle}.

\subsection{K\"ahler potential}
\label{Appendix: Kahler}

In this section we consider coherent states generated by $U[F^{-1}]$ acting on the vacuum $\ket{0}$ and evaluate the K\"ahler potential of the $\Diff^+(\mathbb{R})$ group. The vacuum is defined to satisfy $P_+T\ket{0}=0$.

Given a diffeomorphism
$F \in \mathrm{Diff}_{\mathrm{hol}}^+(\mathbb{R})$, we can decompose it\footnote{The decomposition exists if $F \in \mathrm{Diff}^+(\mathbb{R})$ extends to a univalent holomorphic map on the band $B_\epsilon=\{z\in \mathbb{C}| \Im(z)<\epsilon\}$, and  $\mathrm{Diff}_{\mathrm{hol}}^+(\mathbb{R})\subset \mathrm{Diff}^+(\mathbb{R})$. }
as a product  $F=F_-^{-1}\circ F_+$ where $F_+$ (resp.\ $F_-$) is a univalent\footnote{This means that $F_+: \mathbb{H}_- \to \mathbb{C}$ is holomorphic and injective on the LHP $\mathbb{H}_-$.} map on the lower half plane $\bar{\mathbb{H}}$ (resp.\ upper half plane $\mathbb{H}$).
As we have seen each diffeomorphism defines a unitarily implemented action on the fields
\begin{equation}
  U[F]\Phi U^\dagger[F] = F\triangleright \Phi
\end{equation}
This action can then split into an action of $F_\pm$ and
the holomorphicity condition implies that
\begin{equation}
  P_- (F_+\triangleright) P_+=0, \qquad P_+ (F_-\triangleright )P_-=0.
\end{equation}
The decomposition $F_-^{-1}\circ F_+$ is not unique since we can change $F_\pm \to F_0\circ F_\pm$ while leaving $F$ invariant, for any $F_0$ that is univalent on $\CC$. One can impose some extra conditions to fix this ambiguity and thus uniquely determine $F_\pm$.

One can define an `action' $S(F)$ on the diffeomorphism group by taking the expectation value of $U[F^{-1}]$ in the vacuum:
\begin{equation}
  \exp(-\frac{ic}{24}S(F)) = \expval{U[F^{-1}]}.
\end{equation}
The action can be evaluated in terms of the cocycle by using that the unitary representation of maps univalent on $\mathbb{H}$ fix the vacuum (since such unitaries are generated by $P_+T$):
\begin{equation}
  \expval{U[F]} = \expval{U[F_+^{-1}] U[F_-]}\exp(\frac{ic}{24}C(F_+^{-1},F_-)) = \exp(\frac{ic}{24}C( F_+^{-1}, F_-)),
\end{equation}
so
\begin{equation}
  S(F)= C(F_-^{-1}, F_+)= - C(F_+^{-1},F_-).
  \label{Equation: Vir action}
\end{equation}
The unitaries satisfy $U[F]^{\dagger}= U[\bar{F}^{-1}]$, where $\bar{F}$ is the complex conjugate. Note that $\bar{F}^{-1}=\bar{F}_+^{-1}\circ \bar F_-$ where $\bar{F}_-$ (resp.\ $\bar{F}_+$) is univalent in $\bar{\mathbb{H}}$ (resp.\ $\mathbb{H}$), so one has the reality condition
\begin{equation}
  \overline{S(F)}= -S(\bar{F}^{-1})= C (\bar{F}_-^{-1},\bar{F}_+).
  \label{Equation: Vir action conjugate}
\end{equation}
Note if $F=F_-^{-1}\circ F_+$ is real then
\begin{equation}
  F_-^{-1}\circ F_+ = \bar{F}_-^{-1}\circ\bar{F}_+ \implies \bar{F}_+\circ F_+^{-1} = \bar{F}_-\circ F_-^{-1}.
  \label{Equation: real F pm}
\end{equation}
We can use this to analytically continue $F_-$ as a function of $F_+,\bar{F}_+$ to complex $F$ as well.\footnote{Concretely,
  \begin{equation}
    (F_+\circ \bar F_+^{-1})_+ = \bar{F}_-^{-1}, \qquad (F_+\circ \bar F_+^{-1})_- = F_-^{-1}.
\end{equation}}

Using the decomposition
$F=F_-^{-1}\circ F_+$, the coherent states and dual coherent states are defined as
\begin{equation}
  \ket{F} = U[F_+^{-1}]\ket{0},\qquad
  \langle\bar{F}|= \bra{0}U[\bar{F}_+],
\end{equation}
where we have used $U[F_-]\ket{0}=\ket{0}$. These states are not normalized, but are holomorphic in $F_+$. The K\"ahler potential $K(F)$ is defined via the norm of the coherent states:
\begin{equation}
  \exp(-\frac{c}{24}K(F)) =\langle\bar F| F\rangle.
\end{equation}
$K$ can be expressed in terms of the cocycle as\footnote{The first equality follows from
  \begin{align}
    \bra*{\bar F}\ket{F} &= \expval*{U[\bar F_+]U[F_+^{-1}]} \\
    &= \expval*{U[\bar F_+\circ F_+^{-1}]} \exp(-\frac{ic}{24}C(\bar F_+, F_+^{-1}))\\
    &= \expval*{U[\bar F_-\circ F_-^{-1}]} \exp(-\frac{ic}{24}C(\bar F_+, F_+^{-1}))\\
    &= \expval*{U[\bar F_-]U[F_-^{-1}]} \exp(\frac{ic}{24}C(\bar F_-, F_-^{-1}))\exp(-\frac{ic}{24}C(\bar F_+, F_+^{-1}))\\
    &= \exp(\frac{ic}{24}\qty(C(\bar F_-, F_-^{-1})-C(\bar F_+, F_+^{-1})))
\end{align}}
\begin{align}
  K(F,\bar F) & = -i\left(C(\bar{F}_-,F_-^{-1})-C(\bar{F}_+,F_+^{-1})\right)\label{Equation: Kahler cocyle}\\
  &= -i\qty(B(\bar F_-,F_-^{-1})-B(\bar F_+,F_+^{-1}))\\
  &= -\frac{i}{2\pi}\int_{\mathbb{R}}\dd{v}\qty(\ln\partial_v\bar F_+\partial_v\ln\partial_v F_+ - \ln\partial_v \bar F_-\partial_v\ln\partial_v F_-).
\end{align}
where we used, in the second equality, that
$b(F_\pm^{-1})=b(\bar{F}_\pm)=0$ (since the pullback of $\Theta$ to purely holomorphic or antiholomorphic diffeomorphisms vanishes).
The fact that $K(F)$ is real and positive  follows from  analytical extension.  We denote $z= v+i y$ and use that $F_+$ can be analytically be continued into the lower half plane $\bar{\mathbb{H}}$ where $y\leq 0$, while $F_-$ can be analytically be continued into the upper half plane $\mathbb{H}$ where $y\geq 0$.
We denote the complex integration measure by $\rd^2 z:= -i \rd z\wedge \rd \bar{z} =  2 \rd y \wedge \rd v$ and therefore, we have
\begin{equation}
  K(F,\bar F) = \frac{1}{2\pi}\qty(\int_{\bar{\mathbb{H}}}\dd[2]{z}\abs{\frac{\partial_z^2F_+}{\partial_zF_+}}^2+\int_{\mathbb{H}}\dd[2]{z}\abs{\frac{\partial_z^2F_-}{\partial_zF_-}}^2).
\end{equation}
This matches the expression of the  Teo-Takhtajan energy \cite{Takhtajan:2003hm, Alekseev:2022efp}. Our derivation is much more direct than the original one.\footnote{Our boundary conditions assume that the poles at $\pm i\infty$ vanish.} It is clear from this expression that  $K$ is a \emph{positive} quantity.\footnote{To check the validity of the overall sign, let $\dd[2]{z}=-i\dd{z}\wedge\dd{\bar z}$, and note that $\int_C\dd{\alpha}=\int_{\partial C}\alpha$ with $\partial C$ oriented anti-clockwise around $C$. Let $\eta_\pm=\ln\partial_vF_\pm$. By construction, $\eta_+$ is holomorphic in $\bar{\mathbb{H}}$, and therefore
  \begin{align}
    \int_{\bar{\mathbb{H}}}\dd[2]{z}\abs{\partial_z\eta_+}^2 &= -i\int_{\bar{\mathbb{H}}}\dd{z}\partial_z\eta_+\wedge\dd{\bar z}\partial_{\bar z}\bar\eta_+ = i\int_{\bar{\mathbb{H}}}\dd(\bar\eta_+\dd{\eta_+})= -i\int_{\RR} \bar\eta_+\dd{\eta_+} = -i\int_{\RR}\dd{v}\bar\eta_+\partial_v\eta_+.
  \end{align}
  Similarly  $\eta_-$ is holomorphic in $\mathbb{H}$ and:
  \begin{align}
    \int_{{\mathbb{H}}}\dd[2]{z}\abs{\partial_z\eta_-}^2
    &= i\int_{{\mathbb{H}}}\dd(\bar\eta_-\dd{\eta_-})
    = i\int_{\RR}\dd{v}\bar \eta_-\partial_v\eta_-.
  \end{align}
  Therefore
  \begin{equation}
    \int_{\bar{\mathbb{H}}}\dd[2]{z}\abs{\partial_z\eta_+}^2 + \int_{\mathbb{H}}\dd[2]{z}\abs{\partial_z\eta_-}^2 = -i\int\dd{v}\qty(\bar\eta_+\partial_v\eta_+-\bar\eta_-\partial_v\eta_-),
  \end{equation}
  as required.
}

It is interesting to note that we have the remarkable relation between $S$ and $K$ given by
\begin{equation} \label{KS}
  \Im(S(F)) =  K(F).
\end{equation}
The proof of this relationship follows from the evaluation
\begin{align}
  \expval{U[F]} &= \expval{U[F_-^{-1}]U[F_+]}\exp(\frac{ic}{24} S(F))\\
  &= \expval{U[\bar{F}_-]U[F_-^{-1}]U[F_+]U[\bar{F}_+^{-1}]}\exp(\frac{ic}{24} S(F))\\
  &= \expval{U[\bar{F}_-\circ F_-^{-1}]U[F_+\circ \bar{F}_+^{-1}]}\exp(-\frac{ic}{24} \left(C(\bar{F}_-,F_-^{-1}) - C(F_+,\bar{F}_+^{-1})\right) ) \exp(\frac{ic}{24} S(F))\\
  &= \exp(- \frac{c}{24} K(F)  )\exp(\frac{ic}{24} S(F)).
\end{align}
In the first equality we use \eqref{Equation: Vir action}, in the second we use that  $U[\bar{F}_+^{-1}] \ket{0}=\ket{0}$, in the third that $(\bar F_-\circ F_-^{-1})^{-1}=F_+\circ \bar F_+^{-1}$, and in the last equality we use the skew-symmetry property of $C$ and the definition of the K\"ahler potential. From this we have that
\begin{equation}
  \expval{U[F]}
  \overline{\expval{U[F]}}= \exp(-\frac{c}{12}\Im(S(F))) = \exp( - \frac{c}{12} K(F) ),
\end{equation}
which implies \eqref{KS}.

\section{Covariant normal ordering}
\label{Appendix: more on covariant normal ordering}

Let us now provide a general construction of the covariant normal ordering for observables which are polynomial of arbitrary powers in $\Pi$. The main point of this section is just to demonstrate that such a construction exists and is well-defined. We give a summary of the main ideas in Section~\ref{Subsection: partial covariant normal ordering}. Afterwards we provide full technical details, but these details may be skipped on a first reading.

For simplicity, we  start in the case $c_{\text{cl}}=0$ in Section~\ref{Subsection: symbols and covnormord definition}, where we define covariant normal ordering with the use of `symbols'. Then we explain what changes with a non-zero $c_{\text{cl}}$ in Section~\ref{Subsection: anomaly covnormord}. In Section~\ref{Subsection: generating functions} we provide an alternative construction of covariantly normal ordered observables based on generating functions.

\subsection{Definitions and results}
\label{Subsection: partial covariant normal ordering}

The task of covariant normal ordering is to convert a function $O[\Pi,V,\varphi]$ on the classical phase space $\mathcal{P}$  to an operator $\Cov{O}= \covnormord{O[\Pi,V,\varphi]}$ acting on the kinematical space of quantum states $\mathcal{K}_{\text{kin}}$ which satisfies the covariance property
$U[F]\Cov{O}U[F]^\dagger= \covnormord{O[F\triangleright\Pi,F\triangleright V,F\triangleright \varphi]}$.

It is convenient to introduce an intermediate step in this construction, the purpose of which is to separate out the relatively simple stage of covariant normal ordering of the radiative fields from the more involved stage of covariant normal ordering of the spin 0 fields:
\begin{itemize}
  \item First, we convert $O:\mathcal{P}\to\mathbb{C}$ to a function $\mathbf{O}:\mathcal{P}_0\to\mathcal{B}(\mathcal{H}^\text{rad})$ that goes from a point in the spin 0 phase space $\mathcal{P}_0$ (consisting of configurations of $\Pi,V$) to an operator acting on the radiative Hilbert space $\mathcal{H}^\text{rad}$. We refer to $\mathbf{O}[\Pi,V]$ as the `partial covariant normal ordering' of $O$. It is defined as the quantization of the radiative degrees of freedom using normal ordering of the quantum radiative fields $\varphi_i$ with respect to the dressing time:
    \begin{equation}
      \mathbf{O}[\Pi,V] = \normord{O[\Pi,V, \varphi_i]}^{\rad}_V.
    \end{equation}
    On the right-hand side of this equation the normal ordering quantization is applied only to the radiative fields, i.e.\ it is defined with $P_+^V\hat\varphi_i$ always appearing to the right of $P_-^V\hat\varphi_i$. On both sides, $\Pi$ and $V$ are classical field configurations.
  \item Then, we convert $\mathbf{O}[\Pi,V]$ to the operator $\Cov{O} = \covnormord{O[
    \Pi,V,\varphi_i]}$ acting on $\mathcal{K}_{\text{kin}}$. It is convenient to use the notation
    \begin{equation}
      \covnormord{\mathbf{O}[\Pi,V]} := \covnormord{O[\Pi,V,\varphi_i]}.
    \end{equation}
    On the left-hand side the covariant normal ordering is understood as only being carried out on $\Pi,V$, while on the right-hand side it is understood as being carried out on all of $\Pi,V,\varphi$.
\end{itemize}

If we do a reparametrization on the classical observable it changes to $O[F\triangleright \Pi,F\triangleright V,F\triangleright\varphi]$, and so $\mathbf{O}$ changes to
\begin{equation}
  \mathbf{O}[\Pi,V] \to F\triangleright \mathbf{O}[\Pi,V] := \normord{O[F\triangleright\Pi,F\triangleright V,F\triangleright\varphi]}^{\rad}_V = U_{\rad}[F]\mathbf{O}[F\triangleright \Pi,F\triangleright V]U_{\rad}[F]^\dagger.
\end{equation}
So the partially covariantly normal ordered operator changes according to the combination of a classical spin 0 reparametrization $(\Pi,V)\mapsto (F\triangleright\Pi,F\triangleright V)$ with a quantum reparametrization on the system implemented by conjugation by $U_{\rad}[F]$. For the second step in covariant normal ordering, from $\mathbf{O}[\Pi,V]$ to $\Cov{O}$, to be a genuinely covariant quantization prescription, therefore we require
\begin{equation}
  U[F]\covnormord{\mathbf{O}[\Pi,V]}U[F]^\dagger = \covnormord{F\triangleright\mathbf{O}[\Pi,V]} = \covnormord{U_{\rad}[F]\mathbf{O}[F\triangleright \Pi,F\triangleright V]U_{\rad}[F]^\dagger}
\end{equation}
It is worth noting that the construction of the covariant normal order on the spin-$0$ variables is agnostic of the exact nature of the radiative `system' degrees of freedom. The construction from this point forward thus highlights the special role played by the spin 0 degrees of freedom as a distinguished reference frame. For concreteness in this paper the system has consisted of $M$ free massless scalar fields $\varphi_i$ -- but all that is required now is that the system has a Hilbert space carrying a projective representation of reparametrizations. The system does not even need to have a good classical limit. In this way, the following construction generalizes to other kinds of radiative degrees of freedom, including strongly interacting ones (although one must still assume that the interactions with the spin 0 fields are negligible).

Let us summarize the main results obtained in the rest of the section: we have that
\begin{equation}
  \covnormord{\mathbf{O}[\Pi,V]} = \normord{\mathcal{D}\mathbf{O}[\Pi,V]}^0.
  \label{covnormord0}
\end{equation}
where $\normord{O[\Phi]}^0$ is the normal ordering quantization applied only to the spin-$0$ fields $(\Pi,V)$ and $\mathcal{D}$ is the operator
\begin{equation}
  \mathcal{D} := \exp(\frac\hbar{2\pi}\int\dd{u}\dd{v}\partial_vG_{\rec V}(v,u)\fdv{\Pi(v)}\accentset{\rightarrow}{\fdv{V(u)}}),
  \label{Equation: spin 0 covnormord operator}
\end{equation}
where $\rec V$ is defined recursively as:
\begin{equation}
  \rec V(u) := V(u) + \frac\hbar{2\pi}\int\dd{v}\partial_vG_{\rec V}(v,u)\fdv{\Pi(v)}.
  \label{Equation: rec V}
\end{equation}
and the notation $\accentset{\rightarrow}{\fdv{V(u)}}$ means that this functional derivative is moved all the way to the right in the exponential, without acting on $V$ where it appears in powers of the exponent.\footnote{Meaning we have an expansion of the form $\exp(K[V]\cdot \accentset{\rightarrow}{\fdv{V}}) = \sum_n\frac1{n!}K[V]^n\qty\big(\fdv{V})^n$.}
The operator $\mathcal{D}$ is the formal inverse\footnote{In the sense of a formal power series expansion in $\hbar$.} of the operator
\begin{equation}
  \mathcal{C} := \exp(-\frac\hbar{2\pi}\int\dd{u}\dd{v}\partial_vG_V(v,u)\fdv{\Pi(v)}\accentset{\rightarrow}{\fdv{V(u)}}).
  \label{Equation: normal ordering to symbol0}
\end{equation}
In the presence of a classical central charge $c_{\text{cl}}$ the covariant normal ordered product is deformed into
\begin{equation}\label{covnormordc}
  \covnormord{\mathbf{O}[\Pi,V]}=\normord{\mathcal{D}_{c_{\text{cl}}}\mathbf{O}[\Pi,V]}^0,
\end{equation}
where $\mathcal{D}_{c_{\text{cl}}}$ is given by
\begin{align}
  \mathcal{D}_{c_{\text{cl}}} &= \exp(-\frac{c_{\text{cl}}\hbar}{48\pi}\int\dd{v}\frac{\partial_v^2\rec V}{\partial_v\rec V}\qty(\frac1{i\hbar}\ln(\tfrac{\partial_v(V+i\hbar P_+\fdv{\Pi})}{\partial_v(V-i\hbar P_-\fdv{\Pi})})-\tfrac{\partial_v\fdv{\Pi}}{\partial_vV}))\mathcal{D} \\
  &= \exp\Bigg(
    \frac\hbar{2\pi}\int_{\RR^2}\dd{u}\dd{w}{\textstyle G_{\rec V}(u,w)\fdv{\Pi(u)}\accentset{\rightarrow}{\fdv{V(w)}}}
    -\frac{c_{\text{cl}}\hbar}{48\pi}\int\dd{v}{\textstyle\frac{\partial_v^2\rec V}{\partial_v\rec V}\qty(\frac1{i\hbar}\ln(\tfrac{\partial_v(V+i\hbar P_+\fdv{\Pi})}{\partial_v(V-i\hbar P_-\fdv{\Pi})})-
        \tfrac{\partial_v\fdv{\Pi}}{\partial_vV}
  )}\Bigg).\label{gentau}
\end{align}
When applied to gauge-invariant operators, covariant normal ordering can be expressed in a simple manner in terms of a generating functional.
One finds, quite remarkably, that
\begin{align}
  \covnormord{\exp(\frac{i}{\hbar}\int\dd{\mv} \qty(\tilde\tau(G_--G_+)))\tilde{O}_S[\tilde \varphi_i]}
  = \exp(-\frac{i}{\hbar}\int\dd{\mv}\Cov{\tilde\tau} g_+)
  \Cov{\tilde{O}}_S
  \exp(\frac{i}{\hbar}\int\dd{\mv}\Cov{\tilde\tau} g_-),
  \label{Equation: covnormord gauge invariant}
\end{align}
Here $g_\pm$ satisfy $P_\mp g_\pm=0$, and we set $G_\pm=\exp(g_\pm\partial_v)$.

The rest of this section is devoted to the proof of the statements (\ref{covnormord0}, \ref{covnormordc}, \ref{Equation: covnormord gauge invariant}). It is a technical remainder that  can be skipped at first reading.

\subsection{Symbols and spin 0 covariant normal ordering}
\label{Subsection: symbols and covnormord definition}

Suppose $\hat{O}\in\mathcal{B}(\mathcal{K}_{\text{kin}})$ is any operator acting on the kinematical degrees of freedom. We define the \emph{symbol} of $\hat{O}$ to be the functional $\mathbf{O}[P,F]$ of a function $P$ and diffeomorphism $F$ given by
\begin{equation}
  \mathbf{O}[P,F] = \expval{U_R[F]Y[P]\hat{O}Y[P]^\dagger U_R[F]^\dagger}_0, \qq{where} Y[P] = \exp(\frac1{i\hbar}\int\dd{v}P V),
\end{equation}
with the expectation value taken in the spin 0 vacuum, so that $\mathbf{O}[P,F]$ is an operator acting on $\mathcal{H}_S$. Hence, $\mathbf{O}$ is a functional from the spin 0 phase space $\mathcal{P}_R$ to $\mathcal{B}(\mathcal{H}_S)$.

Suppose we write $\hat O=\normord{\mathbf{O}'[\Pi,V]}^0$, where $\mathbf{O}'$ is some other functional from $\mathcal{P}_R$ to $\mathcal{B}(\mathcal{H}_R)$, and the normal ordering here is done only on the spin-$0$ fields $\Pi,V$. Then one may use $Y[P]\normord{\mathbf{O}'[\Pi,V]}^0Y[P]^\dagger = \normord{\mathbf{O}'[P+\Pi,V]}^0$ and~\eqref{Equation: normal ordering anomaly} (restricted to the spin 0 fields) to show that the symbol is
\begin{equation}
  \mathbf{O}[P,F] = \mathcal{C}\mathbf{O}'[P,F], \qq{where} \mathcal{C} = \exp(-\frac\hbar{2\pi}\int\dd{u}\dd{v}\partial_vG_F(v,u)\accentset{\rightarrow}{\fdv{F(u)}}\fdv{P(v)}).
  \label{Equation: normal ordering to symbol}
\end{equation}
The notation $\accentset{\rightarrow}{\fdv{F(u)}}$ means that this functional derivative is moved all the way to the right in the exponential, without acting on the powers of $\partial_vG_F(v,u)$. From~\eqref{Equation: normal ordering to symbol} it is clear that an operator that is polynomial in $\Pi$ of degree $n$ has a symbol that is polynomial in $P$ of degree $n$.

It is straightforward to show that $\mathcal{F}[F]$ is the symbol of $\covnormord{\mathcal{F}[V]}$, and $P(v)\mathcal{F}[F]$ is the symbol of $\covnormord{\Pi(v)\mathcal{F}[V]}$, using the definitions~\eqref{Equation: covariant normal ordered F[V]} and~\eqref{Equation: linear covariant normal ordering}. In this way, one may recover the original classical observable as the symbol of the corresponding covariantly normal ordered operator (when the observable is at most linear in $\Pi$).

We directly extend this by \emph{defining} the covariant normal ordering of any functional $\mathbf{O}:\mathcal{P}_R\to\mathcal{B}(\mathcal{H}_S)$ such that $\mathbf{O}[\Pi,V]$ is polynomial in $\Pi$ (and depends arbitrarily on $V$) to be the operator acting on $\mathcal{K}_{\text{kin}}$ whose symbol gives back the original functional:
\begin{equation}
  \mathbf{O}[P,F] = \expval{U_R[F]Y[P]\covnormord{\mathbf{O}[\Pi,V]}Y[P]^\dagger U_R[F]^\dagger}_0.
  \label{Equation: covariant normal ordering via symbol}
\end{equation}
We show below that this is a good definition, in the sense that there always exists a unique such $\covnormord{\mathbf{O}[\Pi,V]}$. This immediately implies covariance of the prescription, since for any diffeomorphism $G$ we have
\begin{equation}
  U_{\rad}[G]\mathbf{O}[G\triangleright P,G\triangleright F]U_{\rad}[G]^\dagger = \expval{U_R[F]Y[P]\covnormord{U_{\rad}[G]\mathbf{O}[G\triangleright \Pi,G\triangleright V]U_{\rad}[G]^\dagger}Y[P]^\dagger U_R[F]^\dagger}_0,
\end{equation}
but also
\begin{align}
  \MoveEqLeft U_{\rad}[G]\mathbf{O}[G\triangleright P,G\triangleright F]U_{\rad}[G]^\dagger\\
  &= U_{\rad}[G]\expval{U_R[F\circ G]Y[G\triangleright P]\covnormord{\mathbf{O}[\Pi,V]}Y[G\triangleright P]^\dagger U_R[F\circ G]^\dagger}_0U_{\rad}[G]^\dagger \\
  &= \expval{U_R[F]Y[P]U[G]\covnormord{\mathbf{O}[\Pi,V]}U[G]^\dagger Y[P]^\dagger U_R[F\circ G]^\dagger}_0,
\end{align}
where we have used
\begin{equation}
  Y[G\triangleright P] = U_R[G]^\dagger Y[P] U_R[G].
\end{equation}
The above demonstrates that
\begin{equation}
  \covnormord{U_{\rad}[G]\mathbf{O}[G\triangleright \Pi,G\triangleright V]U_{\rad}[G]^\dagger}
  =
  U[G]\covnormord{\mathbf{O}[\Pi,V]}U[G]^\dagger,
  \label{Equation: covnormord diffeo covariance}
\end{equation}
since the operators on the two sides have the same symbol. This is the required covariance. By a similar argument, one has
\begin{equation}
  \covnormord{\mathbf{O}[\Pi+P,V]}
  =
  Y[P]\covnormord{\mathbf{O}[\Pi,V]}Y[P]^\dagger.
  \label{Equation: covnormord Pi shift covariance}
\end{equation}
Let's now prove that~\eqref{Equation: covariant normal ordering via symbol} is a good definition of covariant normal ordering (for observables polynomial in $\Pi$). This is equivalent to the invertibility of the map $\mathcal{C}$, since the uniqueness of (ordinary) normal ordering then implies that
\begin{equation}
  \covnormord{\mathbf{O}[\Pi,V]} = \normord{\mathcal{D}\mathbf{O}[\Pi,V]},
  \label{Equation: covnormord with C inverse}
\end{equation}
where $\mathcal{D}=\mathcal{C}^{-1}$,
is the only solution to~\eqref{Equation: covariant normal ordering via symbol}. To demonstrate that $\mathcal{C}$ is actually invertible, it suffices to note that the leading term (i.e.\ highest order in $P$) in $O[P,F]$ is the same as that in $O'[P,F]$, in~\eqref{Equation: normal ordering to symbol}. Therefore, the kernel of $\mathcal{C}$ is trivial, and so $\mathcal{D}$ exists.

To order $\order{\hbar}$, one has
\begin{equation}
  \mathcal{C} = 1 - \frac\hbar{2\pi}\int\dd{u}\dd{v}\partial_vG_F(v,u)\fdv{F(u)}\fdv{P(v)} + \order{\hbar^2},
\end{equation}
and one can straightforwardly invert this to the same order to obtain
\begin{equation}
  \mathcal{D} = 1 + \frac\hbar{2\pi}\int\dd{u}\dd{v}\partial_vG_F(v,u)\fdv{F(u)}\fdv{P(v)} + \order{\hbar^2},
  \label{Equation: C to order hbar}
\end{equation}
This implies~\eqref{Equation: covnormord to order hbar}.

We can obtain an exact formula for $\mathcal{D}$ to all orders as follows. First, note
\begin{align}
  \mathcal{C}\qty(\mathcal{F}[F]\mathcal{G}[P])=\mathcal{F}\qty[F(u) - \frac\hbar{2\pi}\int\dd{u}\dd{v}\partial_vG_F(v,u)\fdv{P(v)}]\mathcal{G}[P].
\end{align}
Now suppose we act on the functional on the right by changing $F$ to $\rec{F}$ satisfying
\begin{equation}
  F(u) = \rec F(u) - \frac\hbar{2\pi}\int\dd{v}\partial_vG_{\rec F}(v,u)\fdv{P(v)}.
  \label{Equation: F to hat F}
\end{equation}
Then we would recover $\mathcal{F}[F]\mathcal{G}[P]$, so the map implementing $F\to\rec{F}$ is the desired inverse of $\mathcal{C}$. This map can be explicitly written as
\begin{align}
  \mathcal{D} &= \exp(\int\dd{u}\qty(\rec{F}(u)-F(u))\accentset{\rightarrow}{\fdv{F(u)}})\\
  &= \exp(\frac\hbar{2\pi}\int\dd{u}\dd{v}\partial_vG_{\rec F}(v,u)\fdv{P(v)}\accentset{\rightarrow}{\fdv{F(u)}}).
  \label{Equation: C inverse}
\end{align}
Combining this with~\eqref{Equation: covnormord with C inverse} allows one to covariantly normal order any observable:
\begin{equation}
  \covnormord{\mathbf{O}[\Pi,V]} = \normord{\exp(\frac\hbar{2\pi}\int\dd{u}\dd{v}\partial_vG_{\rec V}(v,u)\fdv{\Pi(v)}\accentset{\rightarrow}{\fdv{V(u)}})\mathbf{O}[\Pi,V]}^0,
\end{equation}
where $\rec V$ is defined~\eqref{Equation: rec V} similarly to $\rec F$. This proves~\eqref{covnormord0} (after doing the change of notation $(F,P) \to (V,\Pi)$).

Note that $\rec{V}$ can be obtained as a power series in $\hbar$ using~\eqref{Equation: rec V}.
To check~\eqref{Equation: C to order hbar}, it suffices to use $\rec V = V + \order{\hbar}$. To the next order we have
\begin{equation}
  \rec V(u) = V(u) + \frac\hbar{2\pi}\int\dd{v}\partial_vG_{V}(v,u)\fdv{\Pi(v)} + \order{\hbar^2},
\end{equation}
to the next we have (recalling $G_V(v,u) = \ln\qty(\frac{V(v)-V(u)}{v-u})$)
\begin{multline}
  \rec V(u) = V(u)
  + \frac\hbar{2\pi}\int\dd{v}\partial_v\ln\Bigg(\frac{V(v)-V(u)}{v-u} \\
  + \frac1{v-u}\frac\hbar{2\pi}\int\dd{w}\partial_w\ln\qty(\frac{V(w)-V(v)}{w-v}\frac{w-u}{V(w)-V(u)})\fdv{\Pi(w)}\Bigg)\fdv{\Pi(v)} + \order{\hbar^3},
\end{multline}
and so on.

\subsection{Including the classical anomaly}
\label{Subsection: anomaly covnormord}

Now let's see what changes when we turn on a non-zero $c_{\text{cl}}$. In this case, $\Pi$ has an anomalous transformation law, and it is useful to describe things in terms of
\begin{equation}
  \Pi^c = \Pi - \frac{c_{\text{cl}}\hbar}{48\pi}\partial_v^2\qty((\partial_vV)^{-1}).
\end{equation}
Unlike $\Pi$, the field $\Pi^c$ transforms like a 1-form; indeed one may straightforwardly show
\begin{equation}
  \frac1{i\hbar}\comm{\Pi^c(u)}{T(v)} = \partial_u(\delta(u-v)\Pi^c(u)),
\end{equation}
so that
\begin{align}
  F\triangleright\Pi^c &:= U[F] \Pi^c U[F]^\dagger
  = \partial_v F\Pi^c\circ F.
\end{align}

We define covariant normal ordering with a slight change to before:
\begin{equation}
  \mathbf{O}[P,F] = \expval{U_R[F]Y[P^c]\covnormord{\mathbf{O}[\Pi,V]}Y[P^c]^\dagger U_R[F]^\dagger}_0,
  \label{Equation: covariant normal ordering via symbol, with classical anomaly}
\end{equation}
with
\begin{equation}
  P^c = P - \frac{c_{\text{cl}}\hbar}{48\pi}\partial_v^2\qty((\partial_vF)^{-1})
\end{equation}
the classical functional corresponding to the quantum field $\Pi^c$.
With this definition, covariance follows as before, and the covariant normal ordering for observables at most linear in $\Pi$ is the same as the $c_{\text{cl}}=0$ case, i.e.\ it is given by~\eqref{Equation: covariant normal ordered F[V]} and~\eqref{Equation: linear covariant normal ordering}.

For observables that are more than linear in $\Pi$, covariant normal ordering depends on $c_{\text{cl}}$. To see this, we generalize~\eqref{Equation: normal ordering to symbol}. Consider a normal ordered operator $\normord{\mathbf{O}'[\Pi,V]}^0$. Below we show that
\begin{equation}
  \covnormord{\mathbf{O}[\Pi,V]}=\normord{\mathbf{O}'[\Pi,V]}^0 \quad\iff\quad \mathbf{O}[P,F] = \mathcal{C}_{c_{\text{cl}}}\mathbf{O}'[P,F],
  \label{Equation: Equation: normal ordering to symbol with anomaly}
\end{equation}
where $\mathcal{C}_{c_{\text{cl}}}$ is given in~\eqref{Equation: C c cl}. When the classical anomaly vanishes~\eqref{Equation: Equation: normal ordering to symbol with anomaly} reduces to~\eqref{Equation: normal ordering to symbol}, because $c_{\text{cl}}=0$ implies $\mathcal{C}_{c_{\text{cl}}}=\mathcal{C}$. But for $c_{\text{cl}}$ non-zero, $\mathcal{C}_{c_{\text{cl}}}$ contains corrections involving quadratic and higher order functional derivatives in $P$. As previously, $\mathcal{C}_{c_{\text{cl}}}$ is an invertible map, and so one has the following unique formula for covariant normal ordering:
\begin{equation}
  \covnormord{\mathbf{O}[\Pi,V]}=\normord{\mathcal{D}_{c_{\text{cl}}}\mathbf{O}[\Pi,V]}^0,
\end{equation}
where $\mathcal{D}_{c_{\text{cl}}}=\mathcal{C}_{c_{\text{cl}}}^{-1}$.
With this one can covariantly normal order any observable, even for non-zero $c_{\text{cl}}$.
Below we show that $\mathcal{D}_{c_{\text{cl}}}$ is given by~\eqref{gentau}.

It is worth noting also that
\begin{equation}
  \mathcal{C}_{c_\text{cl}} = \mathcal{C} + \order{\hbar^2}, \qquad
  \mathcal{D}_{c_\text{cl}} = \mathcal{D} + \order{\hbar^2},
\end{equation}
which means~\eqref{Equation: covnormord to order hbar} still applies for non-zero $c_{\text{cl}}$.

Let's now show how to derive $\mathcal{C}_{c_\text{cl}}$ and the inverse~\eqref{gentau}. The first step is to note that~\eqref{Equation: normal ordering anomaly} must be modified at non-zero $c_{\text{cl}}$. Consider the operator
\begin{equation}
  \normord{\exp(\int\dd{v}(g\Pi+hV))} = \exp(\int\dd{v}hP_+V)\exp(\int\dd{v}g\Pi)\exp(\int\dd{v}P_-V).
\end{equation}
Under a reparametrization this transforms to
\begin{multline}
  U[F]\normord{\exp(\int\dd{v}(g\Pi+hV))}U[F]^\dagger \\\
  = \exp(\int\dd{v}hP_+(F\triangleright V))\exp(\int\dd{v}g(F\triangleright\Pi))\exp(\int\dd{v}P_-(F\triangleright V)).
  \label{Equation: normord transform c_cl}
\end{multline}
We would like to write this as a normal-ordered observable according to the time $F^{-1}(v)$. The outer two terms already respect this, but the inside term requires manipulation:
\begin{align}
  \MoveEqLeft\exp(\int\dd{v}g(F\triangleright\Pi)) \\
  &= \exp(\int\dd{v}g\qty(\partial_vF\Pi\circ F - \frac{c_{\text{cl}}\hbar}{48\pi}\partial_v\qty(\frac1{\partial_v(V\circ F)}\frac{\partial_v^2F}{\partial_vF})))\\
  &=\exp(\int\dd{v}g\circ F^{-1}\Pi + \frac{c_{\text{cl}}\hbar}{48\pi}\int\dd{v}\frac{\partial_v^2F}{\partial_vF}\frac{\partial_v g}{\partial_v(V\circ F)})\\
  &=e^{\int\dd{v}g\circ F^{-1}\Pi} \exp(-\frac{ic_{\text{cl}}}{48\pi}\int\dd{v}\frac{\partial_v^2F}{\partial_vF}\ln(1+\frac{i\hbar\partial_vg}{\partial_v(V\circ F)}))\\
  &=e^{\int\dd{v}g_+\circ F^{-1}\Pi} \exp(-\frac{ic_{\text{cl}}}{48\pi}\int\dd{v}\frac{\partial_v^2F}{\partial_vF}\ln(\frac{\partial_v(V\circ F)+i\hbar\partial_vg_+}{\partial_v(V\circ F)-i\hbar\partial_vg_-}))e^{\int\dd{v}g_-\circ F^{-1}\Pi},\label{Equation: Pi normord transform c_cl}
\end{align}
where we have used the general formula
\begin{equation}
  \exp(\int \dd{v} g\Pi + \mathcal{F}[V]) = \exp(\int \dd{v} g\Pi)\exp(\int_0^1\dd{s}\mathcal{F}[V+i\hbar s g]).
\end{equation}
By substituting the last line~\eqref{Equation: Pi normord transform c_cl} into~\eqref{Equation: normord transform c_cl} we obtain an expression with the desired normal ordering with respect to $F(v)$, from which we deduce
\begin{align}
  \MoveEqLeft U[F]\normord{\exp(\int\dd{v}(g\Pi+hV))}U[F]^\dagger \\
  &= \normord{\exp(\int\dd{v}h(F\triangleright V)+(g\circ F^{-1})\Pi-\frac{ic_{\text{cl}}}{48\pi}\int\dd{v}\frac{\partial_v^2F}{\partial_vF}\ln(\frac{\partial_v(V\circ F)+i\hbar\partial_vg_+}{\partial_v(V\circ F)-i\hbar\partial_vg_-}))}_{F^{-1}}\\
  &= \normord{\exp(\frac{c_{\text{cl}}\hbar}{48\pi}\int\dd{v}\frac{\partial_v^2F}{\partial_vF}\qty(\frac1{i\hbar}\ln(\tfrac{\partial_v(V\circ F)+i\hbar\partial_vg_+}{\partial_v(V\circ F)-i\hbar\partial_vg_-})-\frac{\partial_vg}{\partial_v(V\circ F)}))e^{\int\dd{v}h(F\triangleright V)+g(F\triangleright \Pi)}}_{F^{-1}}.
\end{align}
By linearity, we can extend this to general functionals:
\begin{align}
  \MoveEqLeft U[F]\normord{\mathbf{O}[\Pi,V]}^0U[F]^\dagger \\
  &= \normord{\exp(\frac{c_{\text{cl}\hbar}}{48\pi}\int\dd{v}\frac{\partial_v^2F}{\partial_vF}\qty(\frac1{i\hbar}\ln(\tfrac{\partial_v(V\circ F)+i\hbar\partial_vP_+(\fdv{\Pi}\circ F)}{\partial_v(V\circ F)-i\hbar\partial_vP_-(\fdv{\Pi}\circ F)})-\tfrac{\partial_v(\fdv{\Pi}\circ F)}{\partial_v(V\circ F)}))\mathbf{O}[F\triangleright\Pi,F\triangleright V]}^0_{F^{-1}}\nonumber\\
  &= \normord{\exp(-\frac{c_{\text{cl}}\hbar}{48\pi}\int\dd{v}\frac{\partial_v^2F^{-1}}{\partial_vF^{-1}}\qty(\frac1{i\hbar}\ln(\frac{\partial_vV+i\hbar\partial_vP^F_+\fdv{\Pi}}{\partial_vV-i\hbar\partial_vP^F_-\fdv{\Pi}})-\frac{\partial_v\fdv\Pi}{\partial_vV}))\mathbf{O}[F\triangleright\Pi,F\triangleright V]}^0_{F^{-1}}
  \label{Equation: normord transform c_cl processed}
\end{align}
Now we can use~\eqref{Equation: change of time in normal ordering} to rewrite this as an observable normal-ordered with respect to the original background time $v$. One finds that $U[F]\normord{\mathbf{O}[\Pi,V]}^0U[F]^\dagger=\normord{\mathbf{O}_F[\Pi,V]}^0$ where
\begin{align}
  \mathbf{O}_F[\Pi,V] &= \exp(\frac\hbar{2\pi}\int_{\RR^2}\dd{u}\dd{w}\partial_uG_F(u,w)\fdv{\Pi(u)}\fdv{V(w)})\\
  &\hspace*{2em}\exp(-\frac{c_{\text{cl}}\hbar}{48\pi}\int\dd{v}\frac{\partial_v^2F^{-1}}{\partial_vF^{-1}}\qty(\frac1{i\hbar}\ln(\frac{\partial_vV+i\hbar\partial_vP^F_+\fdv{\Pi}}{\partial_vV-i\hbar\partial_vP^F_-\fdv{\Pi}})-\frac{\partial_v\fdv\Pi}{\partial_vV}))\mathbf{O}[F\triangleright\Pi,F\triangleright V]\nonumber\\
  &= \exp(-\frac{c_{\text{cl}}\hbar}{48\pi}\int\dd{v}\frac{\partial_v^2F^{-1}}{\partial_vF^{-1}}\qty(\frac1{i\hbar}\ln(\tfrac{\partial_vV+i\hbar\partial_vP_+\fdv{\Pi}}{\partial_vV-i\hbar\partial_vP_-\fdv{\Pi}})-\tfrac{\partial_v\fdv{\Pi}}{\partial_v(V+i\hbar P_+\fdv{\Pi}-i\hbar P_+^F\fdv{\Pi})}))\\
  &\hspace*{2em}\exp(\frac\hbar{2\pi}\int_{\RR^2}\dd{u}\dd{w}\partial_uG_F(u,w)\fdv{\Pi(u)}\fdv{V(w)})\mathbf{O}[F\triangleright\Pi, F\triangleright V].
\end{align}
Using this, the symbol of a normal ordered operator $\normord{\mathbf{O}'[\Pi,V]}^0$ may be found to be
\begin{align}
  \mathbf{O}[P,F]&=\exp(-\frac{c_{\text{cl}}\hbar}{48\pi}\int\dd{v}\frac{\partial_v^2F^{-1}}{\partial_vF^{-1}}\qty(\frac1{i\hbar}\ln(\tfrac{1+i\hbar\partial_vP_+(\fdv{P}\circ F)}{1-i\hbar\partial_vP_-(\fdv{P}\circ F)})-\tfrac{\partial_v(\fdv{P}\circ F)}{1+i\hbar \partial_v(P_+(\fdv{P}\circ F)-(P_+\fdv{P})\circ F)}))\\
  &\hspace*{2em}\exp(-\frac\hbar{2\pi}\int_{\RR^2}\dd{u}\dd{w}\partial_uG_F(u,w)\fdv{P(u)}\accentset{\rightarrow}{\fdv{F(w)}})\mathbf{O}'[P,F]\\
  &=\exp(\frac{c_{\text{cl}}\hbar}{48\pi}\int\dd{v}\frac{\partial_v^2F}{\partial_vF}\qty(\frac1{i\hbar}\ln(\tfrac{\partial_v(F+i\hbar P^F_+\fdv{P})}{\partial_v(F-i\hbar P^F_-\fdv{P})})-\tfrac{\partial_v\fdv{P}}{\partial_v(F+i\hbar P_+^F\fdv{P}-P_+\fdv{P})}))\\
  &\hspace*{2em}\exp(-\frac\hbar{2\pi}\int_{\RR^2}\dd{u}\dd{w}\partial_uG_F(u,w)\fdv{P(u)}\accentset{\rightarrow}{\fdv{F(w)}})\mathbf{O}'[P,F]\\
  &=\exp(-\frac\hbar{2\pi}\int_{\RR^2}\dd{u}\dd{w}\partial_uG_F(u,w)\fdv{P(u)}\accentset{\rightarrow}{\fdv{F(w)}})\\
  &\hspace*{2em}\exp(\frac{c_{\text{cl}}\hbar}{48\pi}\int\dd{v}\frac{\partial_v^2\rec F}{\partial_v\rec F}\qty(\frac1{i\hbar}\ln(\tfrac{\partial_v(F+i\hbar P_+\fdv{P})}{\partial_v(F-i\hbar P_-\fdv{P})})-\tfrac{\partial_v\fdv{P}}{\partial_vF}))\mathbf{O}'[P,F]
\end{align}
Here we have used the defining equation~\eqref{Equation: F to hat F} for $\rec F$. The operator in front of $\mathbf{O}'[P,F]$ is
\begin{equation}
  \mathcal{C}_{c_{\text{cl}}} = \mathcal{C}
  \exp(\frac{c_{\text{cl}}\hbar}{48\pi}\int\dd{v}\frac{\partial_v^2\rec F}{\partial_v\rec F}\qty(\frac1{i\hbar}\ln(\tfrac{\partial_v(F+i\hbar P_+\fdv{P})}{\partial_v(F-i\hbar P_-\fdv{P})})-\tfrac{\partial_v\fdv{P}}{\partial_vF})),
  \label{Equation: C c cl}
\end{equation}
which has inverse
\begin{align}
  \mathcal{D}_{c_{\text{cl}}} &= \exp(-\frac{c_{\text{cl}}\hbar}{48\pi}\int\dd{v}\frac{\partial_v^2\rec F}{\partial_v\rec F}\qty(\frac1{i\hbar}\ln(\tfrac{\partial_v(F+i\hbar P_+\fdv{P})}{\partial_v(F-i\hbar P_-\fdv{P})})-\tfrac{\partial_v\fdv{P}}{\partial_vF}))\mathcal{D} \\
  &= \exp(-\frac{c_{\text{cl}}\hbar}{48\pi}\int\dd{v}\frac{\partial_v^2\rec F}{\partial_v\rec F}\qty(\frac1{i\hbar}\ln(\tfrac{\partial_v(F+i\hbar P_+\fdv{P})}{\partial_v(F-i\hbar P_-\fdv{P})})-\tfrac{\partial_v\fdv{P}}{\partial_vF}))\\
  &\hspace*{9em}\exp(\frac\hbar{2\pi}\int_{\RR^2}\dd{u}\dd{w}\partial_uG_{\rec F}(u,w)\fdv{P(u)}\accentset{\rightarrow}{\fdv{F(w)}})
  ,
\end{align}
Since all $F$ functional derivatives are on the right, the exponentials can now be directly combined to obtain~\eqref{gentau} (after changing notation $(F,P) \to (V,\Pi)$).

\subsection{Generating function approach}
\label{Subsection: generating functions}

Here is another way to compute covariant normal ordering. One has
\begin{multline}
  \covnormord{\exp(\frac{i}{\hbar}\int\dd{\mv} \qty(\tilde\tau(G_--G_+)+hX))O_S} \\
  = \exp(-\frac{i}{\hbar}\int\dd{\mv}\Cov{\tilde\tau} g_+) \exp(\frac{i}{\hbar}\int\dd{\mv} h X)O_S \exp(\frac{i}{\hbar}\int\dd{\mv}\Cov{\tilde\tau} g_-),
  \label{Equation: generating function}
\end{multline}
where $h,g_\pm$ are functions satisfying $P_\mp g_\pm=0$, and $G_\pm=\exp(g_\pm\partial_\mv)$, and $O_S$ is a system operator.
Here we consider for simplicity the covariant normal ordering of a symbol $\mathcal{P}_R\to \mathcal{B}(\mathcal{H}_S)$. We can convert this into a statement about covariant normal ordering of an observable $\mathcal{P}\to \mathbb{C}$ by partially covariantly normal ordering $O_S=\exp(\frac{i}\hbar\int\dd{v}\sum_i f_i\varphi_i)$. One finds that the above is then equivalent to
\begin{multline}
  \covnormord{\exp(\frac{i}{\hbar}\qty(\int\dd{\mv} \qty(\tilde\tau(G_--G_+)+hX)+\int\dd{v}\sum_i f_i\varphi_i))} \\
  = \exp(-\frac{i}{\hbar}\int\dd{\mv}\Cov{\tilde\tau} g_+) \exp(\frac{i}{\hbar}\int\dd{\mv} h X)\normord{\exp(\frac{i}{\hbar}\int\dd{v}\sum_i f_i\varphi_i)}_V\exp(\frac{i}{\hbar}\int\dd{\mv}\Cov{\tilde\tau} g_-),
  \label{Equation: generating function with system}
\end{multline}
By taking functional derivatives of~\eqref{Equation: generating function with system}, we can compute the covariant normal ordering of arbitrary functionals of $\tilde\tau,X,\varphi$, or equivalently arbitrary functionals of $\Pi,V,\varphi$. Restricting to gauge-invariant observables gives~\eqref{Equation: covnormord gauge invariant}.

To show~\eqref{Equation: generating function}, first note that $\Cov{\tilde\tau}$ commutes with reparametrizations $U[F]$, since $\tilde\tau$ is a classically gauge-invariant operator:
\begin{equation}
  U[F]\Cov{\tilde\tau}U[F]^\dagger = \covnormord{F\triangleright\tilde\tau} = \Cov{\tilde\tau}.
\end{equation}
Also, the positive frequency part (in the background time $v$) of $\Cov{\tilde\tau}$ annihilates the vacuum; indeed, for any function $f$ we have
\begin{align}
  \MoveEqLeft\int\dd{\mv}\normord{\qty(\partial_v X\Pi\circ X+\frac{(c_{\text{cl}}-12)\hbar}{48\pi}\partial_\mv\qty(\frac{\partial_\mv^2X}{\partial_\mv X}))(P_-g)}\ket{0}\\
  &= \int\dd{v}\normord{\qty(\Pi-\frac{(c_{\text{cl}}-12)\hbar}{4\pi}\partial_v\qty(\frac{\partial_v^2V}{(\partial_vV)^2}))(P_-g)\circ V}\ket{0}\\
  &= \int\dd{v}\qty[P_-\Pi-\frac{(c_{\text{cl}}-12)\hbar}{4\pi}\partial_v\qty(\frac{\partial_v^2P_-\Vv}{(1+\partial_vP_-\Vv)^2})]P_-g(v+P_-\Vv(v))\ket{0}.
\end{align}
All the terms on the last line only contain negative frequencies, so the integral vanishes, and we can conclude $\int\dd{\mv}\Cov{\tilde\tau}P_-g$ annihilates $\ket{0}$ for all $g$, or equivalently $P_+\Cov{\tilde\tau}\ket{0}=0$. Similarly, $\bra{0}P_-\Cov{\tilde\tau}=0$. Finally, as discussed in Section~\ref{Subsection: reorientation algebra}, $\Cov{\tilde\tau}$ generates a representation of diffeomorphisms. In particular, one has
\begin{equation}
  \comm{\int\dd{v}\Cov{\tilde\tau}g}{V(u)} = -i\hbar g(V(u)),
\end{equation}
so that
\begin{equation}
  \exp(\frac{i}{\hbar}\int\dd{v}\Cov{\tilde\tau}f)V\exp(-\frac{i}{\hbar}\int\dd{v}\Cov{\tilde\tau}g) = G\circ V,
\end{equation}
where $G=\exp(g\partial_\mv)$. This implies furthermore that
\begin{equation}
  \exp(\frac{i}{\hbar}\int\dd{v}\Cov{\tilde\tau}g)Y[P]\exp(-\frac{i}{\hbar}\int\dd{v}\Cov{\tilde\tau}g) = \exp(\frac1{i\hbar}\int\dd{v}PG(V))
\end{equation}

Let's now show~\eqref{Equation: generating function} by computing the symbol of the right-hand side. With the above observations, we have
\begin{align}
  \MoveEqLeft\expval{U[F]Y[P]\exp(-\frac{i}{\hbar}\int\dd{\mv}\Cov{\tilde\tau} g_+) \exp(\frac{i}{\hbar}\int\dd{\mv} h X) O_S\exp(\frac{i}{\hbar}\int\dd{\mv}\Cov{\tilde\tau} g_-)Y[P]^\dagger U[F]^\dagger}\\
  &= \expval{U[F]\exp(\frac1{i\hbar}\int\dd{v}PG_+(V))\exp(\frac{i}{\hbar}\int\dd{\mv} h X)O_S\exp(-\frac1{i\hbar}\int\dd{v}PG_-(V))U[F]^\dagger}\\
  & = \exp(\frac{i}{\hbar}\qty(\int\dd{v}\qty(P(G_-(F)-G_+(F))+\int\dd{\mv} h F^{-1})))O_S\\
  & = \exp(\frac{i}{\hbar}\qty(\int\dd{\mv}\qty(\partial_\mv F^{-1}P\circ F^{-1}(G_--G_+)+hF^{-1})))O_S.
\end{align}
Recognizing $\partial_\mv F^{-1}P\circ F^{-1}$ and $F^{-1}$ as $\tilde\tau,X$ respectively (with the identifications $P\sim\Pi$ and $F\sim V$), and using the one-to-one correspondence between covariantly normal ordered observables and their symbols, we have~\eqref{Equation: generating function} as required.

\subsubsection{Covariant star product}

It is also worth noting that one may derive the covariant star product of any observables to all orders using the generating function~\eqref{Equation: generating function}. For simplicity let's set $O_S=1$ (system observables can be included similarly). Then, defining
\begin{equation}
  A[h,G] = \covnormord{\exp(\frac{i}{\hbar}\int\dd{\mv} \qty(\tilde\tau(G_--G_+)+hX))} =  \tilde{U}[G_+]\exp(\frac{i}{\hbar}\int\dd{\mv} h X)\tilde{U}[G_-^{-1}],
\end{equation}
where $G=G_+\circ (G_-)^{-1}$ and
\begin{equation}
  \tilde{U}[\exp(f\partial_\mv)] = \exp(\frac{i}\hbar\int\dd{\mv}\covnormord{\tilde\tau}f),
\end{equation}
one finds
\begin{multline}
  A[h,G]A[h',G']
  = A[(G_-^{-1}\circ G_+')_+\triangleright h + (G_-^{-1}\circ G_+')_+\triangleright h', G\circ G'] \\
  \exp(i\frac{c_{\tilde\tau}}{24\pi}\qty\Big(S(G)+S(G')-S(G\circ G')+C(G,G'))),
  \label{Equation: generating function composition}
\end{multline}
where $c_{\tilde\tau}$ is the central charge of the Virasoro representation generated by $\Cov{\tilde\tau}$,  $F\triangleright h = (\partial_vF)h\circ F$, and $S$ and $C$ are defined in Appendix~\ref{Appendix: cocycle}.
As a result, we can read off the following covariant star product:
\begin{multline}
  \exp(\frac{i}{\hbar}\int\dd{\mv} \qty(\tilde\tau(G_--G_+)+hX))
  \star
  \exp(\frac{i}{\hbar}\int\dd{\mv} \qty(\tilde\tau(G'_--G'_+)+h'X))\\
  =
  \exp(\frac{i}{\hbar}\int\dd{\mv} \qty(\tilde\tau\qty\Big((G\circ G')_--(G\circ G')_+)+\qty((G_-^{-1}\circ G_+')_+\triangleright h+(G_-^{-1}\circ G_+')_+\triangleright h')X))\\
  \exp(i\frac{c_{\tilde\tau}}{24\pi}\qty\Big(S(G)+S(G')-S(G\circ G')+C(G,G'))).
  \label{Equation: generating function covariant star product}
\end{multline}
The covariant star products of all other spin 0 observables can be obtained by taking functional derivatives of~\eqref{Equation: generating function covariant star product}.

\printbibliography

\end{document}